\documentclass[10pt]{article}
\pdfoutput=1

\usepackage[usenames,dvipsnames,svgnames,table]{xcolor}
\usepackage[obeyspaces,hyphens,spaces]{url}
\usepackage{jcapmod}

\usepackage{mathtools}
\usepackage{booktabs}
\usepackage[english]{babel}
\usepackage{amsmath,amssymb,amsbsy,amstext, amsthm, simplewick, amsfonts,braket}
\usepackage{hyperref}
\usepackage{graphicx}
\usepackage[small]{caption}
\usepackage{slashed}
\usepackage[separate-uncertainty=true,multi-part-units=single]{siunitx}
\usepackage{upgreek}
\usepackage{framed}
\usepackage{wrapfig}
\usepackage{multirow}
\usepackage{bbm}

\usepackage[title]{appendix}

\usepackage[utf8x]{inputenc}
\usepackage{selinput}
\usepackage{bm}
\usepackage{float}
\usepackage{geometry}
\usepackage{yfonts}
\usepackage{caption}
\usepackage{subcaption}
\usepackage{sidecap}
\usepackage{longtable}
\usepackage{anyfontsize}
\usepackage{dsfont}
\usepackage{tikz}
\usepackage{relsize}
\usepackage{tcolorbox}

\makeatletter
\newlength{\apb@width}
\makeatother


\allowdisplaybreaks[1]
\newcommand{\Eq}[1]{Eq.~\ref{eq:#1}}
\newcommand{\Sec}[1]{Sec.~\ref{sec:#1}}
\newcommand{\App}[1]{Appendix~\ref{app:#1}}
\newcommand{\Fig}[1]{Fig.~\ref{fig:#1}}

\newcommand{\be}{\begin{equation}}
\newcommand{\ee}{\end{equation}}
\newcommand{\nl}{\nonumber \\}
\newcommand{\np}{\vskip 4 pt}

\newcommand{\km}{\text{km}}

\newcommand{\V}{\text{V}}

\newcommand{\cm}{\text{cm}}
\newcommand{\TeV}{\text{TeV}}
\newcommand{\GeV}{\text{GeV}}
\newcommand{\MeV}{\text{MeV}}
\newcommand{\keV}{\text{keV}}
\newcommand{\eV}{\text{eV}}
\newcommand{\Gyr}{\text{Gyr}}
\newcommand{\K}{\text{K}}

\newcommand{\xvir}{x_{_\text{FTS}}}
\newcommand{\xth}{x_{_\text{LSS}}}

\newcommand{\jcr}{j_\text{cr}}
\newcommand{\jvir}{j_\text{vir}}
\newcommand{\nvir}{n_\text{vir}}
\newcommand{\Vwind}{V_\text{wind}}
\newcommand{\x}{\chi}

\newcommand{\qx}{q_\chi}
\newcommand{\muxN}{\mu_{\x N}}

\newcommand{\Ap}{A^\prime}
\newcommand{\mAp}{m_{A^\prime}}
\newcommand{\rhodm}{\rho_{_\text{DM}}}
\newcommand{\lmfp}{\ell_\text{mfp}}

\newcommand{\eps}{\epsilon}

\newcommand{\p}{\prime}

\newcommand{\grad}{\nabla}

\newcommand{\order}[1]{\mathcal{O}{(#1)}}

\newcommand{\bebox}{\begin{empheq}[box=\fcolorbox{light-gray}{light-gray}]{align}}
\newcommand{\eebox}{\end{empheq}}

\newcommand{\g}{\gamma}
\newcommand{\const}{\text{constant}}

\newcommand{\gv}{\boldsymbol{g}}

\newcommand{\jv}{\boldsymbol{j}}

\newcommand{\xv}{{\bf x}}
\newcommand{\vv}{{\bf v}}
\newcommand{\Vv}{{\bf V}}
\newcommand{\pv}{{\bf p}}

\begin{document}

\pagenumbering{roman}
\begin{titlepage}
\baselineskip=15.5pt \thispagestyle{empty}

\phantom{h}
\vspace{1cm}
\begin{center}
{\fontsize{15}{18}\selectfont  \bfseries  The Terrestrial Density of Strongly-Coupled Relics}
\end{center}

\vspace{0.5cm}
\begin{center}
{\fontsize{12}{22}\selectfont Asher Berlin,$^{1}$  Hongwan Liu,$^{2,3}$ Maxim Pospelov,$^{4,5}$ and  Harikrishnan Ramani$^{6}$} 
\end{center}

\vspace{-0.1cm}
\begin{center}
\vskip 5pt
\textit{$^1$ Theoretical Physics Division, Fermi National Accelerator Laboratory, Batavia, IL 60510, USA}

\vskip 8pt
\textit{$^2$  Center for Cosmology and Particle Physics, Department of Physics, \\ New York University, New York, NY 10003, USA}

\vskip 8pt
\textit{ $^{3}$ Department of Physics, Princeton University,  Princeton, New Jersey, 08544, USA} 

\vskip 8pt
\textit{$^4$  School of Physics and Astronomy, University of Minnesota, Minneapolis, MN 55455, USA}

\vskip 8pt
\textit{$^5$  William I. Fine Theoretical Physics Institute, School of Physics and Astronomy, \\ University of Minnesota, Minneapolis, MN 55455, USA}

\vskip 8pt
\textit{$^6$  Stanford Institute for Theoretical Physics, Stanford University, Stanford, CA 94305, USA}
\end{center}


\vspace{0.8cm}
\hrule \vspace{0.3cm}
\noindent {\bf Abstract}\\[0.1cm]
The simplest cosmologies motivate the consideration of dark matter subcomponents that interact significantly with normal matter. Moreover, such strongly-coupled relics may have evaded detection to date if upon encountering the Earth they rapidly thermalize down to terrestrial temperatures, $T_\oplus \sim 300 \ \K \sim 25 \ \text{meV}$, well below the thresholds of most existing dark matter detectors. This shedding of kinetic energy implies a drastic enhancement to the local density, motivating the consideration of alternative detection techniques sensitive to a large density of slowly-moving dark matter particles. In this work, we provide a rigorous semi-analytic derivation of the terrestrial overdensities of strongly-coupled relics, with a particular focus on  millicharged particles (MCPs). We go beyond previous studies by incorporating improved estimates of the MCP-atomic scattering cross section, new contributions to the terrestrial density of sub-GeV relics that are independent of Earth's gravitational field, and local modifications that can arise due to the cryogenic environments of precision sensors. We also generalize our analysis in order to estimate the terrestrial density of thermalized MCPs that are produced from the collisions of high-energy cosmic rays and become bound by Earth's electric field. 
\vskip10pt
\hrule
\vskip10pt


\end{titlepage}

\thispagestyle{empty}
\setcounter{page}{2}
\tableofcontents


\newpage
\pagenumbering{arabic}
\setcounter{page}{1}

\clearpage

\section{Introduction}

There is compelling gravitational evidence to conclude that Standard Model (SM) particles make up less than 5\% of the energy budget of the Universe. The rest is made up of dark matter (DM) and dark energy, both of whose fundamental nature are unknown despite decades of theoretical and experimental scrutiny. While this program has set stringent limits on non-gravitational interactions between DM/dark energy and the SM, current experimental data does not preclude the existence of cosmological relics (denoted as $\x$) that interact significantly with normal matter; accelerator experiments provide only a weak constraint on such couplings~\cite{Alexander:2016aln,Battaglieri:2017aum}, while cosmological observables are unable to constrain dark relics that make up only a small fraction of the DM abundance~\cite{Dubovsky:2003yn,dePutter:2018xte,Kovetz:2018zan,Buen-Abad:2021mvc}. In fact, thermal relics that are strongly-coupled to the SM should be produced with small abundances~\cite{Lee:1977ua}, providing strong theoretical motivation for such a scenario. Furthermore, provided that their mass $m_\x$ is not too large, strongly-coupled relics thermalize in the overburden consisting of Earth's atmosphere and crust  before reaching terrestrial direct detection experiments, such that their kinetic energy after thermalization, set roughly by room temperature $T_\oplus \sim 300 \ \K \sim 25 \ \text{meV}$, is far below the threshold of conventional direct detection experiments. 

\np
Upon thermalizing in Earth's environment, the $\x$ phase space is drastically modified compared to the virialized galactic population. In particular, the loss of kinetic energy typically implies a significant enhancement to the local terrestrial density. For instance, since $\x$'s thermal velocity $v_\text{th} \sim \sqrt{300 \ \K / m_\x} \sim 5 \times 10^{-6} \sqrt{1 \ \GeV / m_\x}$ is well below Earth's escape velocity $v_\text{esc} \simeq 4 \times 10^{-5}$ for $m_\x \gg 1 \  \GeV$, such particles remain gravitationally bound as a hydrostatic gas, accumulating over the entire age of the Earth $t_\oplus \simeq 4.5 \ \Gyr$ to extremely large densities~\cite{Neufeld:2018slx}. However, this hydrostatic piece is only one contribution to their total terrestrial density. An additional component arises as our solar system sweeps through the galactic halo, giving rise to a dynamic influx of $\x$'s currently entering Earth's environment~\cite{Wallemacq:2013hsa,Wallemacq:2014lba,Wallemacq:2014sta,Laletin:2019qca,Pospelov:2019vuf,Pospelov:2020ktu,Leane:2022hkk}. This component has not yet reached hydrostatic equilibrium and is thus conceptually distinct from the accumulated density of gravitationally-bound particles. Although this population does not accumulate in time and so is not enhanced in its density by the large timescale $t_\oplus$, conservation of flux implies a large local overdensity that can exist even in the absence of Earth's gravitational field. As we show below, the evolution of particles in this terrestrial ``traffic jam'' is well-described by the physics of diffusion as a gravitationally-biased random walk. Although the total number of such particles currently flowing into Earth's atmosphere is potentially minuscule compared to the accumulated population, the former is relevant for particles of any mass and in fact may dominate over the hydrostatic piece near Earth's surface across a large range of model space. 

\np
In this work, we provide a model-independent framework to determine the terrestrial population of strongly-coupled relics. The formation of large terrestrial overdensities of strongly-coupled relics has been investigated in the literature previously. For instance, the gravitationally-bound hydrostatic population was considered in Ref.~\cite{Neufeld:2018slx}, while the dynamic traffic jam contribution was investigated in, e.g., Refs.~\cite{Wallemacq:2013hsa,Wallemacq:2014lba,Wallemacq:2014sta,Laletin:2019qca,Pospelov:2019vuf,Pospelov:2020ktu,Leane:2022hkk}. As we will show below, we go beyond previous calculations in several ways: 1) we include additional contributions to the terrestrial density of sub-GeV particles in the well-motivated case that the scattering cross section is enhanced at low momentum transfer, 2) we provide simple semi-analytic expressions that can be used to determine the detailed spatial density profile of the traffic jam overdensities, and 3) we incorporate local effects stemming from the presence of laboratory detectors and inhomogeneities in the density of normal matter. This formalism can also be generalized to incorporate additional effects arising from Earth's electromagnetic fields, which may be relevant for certain models of millicharged relics. 

\np
A natural class of strongly-coupled particles that quickly thermalize to terrestrial temperatures---thus evading limits from underground detectors---are those in which the interaction is mediated by a light force-carrier. In particular, upon encountering Earth's crust, the mean free path of $\x$ is much smaller than a kilometer---the characteristic depth of underground detectors---only if the mass of this force-carrier is much below $\order{1} \ \GeV$. However, the existence of a new light mediator is typically constrained to be extremely feebly coupled to the SM by, e.g., direct searches for new forces at low-energy accelerators, fifth-force experiments, and considerations of stellar cooling~(see, e.g., Ref.~\cite{Knapen:2017xzo} and references therein). An exception to this arises if $\x$ feebly couples to the SM through a \emph{known} long-range force, such as electromagnetism. This naturally arises if the DM-SM interaction is mediated by an ultralight, kinetically-mixed dark photon, since direct bounds on dark photons decouple in the small-mass limit~\cite{Holdom:1985ag}. In this case, $\x$ appears as effectively charged under electromagnetism over length scales smaller than the dark photon Compton wavelength, which can be macroscopic. Even for small effective charges $q_\x \ll 1$, such millicharged particles (MCPs) can efficiently scatter with nuclei many times in Earth's atmosphere or crust, shedding their kinetic energy before encountering underground detectors~\cite{Emken:2019tni}. For this reason, we will focus exclusively on MCPs in this work, although our formalism can be easily adapted to any theory of strongly-coupled relics. Compared to previous analyses~\cite{Pospelov:2020ktu}, we find qualitative agreement in the formalism developed to calculate the local overdensity of MCPs with mass $m_\x \gg 1 \ \GeV$, although we show below that a careful treatment of $\x$-nuclear scattering leads to an overdensity a few orders of magnitude larger than previously estimated.  For MCPs of mass $m_\x \ll 1 \ \GeV$, we also point out a new contribution to the terrestrial density that leads to large overdensities even for MCPs so light that Earth's gravitational field can be safely ignored. The effect of ``thermal diffusion,'' pointed out in the context of DM in Ref.~\cite{Gould:1989hm} and reemphasized recently in Ref.~\cite{Leane:2022hkk}, is also explicitly included in our calculations. We find that this leads to a relatively minor numerical change in the local distribution of strongly-coupled relics.

\np
We also generalize our formalism to determine the terrestrial density of sub-GeV MCPs that are produced from the collision of high-energy cosmic rays in the atmosphere and then rapidly thermalize in Earth's crust. This constitutes an irreducible contribution to the local MCP density independent of any additional cosmological population generated in the early Universe. We demonstrate that for certain classes of MCP models, the atmospheric voltage can efficiently trap such MCPs, potentially leading to irreducible densities as large as $n_\x \sim 1 \ \cm^{-3}$ for certain ranges of masses and couplings. Throughout most of this work, we focus predominantly on MCP cosmological relics. It will be stated explicitly when we generalize our analysis to include MCPs produced from cosmic rays.

\np
\paragraph{Outline} The outline of the paper is as follows: Secs.~$\ref{sec:overview} - \ref{sec:results}$ focus exclusively on the terrestrial abundance of millicharged relics. In particular, \Sec{overview} begins by reviewing the relic MCP model space and presenting a conceptual overview of the development of large terrestrial overdensities, in order to provide intuition for our main results. In \Sec{scattering}, we briefly outline details relevant for determining terrestrial MCP-nuclear scattering, such as the calculation of the transfer cross section and modeling of the terrestrial environment. In \Sec{EMfields}, we discuss model-dependent modifications to our analysis that can arise from the effect of Earth's electric and magnetic fields. In \Sec{formalism}, we introduce the general formalism used to evaluate overdensities. In \Sec{static}, we review the accumulation of the gravitationally-bound hydrostatic population. In \Sec{traffic}, we apply our general formalism to the dynamic traffic jam population. In \Sec{results}, we present our final estimates for the terrestrial density of MCP relics. In \Sec{cosmicray}, we generalize this analysis to include MCPs that are produced from the collisions of high-energy cosmic rays. Finally, we state our conclusions and discuss future lines of research in \Sec{conclusion}. We also provide an appendix that contains additional technical details.

\section{Model Space and Conceptual Overview}
\label{sec:overview}

%
\begin{figure}[t]
\centering
\includegraphics[width=0.9\columnwidth]{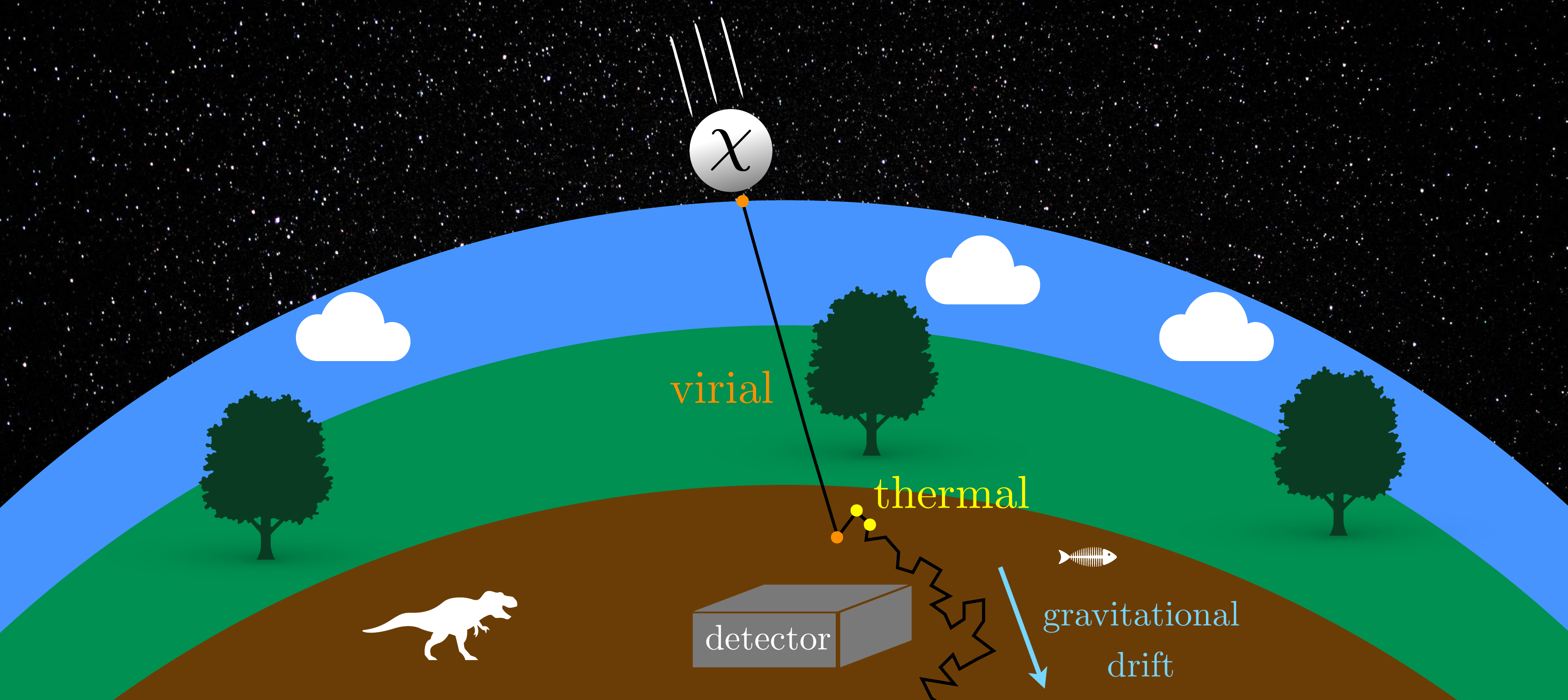}
\caption{A cartoon depiction of a millicharged relic particle $\x$ entering Earth's atmosphere with an initially virialized effective temperature  $T_\text{vir} \sim 10^{-6} \, m_\x$. Before reaching an underground detector, $\x$ sheds its kinetic energy by scattering with nuclei in Earth's atmosphere and crust, thermalizing down to terrestrial temperatures $T_\oplus \sim 300 \ \K$. Since the interaction rate is strongly-peaked towards smaller momentum transfer, thermalization significantly shortens the mean free path of $\x$. After thermalizing, $\x$ diffuses throughout the Earth and may become gravitationally bound depending on its mass.}
\label{fig:cartoon}
\end{figure}

MCPs possess a small effective charge $\qx \ll 1$ under standard electromagnetism. Most classes of MCP models fall under one of two categories. In the first category, small charges arise \emph{indirectly} if $\x$ has a dark charge  $Q_\x \sim 1$ under a dark photon $\Ap$ that has a small kinetic mixing with the SM photon. In this case, $\x$ couples to normal electromagnetism on length scales smaller than the dark photon's Compton wavelength with an effective charge $\qx = \eps \, e^\p \, Q_\x  / e$, where $\eps \ll 1$ is the small kinetic mixing parameter, $e^\p$ is the $\Ap$ gauge coupling, and $e$ is the SM electric charge~\cite{Holdom:1985ag}. We will refer to such particles as ``effective MCPs." Alternatively, $\x$ may \emph{directly} couple to the SM photon with a small electric charge on all length scales. Although, this second possibility appears unnatural from top-down considerations of gauge-coupling unification, it is technically natural to posit the existence of new particles with arbitrarily small electric charge. We will refer to such particles in this latter scenario as ``pure MCPs." 

\np
Over a large range of masses and couplings, relic MCPs $\x$ rapidly thermalize to terrestrial temperatures upon scattering with nuclei in Earth's environment (this will be made more precise in later sections). Here, we briefly describe the effects that thermalization has on the terrestrial overdensities of strongly-coupled relics. After thermalizing, $\x$ diffuses throughout the Earth, which can be modeled as a random walk biased in the direction of Earth's gravitational field, as shown schematically in \Fig{cartoon}. The ultimate fate of such particles depends on their mass $m_\x$. For $m_\x \gg 1 \ \GeV$, an $\order{1}$ fraction of the incoming galactic flux becomes gravitationally bound. Such particles thus contribute to two distinct terrestrial populations: a gradually accumulating population in hydrostatic equilibrium, as well as a dynamic in-flowing (or ``traffic jam") population as new particles are constantly raining in from the galactic halo. Alternatively, MCPs much lighter than $1 \ \GeV$ do not remain gravitationally bound and thus solely contribute to the traffic jam population. Since the hydrostatic overdensity has already been worked out in detail in Ref.~\cite{Neufeld:2018slx}, we briefly review it in \Sec{static} and continue in this section with an introduction to the traffic jam population which provides a qualitative explanation of this phenomenon, leaving a more detailed explanation to Sec.~\ref{sec:traffic}.

\np
A broad understanding of the traffic jam overdensity follows straightforwardly from conservation of flux. In Earth's frame, approximating Earth's surface as extending across the $y-z$ plane, we characterize the incoming bulk flow of galactic virialized (i.e., not-yet-thermalized to terrestrial temperatures) MCPs as $\jvir \sim \nvir \, \Vwind $, where $\nvir$ is the number density of MCPs in the galaxy. The velocity of virialized DM should be calculated as an average incoming velocity transverse to Earth's surface, i.e., $\Vwind = \int_{v_x>0} d^3 \vv \, v_x\, f_\x (\vv) $, where the DM velocity distribution $f_\x (\vv)$ is approximately a Maxwell-Boltzmann truncated at the galactic escape velocity and boosted by Earth's relative velocity $\sim 10^{-3}$ through the Milky Way halo. Therefore, $\Vwind $ varies as a function of time and geographical position, but we ignore these subleading effects. At a sufficient depth below Earth's atmosphere, most strongly-coupled MCPs thermalize to terrestrial temperatures. In this region, the MCP flux is solely determined by the combined effect of Earth's gravitational field and diffusion. In particular, $\x$ is biased to drift radially inwards, due to the attractive pull of Earth's gravitational field $\gv$. This attraction is resisted by the friction imparted by rapid collisions with surrounding nuclei $N$. Diffusion, on the other hand, is dictated simply by the temperature and collision rate between $\x$ and the surrounding medium, and biases the flow of particles against large density and temperature gradients. As discussed below in \Sec{formalism}, in the absence of large temperature gradients, both effects combine to yield a thermalized flow given by the bulk number-current density
\be
\label{eq:jx1}
\jv_\x \simeq n_\x \, \Vv_g - D_\x \, \grad n_\x
~,
\ee
where $n_\x$ is the terrestrial number density of $\x$, $\Vv_g$ the gravitational drift velocity (precisely the terminal velocity of $\x$, with gravitational and collision-induced frictional forces balancing out), and $D_\x$ the diffusion coefficient. For thermalized MCPs diffusing with a characteristic step size $\ell_\oplus$,\footnote{$\ell_\oplus$ is the typical length traversed by an MCP before exchanging an $\order{1}$ fraction of its momentum, i.e., the distance it travels before turning around. This is in general distinct from the mean free path, the latter being the typical length traversed between scattering events.} $V_g$ and $D_\chi$ are parametrically set by $V_g \sim g \, \ell_\oplus \, / \, \bar{v}_0$ and $D_\x \sim  \ell_\oplus \, \bar{v}_0$, where $\bar{v}_0 \sim \sqrt{T_\oplus / m_\x + T_\oplus / m_N}$ is the relative thermal velocity between $\x$ and nuclei $N$ at terrestrial temperature $T_\oplus$. 

\begin{figure}[t]
\centering
\includegraphics[width=0.9\columnwidth]{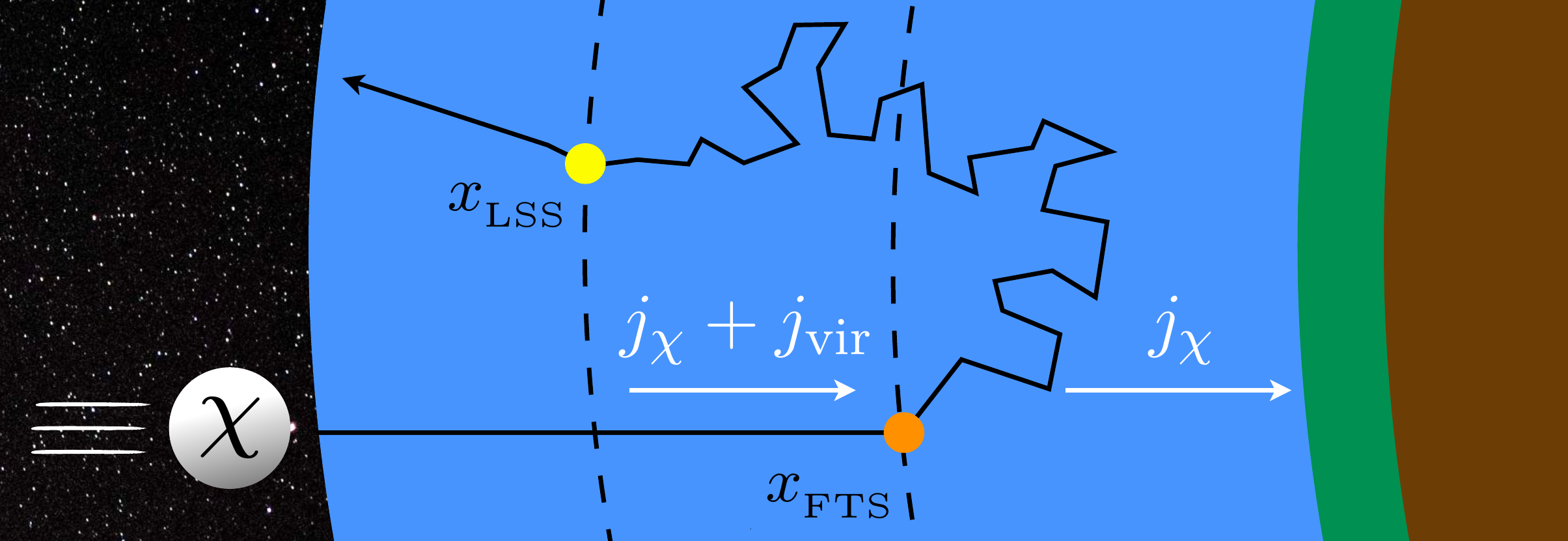}
\vspace{0.3cm}
\caption{A schematic cartoon outlining the journey of a millicharged relic particle $\x$ thermalizing in Earth's environment. A virialized MCP $\x$ enters Earth's atmosphere from the left, traveling to the point $\xvir$ until it suddenly thermalizes with terrestrial matter (orange dot). At terrestrial temperatures, the mean free path of $\x$ is greatly reduced as it diffuses throughout the Earth. Once an outwardly traveling thermalized MCP passes a surface located at $\xth$ (yellow dot), it is able to free stream out of the atmosphere or remains gravitationally or electrically bound and begins the entire process again, depending on the MCP mass and charge. At positions to the left of $\xvir$, the flow of MCPs is governed by both the galactic flux $\jvir$ as well as the thermal diffusion flux $j_\x$. The boundary condition involving $j_\x$ and $j_\text{vir}$ imposed at $\xvir$ and $\xth$ will be discussed later in \Sec{traffic}. In particular, at positions to the right of $\xvir$, the flow is solely governed by the thermal diffusion flux $j_\x$.}
\label{fig:current}
\end{figure}

\np
The resulting terrestrial density $n_\x$ can be determined by demanding continuity of MCP number. In the low-mass limit ($m_\x \ll 1 \ \GeV$), Earth's gravitational field plays little role, and so the first term of \Eq{jx1} can be ignored. Furthermore, we assume that upon encountering Earth's environment, a galactic MCP travels at a speed $\Vwind$ before cooling down to a temperature of $T_\oplus \sim 300 \ \K$ upon reaching the ``first thermalization surface" located at $\xvir$ (see \Eq{xstart}). After thermalizing, $\x$ then remains in the Earth until it diffuses outwards past the ``last scattering surface" located at $\xth$ (see \Eq{xth}). Here, the positions $\xvir$ and $\xth$ mark the beginning and ending of the MCP's random walk through the Earth and are shown as the orange and yellow circles in \Fig{current}, respectively (see \Sec{draglength} for a more precise definition). Hence, escaping the Earth involves diffusing outward  a distance of $\xvir - \xth$, which takes a time of roughly $t _\text{diff} \sim (\xvir - \xth)^2 / D_\x$. Hence, while the rate of deposited MCPs is $\jvir$, the outgoing diffusion flux diminishing the terrestrial density is $\sim n_\x \, (\xvir - \xth) / t_\text{diff} \sim n_\x \, D_\x / (\xvir - \xth)$. Steady state is reached after a time $t \gg t_\text{diff}$ when the terrestrial density $n_\x$ builds up to a point at which the outgoing flux (i.e., the rate at which thermalized MCPs diffuse from $\xvir$ to $\xth$) is comparable to the incoming galactic flux (i.e, the rate that MCPs are deposited at $\xvir$), which occurs for a terrestrial overdensity of
\be
\label{eq:overdensity1}
\lim_{m_\x \ll \GeV} ~ \frac{n_\x}{\nvir} \sim \frac{\Vwind \, (\xvir - \xth)}{D_\x} \sim \frac{\Vwind}{\bar{v}_0} ~ \frac{\xvir - \xth}{\ell_\oplus}
~.
\ee
Note that the characteristic step size is much smaller at lower temperatures (i.e., $\ell_\oplus \ll \xvir - \xth$) since the MCP-nuclear scattering rate is enhanced at small momentum transfer (see \Fig{sigmaT} below). Hence, both factors in the last equality of \Eq{overdensity1} contribute to a large terrestrial MCP overdensity. 

\np
On the other hand, in the large-mass limit ($m_\x \gg 1 \ \GeV$), the first term in \Eq{jx1} dominates and the MCP density is simply dictated by equating $\jvir$ and  $\sim n_\x \, V_g$, which gives
\be
\label{eq:overdensity2}
\lim_{m_\x \gg \GeV} ~ \frac{n_\x}{\nvir} \sim \frac{\Vwind}{V_g} 
~.
\ee
As an example, for MCP masses and couplings of $m_\x \lesssim 10 \ \TeV$ and $\qx \gtrsim 10^{-9}$, the gravitational drift velocity $V_g \lesssim 10^{-6}$ is much smaller than $V_\text{wind} \sim 10^{-3}$ (see \Sec{formalism}), implying a density $n_\x$ many orders of magnitude larger than the galactic density. 

\np
In later sections, we will carefully derive Eqs.~\ref{eq:overdensity1} and \ref{eq:overdensity2} in more detail. At that point, we will also see how to generalize these results to the case where the density or temperature of terrestrial nuclear scatterers (and, hence, the MCP scattering rate) are not uniform in space. First, though, in Secs.~\ref{sec:scattering} and \ref{sec:EMfields} we briefly discuss the determination of the scattering rate between MCPs and terrestrial atoms and the effect of MCP interactions with Earth's electromagnetic fields, respectively.

\section{Terrestrial Scattering}
\label{sec:scattering}

In this section, we evaluate the MCP-atomic momentum-transfer scattering cross section $\sigma_T$ at an MCP temperature $T_\x$ before and after thermalization, and we discuss other inputs to determine the scattering rate, such as terrestrial compositions, densities, and temperatures in the atmosphere, crust, mantle, and core. The crust, mantle, and core are described using the preliminary reference Earth model (PREM)~\cite{Dziewonski:1981xy}, incorporating the most abundant elements: oxygen, silicon, aluminum, and iron. For the atmosphere consisting mostly of oxygen, nitrogen, and helium, we employ the MSISE model~\cite{atmosphere}.  In our analysis, we only incorporate MCP-atomic nuclear elastic scattering since this has been found to dominate over other processes such as electronic recoils and Rutherford scattering with rare ions/electrons~\cite{Emken:2019tni}. We also make the simplifying approximation to treat the atoms of terrestrial molecules as individual target scatterers. We do not expect these approximations to introduce more than an $\order{1}$ error in our final results. 

\np
To account for the screening of the nuclear charge by electrons in terrestrial atoms, we follow Ref.~~\cite{Emken:2019tni} and model the electrostatic potential of the nucleus $N$ using the Thomas-Fermi model. In the case that the $\Ap$ is much longer-ranged than the size of the atom, the MCP interacts with the modified nuclear Coulomb potential $(Z e / r)  \, e^{-r / a_Z}$, where $Z$ is the atomic number of the nucleus, $a_Z$ the Thomas-Fermi radius $a_Z = ( 9 \pi^2 / 2 Z)^{1/3} (a_0 / 4)$, and $a_0 = 1/ (\alpha_\text{em} \, m_e)$ the Bohr radius. In certain limits, further approximations can be made to simplify the calculation of the scattering cross section. For instance, in the perturbative ($\alpha_\text{em} \, Z \, \qx \, \muxN \, a_Z \ll 1$) or large momentum transfer ($\muxN \, v_\text{rel} \, a_Z \gg 1$) regimes, the Born or classical approximations apply, respectively, where $\muxN$ is the MCP-nuclear reduced mass and $v_\text{rel}$ the relative velocity. However, no such approximation is valid across the entirety of the mass, coupling, and momentum-transfer ranges considered in this work. We thus choose to employ semi-analytic solutions to the Schr\"{o}dinger equation incorporating the Thomas-Fermi potential, adapting the results for the transfer cross section $\sigma_T$ from Refs.~\cite{Tulin:2013teo,Colquhoun:2020adl}.\footnote{These studies quote results for  scattering through a potential $(\alpha_X / r)  \, e^{-m_\phi \, r}$ within the context of self-interacting DM $X$. In adapting these results to MCP-nuclear scattering, we therefore make the replacements $m_\phi \to a_Z^{-1}$, $\alpha_X \to Z \, \alpha_\text{em} \, \qx$, and $m_X \to 2 \muxN$.} 

\begin{figure}[t]
\centering
\includegraphics[width=0.6\columnwidth]{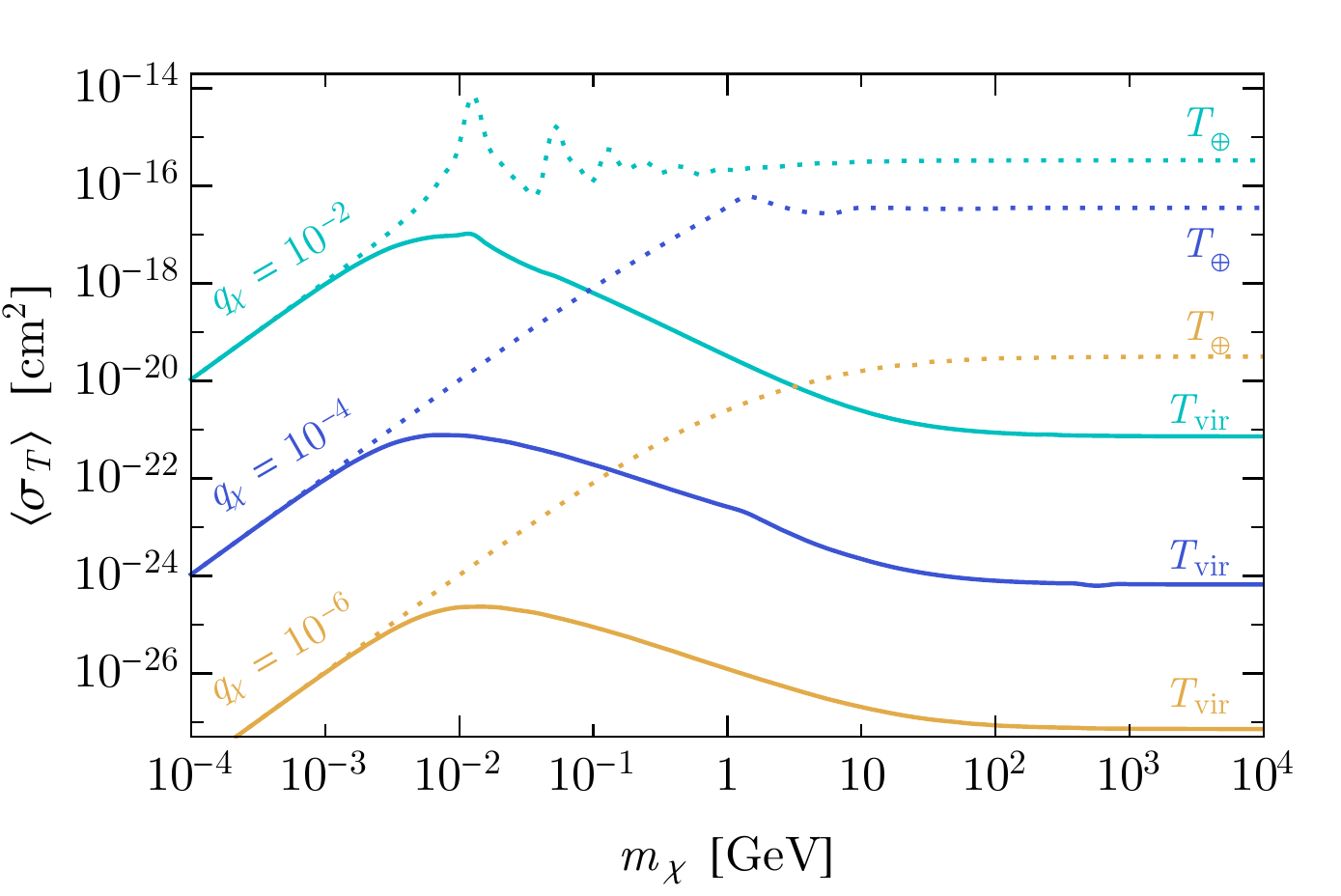}
\caption{The thermally-averaged transfer cross section $\langle \sigma_T \rangle$ for MCP scattering off of room-temperature nitrogen (averaged between positively- and negatively-charged MCPs), as a function of the MCP mass $m_\x$ for various values of the effective charge $q_\x$. The solid lines take the MCP phase space to track that of the virialized galactic population, corresponding to MCPs that have not yet thermalized with the Earth, whereas the dotted lines assume that MCPs have equilibrated to room temperature $T_\oplus \simeq 300 \ \K$.}
\label{fig:sigmaT}
\end{figure}

\np
\Fig{sigmaT} shows the thermally-averaged transfer cross section $\langle \sigma_T \rangle$ for $\x$ scattering off of a nitrogen atom (the most abundant element in the atmosphere), as a function of the MCP mass $m_\x$ and for different choices of charge $\qx$ and MCP temperature, using the full semi-analytic results as given in Refs.~\cite{Tulin:2013teo,Colquhoun:2020adl}. In this section, we average the scattering rate for positively- and negatively-charged MCPs, in order to simplify the presentation. However, in our final numerical results, we separately present the MCP density for either charge. The brackets correspond to a thermal average over the $\x-N$ relative velocity $v_\text{rel}$~\cite{Dvorkin:2013cea,Boddy:2018wzy,Becker:2020hzj,Dvorkin:2020xga} 
\be
\label{eq:brackets}
\langle \cdots \rangle \equiv \int_0^\infty dv_\text{rel} ~ (\cdots) ~ \sqrt{\frac{2}{\pi}} \, \frac{v_\text{rel}^2}{\bar{v}_0^3} \, e^{-v_\text{rel}^2 / 2 \bar{v}_0^2}
~~,~~
\bar{v}_0 \equiv \sqrt{T_\x / m_\x + T_N / m_N}
~.
\ee
In \Fig{sigmaT}, the solid lines show the transfer cross section for an MCP phase space that tracks the galactic distribution, i.e., with an effective temperature of $T_\text{vir} \simeq m_\x \, V_\text{wind}^2 / 3$, whereas the dotted lines assume that the MCPs have equilibrated to room temperature $T_\oplus \simeq T_N \simeq 300 \ \K$.  

\np
For $m_\x \ll 1 \ \GeV$, the momentum transfer in a single scatter is very small and is thus in the ``quantum regime"~\cite{Tulin:2013teo}. In this case, when solving the Schr\"{o}dinger equation the interaction potential is often approximated with the so-called Hulth\'en potential, since it admits simple analytic solutions. The transfer cross section is then determined to be~\cite{Tulin:2013teo} 
\be
\lim_{m_\x \ll \GeV} \langle \sigma_T \rangle \sim \min{\bigg( \frac{30 \, Z^{2/3} \, \qx^2 \, \muxN^2}{\alpha_\text{em}^2 \, m_e^4} ~,~ \frac{4 \pi}{\muxN^2 \, \bar{v}_0^2}  \bigg)}
~.
\ee
Note that in the expression above, the rate is independent of the characteristic relative thermal velocity $\bar{v}_0$ for small charge $q_\x$ or mass $m_\x$, which arises from the screening of the nucleus at small momentum transfer, as characterized by the Thomas-Fermi potential. This is exemplified by the fact that the solid and dotted lines in \Fig{sigmaT} (corresponding to different MCP temperatures and thus different relative velocities) approximately coincide for small MCP masses. Alternatively, for $m_\x \gg 1 \ \GeV$ the momentum transfer is enhanced and is in the ``classical regime"~\cite{Colquhoun:2020adl}. In this case, depending on the coupling strength the limiting forms of $\langle \sigma_T \rangle$ are parametrically~\cite{Colquhoun:2020adl} 
\be
\lim_{m_\x \gg \GeV} \langle \sigma_T \rangle \sim 
\begin{cases}
\frac{10 \, Z^2 \, \alpha_\text{em}^2 \, \qx^2}{\muxN^2 \,  \bar{v}_0^4}  & ,~  Z \, \alpha_\text{em} \, \qx / a_Z \ll \muxN \, \bar{v}_0^2 
\\
\frac{10}{\alpha_\text{em}^2 \, Z^{2/3} \, m_e^2} \, \log{\Big( \frac{Z \, \alpha_\text{em} \, \qx}{a_Z \, \muxN \, \bar{v}_0^2} \Big)}  & ,~ Z \, \alpha_\text{em} \, \qx / a_Z \gg \muxN \, \bar{v}_0^2
~.
\end{cases}
\ee
The first line in the expression above implies that for small coupling, $\sigma_T$ is significantly enhanced at small momentum transfer (as can be seen by comparing the dotted and solid lines for $m_\x \gg 1 \ \GeV$ in  \Fig{sigmaT}). Instead, for large coupling, $\sigma_T$ has a mild logarithmic growth with increasing $q_\x$. This is exemplified in \Fig{sigmaT} when comparing the results for $q_\x = 10^{-4}$ and $q_\x = 10^{-2}$; for large masses and small relative velocities (corresponding to the dotted lines in \Fig{sigmaT}), there is only a mild enhancement in $\langle \sigma_T \rangle$. 

\np
Later in \Sec{formalism}, this scattering rate is incorporated into the fundamental equations governing the evolution of the MCP phase space. In this work, we ignore model-dependent contributions arising from MCP self-interactions. In models of pure MCPs, $\x - N$ scattering is enhanced relative to $\x-\x$ scattering by $Z / q_\x \gg 1$ and the larger density of terrestrial matter. However, in models of effective MCPs, self-scattering can proceed via $\Ap$ exchange. As an example, fixing $e^\prime = e$ and $q_\x \gtrsim 10^{-7}$, we find that $\x -\x$ scattering dominates over $\x - N$ scattering in the crust for $n_\x \gtrsim 10^{10} \ \cm^3$. Since we only ever calculate the relative terrestrial overdensity $n_\x / \nvir$, we implicitly assume that $\nvir$ (and hence $n_\x$) are sufficiently small such that $\x - N$ interactions completely dominate the evolution of the terrestrial MCP population.

\section{Terrestrial Electromagnetic Fields}
\label{sec:EMfields}

In addition to scattering off of normal matter, MCPs may interact with electromagnetic fields sourced by the Earth, Sun, and Milky Way. Interactions with environmental magnetic fields and the solar wind have been considered before~\cite{Chuzhoy:2008zy,Sanchez-Salcedo:2010gfa,Dunsky:2018mqs,Emken:2019tni,Stebbins:2019xjr,Li:2020wyl}. We leave the incorporation of extraterrestrial electromagnetic fields in estimating the terrestrial population to future work. We will assume that these effects can either be incorporated by rescaling $\nvir$ or that they can be entirely ignored due to model-dependent effects that we consider below. 

\np
In this work, we instead focus on the electromagnetic fields present on Earth. These are Earth's magnetic field $B_\oplus \sim 0.5 \ \text{G}$ sourced by its fluid outer core and the radially inward electric field $E_\oplus \sim 1 \ \V / \cm$ present between the crust and ionosphere as a result of the large potential difference $V_\oplus \sim 0.5 \ \text{MV}$ relative to the top of the atmosphere. The voltage difference between Earth's conducting crust and ionosphere layers is maintained by the insulating intermediate region of air in Earth's lower atmosphere. However, the small conductivity of air leads to a radially inward discharging current $I_\oplus \sim 2 \ \text{kA}$ of positively-charged SM ions. This discharging current is approximately offset by the average occurrence of lightning, which acts to recharge the crust and maintain the large atmospheric voltage across the global scale. This mechanism is referred to as the ``global atmospheric electrical circuit" (GAEC)~\cite{Feynman:1494701,Harrison2004,SIINGH200791}. 

\addtocontents{toc}{\protect\setcounter{tocdepth}{1}}
\subsection{Pure MCPs}
\addtocontents{toc}{\protect\setcounter{tocdepth}{2}}
\label{sec:EMfieldsPure}

For models in which MCPs directly couple to electromagnetism across all length-scales (the so-called ``pure MCPs" discussed in \Sec{overview}), terrestrial electromagnetic fields may significantly impact the trajectories and retention of incoming MCP relics on Earth. For instance, the incoming flux of MCPs from the galactic halo is significantly modified if their corresponding gyroradius $r_g \sim (m_\x \Vwind) / (e q_\x B_\oplus)$ is much smaller than Earth's radius $R_\oplus \simeq 6.4 \times 10^3 \ \km$, which is the case for millicharges of $q_\x \gg 10^{-5} \times (m_\x / \GeV)$. For such couplings, galactic MCPs cannot efficiently penetrate into Earth's environment near the equator, but still may gather near the poles, analogous to the Van Allen radiation belt of solar wind particles~\cite{Emken:2019tni}.

\np
Pure MCPs also directly couple to the atmospheric electric field, which can impact the terrestrial accumulation by either accelerating or decelerating positively- or negatively-charged MCPs, respectively. For instance, any thermalized negatively-charged MCPs are ejected from the lower atmosphere for $e q_\x \, V_\oplus \gg T_\oplus \sim 300 \ \K$, corresponding to $q_\x \gg 10^{-7}$. On the other hand, for such couplings positively-charged MCPs become electrically bound to the Earth and remain below the crust. In this case, the incoming flux of positively-charged MCPs amounts to an additional radially inward electrical current $I_\x \sim e q_\x \,(\jvir/2) \, 4 \pi R_\oplus^2$ on Earth. Thus, the existing limit on anomalous charge transport through the atmosphere (i.e., demanding that $I_\x \lesssim I_\oplus$) forbids charges
\be
\label{eq:voltageadiabatic}
q_\x \gtrsim \max \big[ \, 10^{-7} \, , \, 5 \times 10^{-4} \times (m_\x / \GeV) \, (\rhodm / \rho_\x) \, \big]
~,
\ee
 where $\rho_\x$ and $\rhodm$ are the galactic MCP and DM energy densities, respectively. For couplings smaller than \Eq{voltageadiabatic}, the incoming adiabatic flux of MCPs does not modify the GAEC and any backreaction on Earth's electric field is negligible.  In other words, this millicharge separation is compensated for by an opposing SM charge separation, with a large enough mismatch to maintain the global electric field. Conversely, \Eq{voltageadiabatic} constitutes a bound (although one that is weaker than direct limits on MCPs), since MCP couplings greater than this value would significantly modify the mechanism at work that maintains the atmospheric voltage. 

 \addtocontents{toc}{\protect\setcounter{tocdepth}{1}}
\subsection{Effective MCPs}
\addtocontents{toc}{\protect\setcounter{tocdepth}{2}}
\label{sec:EMfieldsAprime}

The situation is modified if MCPs \emph{indirectly} couple to electromagnetism, as in models of ``effective millicharge." For instance, when such interactions are mediated by an ultralight kinetically-mixed dark photon, the MCP trajectory is dictated by the dark magnetic field sourced by the Earth, whose strength is exponentially suppressed if the dark photon's Compton wavelength is shorter than an Earth radius, corresponding to a dark photon mass of $\mAp \gtrsim 10^{-14} \ \eV$ (in this case, the solar and galactic dark magnetic field are also suppressed). Also demanding that the dark photon is sufficiently light to not alter the low-momentum transfer enhancement associated with MCP-nuclear scattering at terrestrial temperatures (see \Sec{scattering}), we restrict our consideration to $\mAp \lesssim \sqrt{\muxN \, T_\oplus} \sim 5 \ \keV  \times (\muxN / \GeV)^{1/2}$. Direct bounds on the existence of dark photons in this mass range $10^{-14} \ \eV \lesssim \mAp \lesssim 10 \ \keV$ restrict the kinetic mixing parameter to be $\eps \lesssim 10^{-6}$~\cite{An:2013yfc,Redondo:2013lna,Williams:1971ms,Mirizzi:2009iz,Caputo:2020bdy}. Thus, in these models, for the largest charges that we consider in the sections below ($q_\x \sim 10^{-2}$), we require $e^\p \, Q_\x \,  = e \, q_\x / \eps \sim 10^{3}$, where $e^\p$ is the dark gauge coupling and $Q_\x$ is the dark charge of $\x$. Hence, the effect of Earth's magnetic field can be screened throughout the entire parameter space of interest if $\x$ is a dark composite state with a correspondingly large charge under the dark photon.

\np
The effect of Earth's electric field is rendered completely negligible if the MCP interaction is mediated by a dark photon, independent of the dark photon's mass. To see this, let's consider the case that the $\Ap$ is exactly massless, in which case SM charged particles source solely the visible massless SM photon, whereas $\x$ couples to both the massless visible and invisible photons~\cite{Holdom:1985ag}. For such a model, positively-charged MCPs accumulate in Earth's crust until they source a dark electric field  of opposite sign that couples to the incoming MCPs just as strongly as the normal terrestrial electric field. This occurs once $N_+ \sim (\eps / e^\p) \, E_\oplus \, R_\oplus^2$ positively-charged MCPs have accumulated on Earth. The timescale for this dark-electric-shorting to occur is typically much smaller than the age of the Earth,
\be
t_\text{short} \sim (\eps / e^\p) \, E_\oplus / \jvir \sim 10^{-2} \ \text{sec} \times (m_\x / \GeV) \, (\eps / e^\p) \,  (\rhodm / \rho_\x) 
~.
\ee
Since $\eps \ll e^\p$, the resulting number density of electrically-bound MCPs $n_+ \sim (\eps / e^\p) \, E_\oplus / R_\oplus \sim 10^{-4} \ \cm^{-3} \times (\eps / e^\p)$ is much smaller than the terrestrial densities that we consider below and has little effect on Earth's normal electric field. Thus, in models involving kinetically-mixed dark photons, the effect of Earth's dark electric field is quickly neutralized. 

\np
From the discussion above, we see that the role of Earth's electromagnetic fields is highly model-dependent. In our analysis below, we investigate various cases. For models of effective MCPs, we take the dark photon mass to be sufficiently heavy such that the strength of local electromagnetic fields is exponentially screened. Alternatively, in models of pure MCPs, we incorporate the effect of the terrestrial electric field in binding positively-charged MCPs to the Earth. In this case, additional effects stemming from the terrestrial magnetic field are assumed to be implicitly incorporated by rescaling the incoming MCP flux, set by the density $n_\text{vir}$.

\section{Strongly-Coupled Fluids and Random Walks}
\label{sec:formalism}

After determining how strongly-coupled relics interact with the terrestrial environment, we can calculate the resulting modifications to their phase space. At the coarse-grained level, the dense terrestrial population of MCPs is most economically described as a classical fluid. In Secs.~\ref{sec:genformalism} and \ref{sec:draglength}, we write down the equations governing this MCP fluid and show how at the particle level this is equivalent to a gravitationally-biased random walk. The starting and ending positions of this random walk are dictated by the thermalization and momentum-transfer drag lengths, which are calculated in \Sec{draglength}. This general formalism will be used in following sections to determine the terrestrial MCP density.

\addtocontents{toc}{\protect\setcounter{tocdepth}{1}}
\subsection{General Formalism for a Strongly-Coupled Fluid}
\addtocontents{toc}{\protect\setcounter{tocdepth}{2}}
\label{sec:genformalism}

A tightly-coupled MCP fluid is completely specified in terms of its number density $n_\x$ and bulk velocity $\Vv_\x$.\footnote{We use notation where capitalized vectors correspond to bulk velocities of a fluid whereas lowercase vectors correspond to the velocities of individual particles.}  Higher moments of the phase space (quadrupole, octupole, etc.) are parametrically suppressed by the large interaction rate between MCPs and terrestrial matter~\cite{landaukinetics}.\footnote{This is analogous to how the baryon-photon fluid in the early Universe is well-described by density and velocity perturbations, with a small quadrupole suppressed by the large Compton scattering rate~\cite{Dodelson:2003ft}.} The fundamental equation is the continuity equation, which expresses conservation of MCP number
\be
\label{eq:continuity}
\partial_t \, n_\x +  \grad \cdot \jv_\x= 0 
~,
\ee
where $j_\x \equiv n_\x \Vv_\x$ is the number-current density associated with the bulk motion of the MCP fluid. 

\np
We proceed by making several simplifying approximations. First, we restrict our analysis to a single spatial dimension. From the perspective of a strongly-coupled MCP, Earth is taken to be a one-dimensional semi-infinite region starting at the top of the atmosphere and extending to deep underground at $x \to \infty$ with gravitational field $\gv = g \, \hat{\xv}$. In this work, we evaluate the MCP density near Earth's surface, i.e., at depths much smaller than an Earth radius, $x \ll R_\oplus$. Hence, treating Earth as a semi-infinite system is well-justified provided that an incoming galactic MCP thermalizes well before traversing an $\order{1}$ fraction of Earth's radius.\footnote{In three spatial dimensions, conservation of flux implies $j_\x = \text{const.} / r^2$ for a spherically-symmetric system, instead of $j_\x = \text{const.}$ as in the case of a single dimension. Hence, provided that the radial coordinate is well-approximated as $r \simeq R_\oplus$, then it suffices to work in a single spatial dimension. } 

\np
Our next assumption involves modeling the thermalized MCPs as an ideal gas\footnote{This is valid provided that the potential energy arising from self-interactions of nearby MCPs is smaller than their thermal kinetic energy. For models of effective MCPs that couple to a dark photon, this corresponds to $(Q_\x  e^\prime)^2 \, n_\x^{1/3} \gtrsim T_\x$, or equivalently to dark Debye lengths greater than the MCP interparticle spacing.} of temperature $T_\oplus (\xv)$ and ignoring terms that depend on spatial gradients of $T_\oplus$. Note that the terrestrial temperature varies by less than two orders of magnitude within the Earth's atmosphere and crust, compared to the nuclear density which varies by more than thirteen orders of magnitude. Hence, for most regions throughout Earth, we do not expect such temperature gradients to play a large role in determining the MCP density. The importance of thermal gradients for $m_\x \lesssim 10 \ \GeV$ was recently stressed in Ref.~\cite{Leane:2022hkk}. In \Sec{tempgrad}, we investigate the role of such effects, finding that they modify the final density near Earth's surface by less than $\order{10 \%}$.

\np
Finally, we take the MCPs to be tightly-coupled to terrestrial matter, such that their mean free path is much smaller than Earth's radius. As shown in \App{Gamma}, with these assumptions the MCP current takes the simplified form
\be
\label{eq:convdiff1}
\jv_\x \simeq  n_\x \, \Vv_g - D_\x \,  \grad n_\x
~.
\ee
The gravitational drift velocity $\Vv_g$ and diffusion coefficient $D_\x$ are defined as
\be
\label{eq:DiffDef}
\Vv_g = \frac{\gv}{\Gamma_p} 
~~,~~
D_\x = \frac{T_\oplus}{\Gamma_p \, m_\x} 
~,
\ee
where $\Gamma_p$ is the ``momentum drag rate," i.e., the rate at which scattering with nuclei changes the MCP momentum by an $\order{1}$ fraction. The momentum drag rate $\Gamma_p$ evaluated at an MCP temperature $T_\x$ is related to the $\x N \to \x N$ transfer cross section $\sigma_T$ by 
\be
\label{eq:Gamma1}
\Gamma_p (T_\x) \simeq
n_N \times
\begin{cases}
3 \bar{v}_0^2  \, / \langle v_\text{rel} / \sigma_T \rangle & (m_\x \ll m_N)
\\
(m_N / m_\x) \, \langle \sigma_T \, v_\text{rel}^3 \rangle / (3 \bar{v}_0^2) & (m_\x \gg m_N)
\end{cases}
 \sim  \frac{\muxN}{m_\x} \, \frac{\bar{v}_0}{\lmfp}
~,
\ee
where $n_N$ is the terrestrial number density of nuclei, the brackets are defined as in \Eq{brackets}, and $\lmfp \sim 1 / (n_N \sigma_T)$ is the mean free path of $\x$.\footnote{More precisely, $\lmfp$ is the distance traveled before a momentum transfer of $\sim \muxN  v_\text{rel}$ occurs.} We use the first and second lines of \Eq{Gamma1} in our calculations, which are approximate expressions that hold in the limit that $m_\x \ll m_N$ and $m_\x \gg m_N$, respectively. Note that the ratio between the diffusion coefficient and gravitational drift velocity, $D_\x / V_g = T_\oplus / (m_\x \, g)$ is independent of the interaction rate $\Gamma_p$ and is effectively the distance over which a particle with kinetic energy $T_\oplus$ can move against a uniform gravitational field. As we will discuss further in \Sec{localtrafficjam}, this implies that in certain cases thermal diffusion can homogenize and wash out spatial structure in terrestrial overdensities on smaller length scales.

\np
In general, forces from terrestrial electromagnetic fields should also be included on the right-hand side of \Eq{convdiff1} if $\x$ also couples to electromagnetism on length scales larger than $\order{100} \ \km$. We omit these effects for now since their contribution is model-dependent, as discussed previously in \Sec{EMfields}. We have also ignored modifications to $j_\x$ resulting from the bulk flow of normal matter in Earth's frame, which amounts to ignoring transient effects from weather and circulation of air in Earth's atmosphere.

\addtocontents{toc}{\protect\setcounter{tocdepth}{1}}
\subsection{Momentum and Thermalization Drag Length}
\addtocontents{toc}{\protect\setcounter{tocdepth}{2}}
\label{sec:draglength}

\begin{figure}[t]
\centering
\includegraphics[width=0.49\columnwidth]{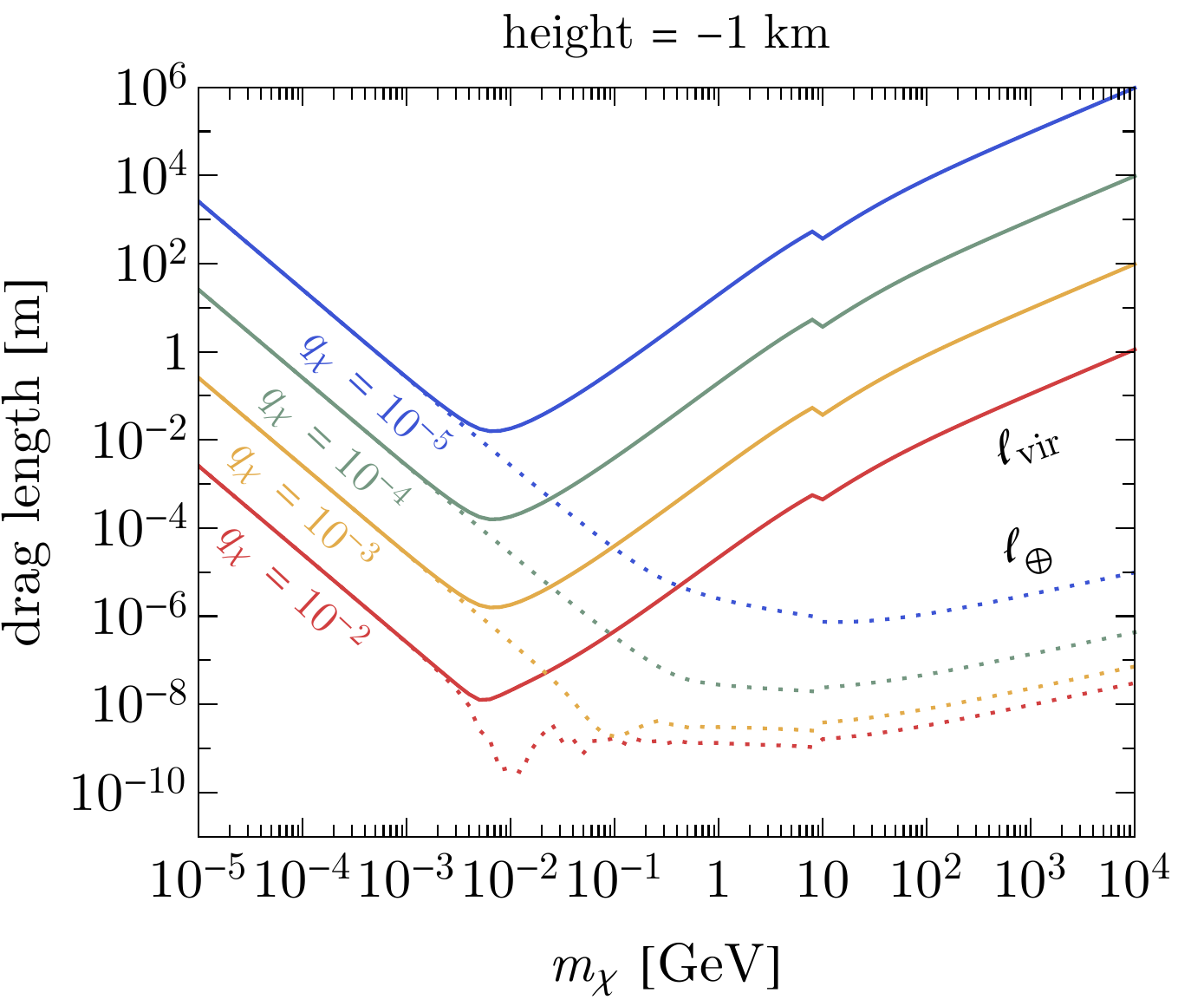}
\includegraphics[width=0.49\columnwidth]{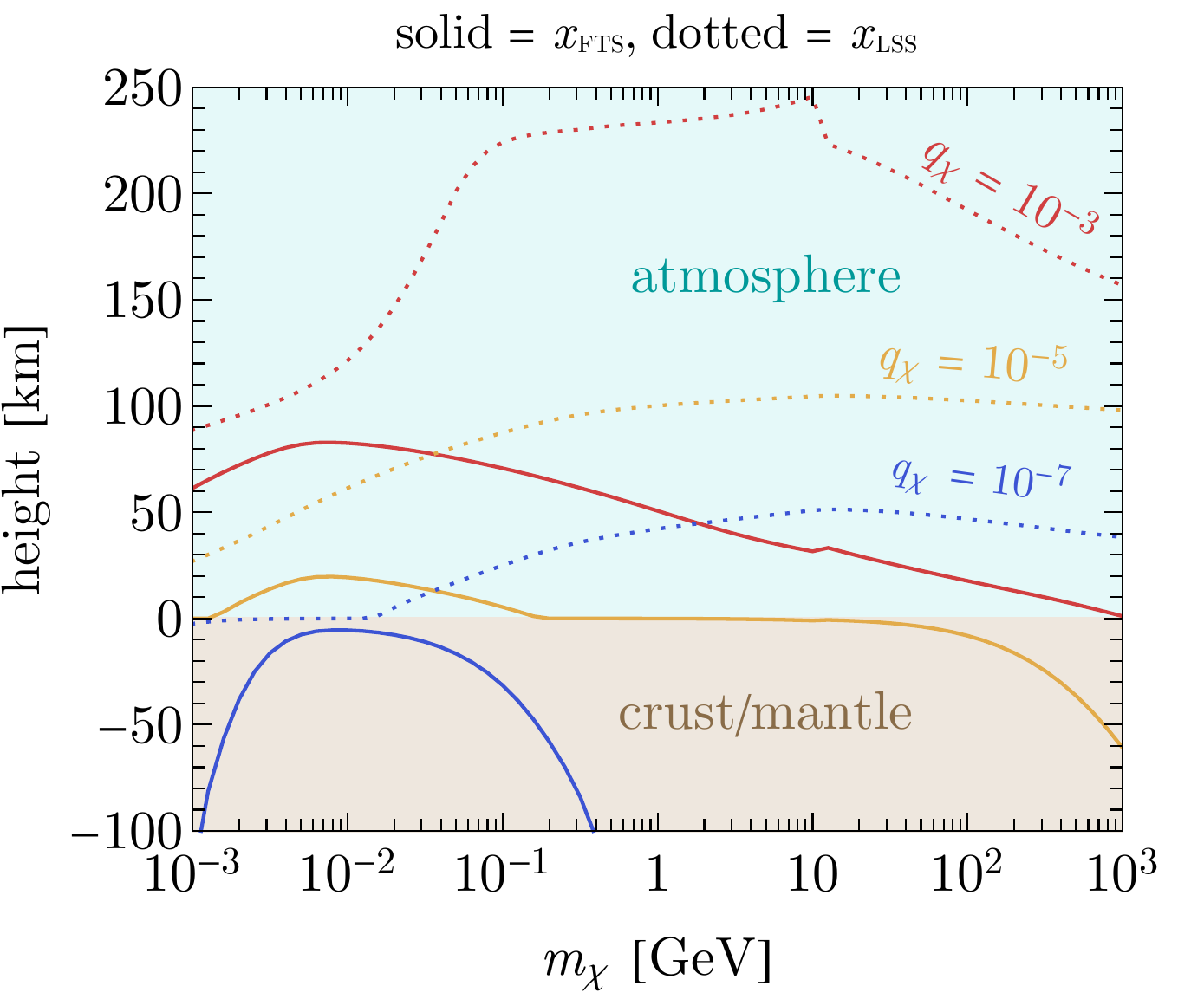}
\caption{\textbf{Left panel}: The MCP-nuclear momentum drag length (see \Eq{defdraglength}) (i.e., the distance traversed by an MCP before exchanging an $\order{1}$ fraction of its momentum due to scattering with terrestrial nuclei), as a function of the MCP mass at a depth of $1 \ \km$ underground and for various choices of the MCP coupling. The solid lines correspond to an MCP phase space that tracks the virialized galactic DM distribution, whereas the dotted lines assume that the MCPs have thermalized to terrestrial temperatures. \textbf{Right panel}: The position $\xvir$ at which a virialized MCP from the galactic halo thermalizes with the Earth (solid lines) and the position $\xth$ at which an outwardly traveling MCP that is thermalized with the Earth free streams out of the atmosphere (dotted lines) as a function of mass for various choices of the MCP coupling. These positions are shown as the corresponding height with respect to the Earth's surface.}
\label{fig:draglength}
\end{figure}

For spatially-uniform $V_g$ and $D_\x$, Eqs.~\ref{eq:continuity} and \ref{eq:convdiff1} reduce to the standard form of the convection-diffusion equation, 
\be
\label{eq:convdiff2}
\partial_t n_\x =  D_\x \grad^2 n_\x - \Vv_g \cdot \grad n_\x
~,
\ee
which is mathematically equivalent to the underlying equation governing the concentration $\mathcal{P}$ of particles undergoing a one-dimensional biased random walk~\cite{redner_2001}. The locations that mark the beginning ($\xvir$) and ending ($\xth$) of this random walk (see the orange and yellow circles in \Fig{current}, respectively) are determined by $\Gamma_p$, which is  the rate for an MCP to lose an $\order{1}$ fraction of its momentum. Thus, a virialized galactic MCP entering Earth's environment will thermalize and begin random walking at the position $x = \xvir$ of the ``first thermalization surface" (FTS) defined by
\be
\label{eq:xstart}
1 \sim \sqrt{\frac{\muxN}{m_N}} \, \int_{- \infty}^{\xvir} dx ~ \frac{\Gamma_p (T_\text{vir})}{V_\text{wind}}
~,
\ee
where $\Gamma_p$ is evaluated at the effective temperature $T_\text{vir} \simeq m_\x V_\text{wind}^2 / 3$ of the galactic population and the integral is over the terrestrial depth $x$ traversed by an MCP. The prefactor on the RHS of \Eq{xstart} accounts for the fact that in the low-mass limit,  $\x$ must exchange its momentum $m_N / m_\x \gg 1$ times before exchanging a significant portion of its energy, and hence the distance it penetrates before thermalizing is enhanced by the square root of this ratio~\cite{Dvorkin:2013cea,Boddy:2018wzy,Becker:2020hzj,Dvorkin:2020xga}. Analogous to \Eq{xstart}, we define the endpoint of the random walk  $x = \xth$ as the point at which an outgoing thermalized MCP is unlikely to undergo further momentum exchange through scattering,
\be
\label{eq:xth}
1 \sim \int_{- \infty}^{\xth} dx ~ \frac{\Gamma_p (T_\oplus)}{\sqrt{3 T_\oplus / m_\x}}
~.
\ee
In this sense, $\xth$ is the ``last scattering surface" (LSS), above which MCPs efficiently free stream out of Earth's environment without significant momentum transfer. Note that for MCPs that are gravitationally or electromagnetically bound to the Earth, $\xth$ is only a \emph{temporary} endpoint, as such particles remain on a bound orbit as they free-stream out, eventually reenter Earth's environment, and begin random walking once again. The integrands of Eqs.~\ref{eq:xstart} and \ref{eq:xth} are the inverse of the virialized (i.e., galactic) and thermalized momentum-exchange length scales,
\be
\label{eq:defdraglength}
\ell_{\text{vir}, \oplus} \equiv \sqrt{3  \, T_{\text{vir},\oplus} / m_\x \, } \, / \, \Gamma_p (T_{\text{vir}, \oplus})
~,
\ee
which are shown as the solid and dotted lines in the left panel of \Fig{draglength}, respectively. The various features of these curves follow the same line of argument as given for \Fig{sigmaT} in \Sec{scattering}, after performing the mapping from $\sigma_T$ to $\Gamma_p$ in \Eq{Gamma1}. Note that $\ell_\oplus$ is the distance a thermalized MCP travels before exchanging an $\order{1}$ fraction of its momentum and turning around and, hence, is effectively the ``step size" of its random walk throughout Earth. 

\np
Eqs.~\ref{eq:xstart} and \ref{eq:xth} only determine the \emph{typical} values of the starting and stopping positions of the MCP's random walk. Fundamentally, $\xvir$ and $\xth$ are distributed according to their probability distribution functions, as not every MCP will equilibrate or escape the Earth at the exact same location. For $m_\x \ll m_N$ or $m_\x \gg m_N$, $\x$ must travel many mean free paths $\lmfp \sim 1 / (n_N \sigma_T)$ before thermalizing to terrestrial temperatures. In this case, the central limit theorem implies that the distribution for $\xvir$ is very narrowly peaked and hence is well approximated by the deterministic expression of \Eq{xstart}. Such a statement cannot be made for the LSS $\xth$ for $m_\x \ll m_N$, but as we show below the dependence on $\xth$ is mild when evaluating the terrestrial density at greater depths $x \gg \xth$, which applies to most of the parameter space of interest. In the right panel of \Fig{draglength}, we use our estimates of $\ell_{\text{vir}, \oplus}$ as well as the modeling of Earth's environment to determine the typical \emph{heights} corresponding to $\xvir$ (solid lines) and $\xth$ (dotted lines). Here, positive or negative values of the ``height" are defined with respect to Earth's surface and hence correspond to Earth's atmosphere or crust/mantle/core, respectively. As expected, for $m_\x \gg 1 \ \GeV$ $\xth$ is located at much larger heights (where the Earth's density is parametrically reduced) than $\xvir$, due to the large enhancement of the scattering rate at low velocities (see \Sec{scattering}). At the position $\xvir$, an incoming galactic MCP exchanges an $\order{1}$ fraction of its energy and begins random walking throughout the terrestrial environment. Above the position $\xth$, thermalized MCPs can traverse the remaining atmosphere without significant exchange of their momentum. Thus, a thermalized MCP diffusing outwards is unlikely to undergo any further significant scatters beyond this point. These locations, $\xvir$ and $\xth$, therefore define the beginning and ending points of an MCP's random walk and will be used as inputs in \Sec{traffic} to calculate the traffic jam density. 

\np
For $m_\x \ll m_N$, the larger number of momentum exchanges required to thermalize implies that only a fraction of low mass MCPs efficiently shed their kinetic energy before possibly turning around and reflecting off of the Earth. We determine the likelihood to successfully thermalize $\text{Pr}_\text{th}$ by first noting that \Eq{convdiff2} also applies to the evolution of the probability per unit length $\mathcal{P}$  of an individual diffusing particle~\cite{redner_2001}. For an MCP that exchanges a significant fraction of its momentum and begins diffusing through the Earth after traversing a distance $\ell_\text{vir}$ at time $t = 0$, the probability that it is found at a position $x$ at time $t > 0$ is given by the following solution to \Eq{convdiff2}~\cite{redner_2001}
\be
\label{eq:Prob0}
\mathcal{P} (x,t) = \frac{1}{\sqrt{4 \pi D_\x \, t}} ~ \Big( e^{-\frac{(x - \ell_\text{vir})^2}{4 D_\x t}} - e^{-\frac{(x + \ell_\text{vir})^2}{4 D_\x t}} \Big)
~.
\ee
We employ a ``sudden thermalization approximation," such that after this initial time $\Delta t_p \sim \Gamma_p (T_\text{vir})^{-1}$ spent traveling a distance $\ell_\text{vir}$, the incoming MCP suddently thermalizes after scattering for an additional time of $\Delta t_\text{th} \simeq (m_N / \muxN - 1) \, \Gamma_p (T_\text{vir})^{-1} = (m_N / m_\x) \, \Gamma_p (T_\text{vir})^{-1}$~\cite{Boddy:2018wzy}. The thermalization probability is thus the integrated probability that the particle has not escaped after time $\Delta t_\text{th}$, which is given by
\be
\label{eq:ProbTherm1}
\text{Pr}_\text{th} = \int_0^\infty dx ~ \mathcal{P} \big( x\, , \Delta t_\text{th} \big) 
\simeq \text{erf} \Big( \sqrt{3 m_\x / 4 m_N} \,  \Big)
~,
\ee
where we used that $D_\x = T_\text{vir} / (\Gamma_p \, m_\x)$ and $\ell_\text{vir} \simeq V_\text{wind} / \Gamma_p$. In the low- and high-mass limits, we thus have $\text{Pr}_\text{th} \simeq \sqrt{(3 / \pi) \, (m_\x / m_N)}$ and $\text{Pr}_\text{th} \simeq 1$, respectively. 
 
\np
Recent numerical study of $\text{Pr}_\text{th} $ conducted in Ref.~\cite{Bramante:2022pmn} shows  a
broad consistency with the estimate of \Eq{ProbTherm1}. We remark that both \Eq{ProbTherm1} and the results of Ref.~\cite{Bramante:2022pmn} are derived for the case of roughly isotropic scattering. The actual MCP-atomic scattering cross sections, depending on the model parameters, may be strongly peaked in the forward direction, in which case the probability of large-angle scattering is relatively suppressed. In this case, the thermalization probability $\text{Pr}_\text{th}$ may be somewhat enhanced, e.g., by a Coulomb logarithm, since energy loss can occur continuously over many low-momentum-transfer scatters. Therefore, \Eq{ProbTherm1} should be viewed as a conservative estimate. We also note that MCPs may lose energy via ionization, especially at larger velocities, but this energy loss mechanism appears to be subdominant to elastic scattering on atoms due to the small velocities of MCPs compared to atomic electron velocities.

\section{The Hydrostatic Population}
\label{sec:static}

Before calculating the dynamic traffic jam density in  \Sec{traffic}, we first review the hydrostatic population bound by Earth's gravitational or electromagnetic fields. In doing so, we follow the approach of Ref.~\cite{Neufeld:2018slx}, finding good agreement with their results for the gravitationally-bound population. The volume-averaged rate at which a wind of virialized galactic MCPs flows into the Earth is
\be
\label{eq:noplusdot}
\dot{n}_\oplus \simeq \jvir  ~ \frac{\pi R_\oplus^2}{(4 \pi / 3) \, R_\oplus^3} = \frac{3}{4} \, \frac{\nvir \, V_\text{wind}}{R_\oplus}
~,
\ee
where a dot denotes a time derivative and $R_\oplus$ is the Earth radius. However, as discussed above, not all incoming MCPs thermalize to terrestrial temperatures. Accounting for this implies that the volume-averaged capture rate is 
\be
\label{eq:ncap}
\dot{n}_\text{cap} \simeq \text{Pr}_\text{th} ~ \dot{n}_\oplus
~,
\ee
with $\text{Pr}_\text{th}$ given in \Eq{ProbTherm1}. Here, we have ignored depletion of the terrestrial density from self-annihilations of MCPs to dark or SM photons. As discussed above, since the importance of such self-interactions is highly model-dependent, we implicitly assume that $\nvir$ (and hence $n_\x$) is sufficiently small so that such processes do not significantly modify the evolution of the MCP population.

\np
Let us begin by focusing on the role of Earth's gravitational field. To remain gravitationally bound, a thermalized MCP must have a radially-outward velocity smaller than the terrestrial escape velocity. The fraction of particles with a speed greater than the escape velocity leads to an outgoing ``evaporation flux" that depletes the terrestrial population, denoted as $j_\text{evap}$. The corresponding volume-averaged depletion rate is related to $j_\text{evap}$ by 
\be
\label{eq:evap1}
\dot{n}_\text{evap} \simeq j_\text{evap}  ~ \frac{4 \pi R_\oplus^2}{(4 \pi / 3) \, R_\oplus^3} = 3 ~ \frac{j_\text{evap}}{R_\oplus}
~.
\ee
The evaporation flux is given by $j_\text{evap} \simeq n_\x \, \langle v_\text{out,esc} \rangle$, where $\langle v_\text{out,esc} \rangle$ is the average outward-going velocity above the escape velocity~\cite{Catling:2017}
\be
\label{eq:evap2}
\langle v_\text{out,esc} \rangle \simeq \int d^3 \vv ~  \frac{e^{- v^2 / v_0^2}}{\pi^{3/2} \, v_0^3} ~ v \cos{\theta} ~ \Theta(v - v_\text{esc})
= \frac{v_0 \, (1 + v_\text{esc}^2 / v_0^2)}{2 \sqrt{\pi}} \, e^{- v_\text{esc}^2 / v_0^2}
~,
\ee
and $v_0 = \sqrt{2 \, T_\oplus (\xth) / m_\x} \, $ is evaluated at the free-streaming surface $\xth$.  This outgoing flux is significant provided that the exponential in the last equality of \Eq{evap2} does not suppress the rate. Taking $T_\oplus (\xth) \simeq 300 \ \K$, $v_\text{esc} / v_0 \lesssim \order{10}$ for $m_\x \lesssim 1 \ \GeV$; thus, we expect thermal evaporation to significantly deplete the gravitationally bound hydrostatic population  for $m_\x \ll 1 \ \GeV$. From Eqs.~\ref{eq:evap1} and \ref{eq:evap2}, the volume-averaged evaporation rate is 
\be
\label{eq:Gammaevap}
\dot{n}_\text{evap} = \Gamma_\text{evap} \, n_\x 
~~,~~
\Gamma_\text{evap} =  \frac{3}{2 \sqrt{\pi}} \,\frac{ v_0 \, (1 + v_\text{esc}^2 / v_0^2)}{R_\oplus} \, e^{- v_\text{esc}^2 / v_0^2}
~.
\ee
The final terrestrial abundance of gravitationally bound MCPs is governed by both capture and evaporation such that $\dot{n}_\x = \dot{n}_\text{cap} - \dot{n}_\text{evap} = \dot{n}_\text{cap} - \Gamma_\text{evap} \, n_\x$. Assuming a negligible initial abundance, this can be solved to find the volume-averaged abundance that has accumulated over the age of the Earth $t_\oplus \simeq 4.5  \ \Gyr$,
\be
\label{eq:nbound1}
\overline{n}_\text{bound} \simeq \dot{n}_\text{cap} \, t_\oplus ~ \frac{1 - e^{- \Gamma_\text{evap} \, t_\oplus}}{\Gamma_\text{evap} \, t_\oplus}
~.
\ee

\np
In \Eq{nbound1}, we have put a bar over the number density to signify that this is the mean density averaged over the entire terrestrial volume. A more  detailed description can be obtained by noting that such a population eventually approaches hydrostatic equilibrium. This corresponds to zero bulk flow of the accumulated MCP fluid. Enforcing zero bulk current $j_\x = 0$ in \Eq{convdiff1} yields $\partial_r \log n_\x \simeq - m_\x \, g / T_\oplus$. Integrating over the terrestrial radial coordinate $r$, we then have
\be
\label{eq:nhydrostatic2}
n_\x(r) = \frac{1}{3} \, \overline{n}_\text{bound} ~ \frac{R_\oplus^3 \, e^{- m_\x \, I(r)}}{\int_0^{R_\oplus} dr^\p ~r^{\p \, 2} ~ e^{- m_\x I (r^\p)}}
~~,~~
I(r) = \int_0^r dr^\p ~ \frac{g(r^\p)}{T_\oplus(r^\p)}
~,
\ee
where the $r$-independent constant for $n_\x (r)$ has been fixed by demanding that the mean density is equal to $\overline{n}_\text{bound}$ of \Eq{nbound1}. 

\begin{figure}[t]
\centering
\includegraphics[width=0.6\columnwidth]{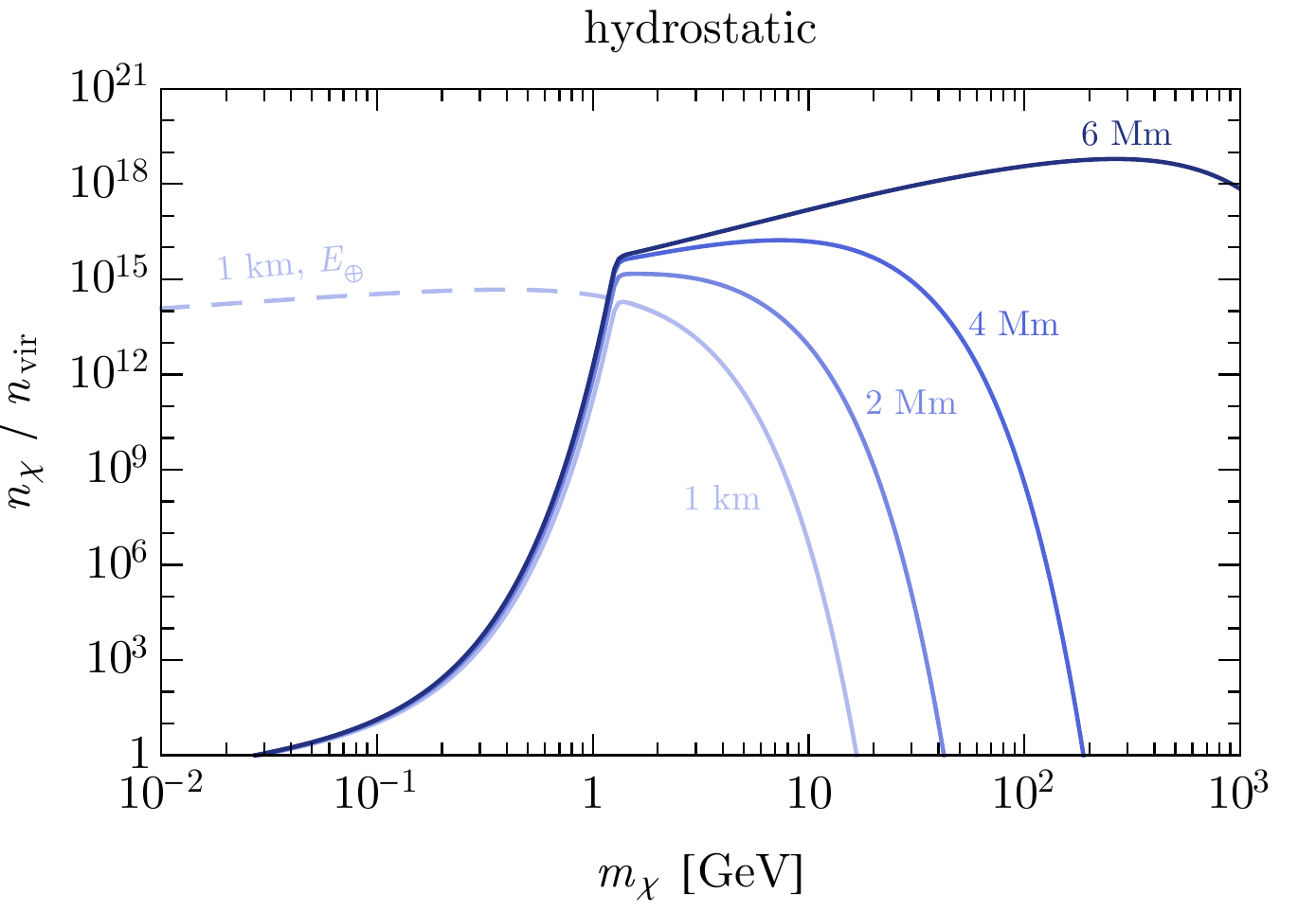}
\caption{The gravitationally bound hydrostatic overdensity (solid lines) of MCP relics as a function of the MCP mass, evaluated at various depths underground (from darkest to lightest blue, $6 \ \text{Mm}$, $4 \ \text{Mm}$, $2 \ \text{Mm}$, and $1 \ \text{km}$ underground). The MCP coupling is assumed to be large enough for MCPs to rapidly thermalize in Earth's environment and the temperature at which these particles evaporate from the Earth is fixed to $T_\oplus = 300 \ \K$. Also shown as the dashed line is an example of the electrically bound hydrostatic population, which tracks the gravitationally bound density at large masses. 
}
\label{fig:hydrostatic}
\end{figure}

\np
\Eq{nhydrostatic2} is the radial profile for the gravitationally bound hydrostatic population of MCPs. As shown as the solid lines in \Fig{hydrostatic}, we use it to determine the terrestrial overdensity $n_\x / \nvir$ of this hydrostatic population at various depths underground, assuming a temperature at the last scattering surface of $T_\oplus (\xth) = 300 \ \K$. Note that this is only a representative example, as the temperature at the last scattering surface depends on the MCP charge and mass (we find that $T_\oplus (\xth) \simeq 300 \ \K$ for $q_\x \sim 10^{-5} /  \min{[1 \ \GeV , m_\x]}$).  For $m_\x \ll 1 \ \GeV$, an $\order{1}$ fraction of the incoming flux evaporates on a timescale much shorter than the age of the Earth, thus considerably reducing the gravitationally bound density. Instead, for $m_\x \gg 1 \ \GeV$, MCPs remain efficiently gravitationally bound until today. However, in this case, hydrostatic equilibrium in the presence of Earth's gravitational field dictates that the density of more massive MCPs is more strongly peaked towards the Earth's inner core, diminishing the density near Earth's surface.  This explains the large hierarchy between the various curves of \Fig{hydrostatic} for large masses. This behavior can be seen explicitly in  the numerator on the RHS of \Eq{nhydrostatic2}, since the exponential is strongly suppressed for large $m_\x$ unless $I(r)$ is particularly small (corresponding to radial positions near Earth's center). Note that in this section, we have not approximated Earth as one-dimensional, and thus our results are valid for arbitrarily large depths. At depths relevant for underground detectors ($\sim 1 \ \km$), we see that the overdensity of such a population of strongly-coupled relics is only significant over a narrow mass-range centered around $m_\x \sim 1 \ \GeV$. In the next section, we show that the dynamical traffic jam population can instead contribute the dominant terrestrial density near Earth's surface. 

\np
Also shown as the dashed line in \Fig{hydrostatic} is the density corresponding to an electromagnetically bound hydrostatic population of MCPs. As discussed in \Sec{EMfields}, this arises if $q_\x \gg 10^{-7}$ and $\x$ couples to Earth's electric field over large distance scales, such as in models of pure MCPs. In this case, even very light thermalized particles can remain efficiently bound in Earth's crust. For $m_\x \gg 1 \ \GeV$, the resulting profile underground is identical to that of the gravitationally bound population, since this electric field is screened in the crust. At the level of our analysis, we have incorporated this effect by simply setting the evaporation flux $j_\text{evap}$ to zero. Accounting for inhomogeneities in Earth's electric field is beyond the scope of this work, but could suppress the efficiency of Earth's electric field in trapping MCPs over terrestrial timescales. We leave the consideration of this to future work.  

\np
We conclude this section by briefly commenting on the nature of the MCP population above the last scattering surface  $\xth$, where the phase space transitions from hydrostatic to free-streaming. A detailed understanding of this region is relevant for studies of indirect signatures arising from the annihilation into SM particles at large radii above the absorptive layers of the Earth and its atmosphere. We can approximate this ``ballistic" free-streaming population at height $h > 0$ above the last scattering surface as
\be
n_\x (h) \sim n_\x (\xth) \times \exp \left( -\frac{m_\x \, g \, h}{T_\oplus} \right)
\simeq n_\chi(\xth) \times \exp \left( -\frac{m_\chi}{\GeV} \times \frac{h}{240 \ \km} \right)
~,
\ee
where we took $T_\oplus \simeq 300 \ \K$ and ignored model-dependent effects arising from terrestrial electromagnetic fields. Thus, substantial MCP densities can persist in regions well above the last scattering surface for certain masses. We leave a dedicated study of this region to future work.

\section{The Traffic Jam Population}
\label{sec:traffic}

In this section, we derive the terrestrial overdensity of MCPs arising from the dynamical traffic jam population, as was previously described parametrically in Eqs.~\ref{eq:overdensity1} and \ref{eq:overdensity2}. This corresponds to MCPs that have only just recently entered Earth's environment and is thus distinct from the accumulated hydrostatic density of the previous section. In the various subsections below, we determine the traffic jam density by solving \Eq{convdiff1}. In order to determine explicit solutions, this equation must be supplemented with the appropriate boundary condition at the points $\xvir$ (see \Eq{xstart}) and $\xth$ (see \Eq{xth}) at which an incoming galactic MCP thermalizes and begins diffusing or an outwardly diffusing MCP stops exchanging momentum with nuclei, respectively. Thus, $\xvir$ and $\xth$ denote the starting point and the minimal depth for an MCP's random walk as it diffuses throughout Earth. 

\np
The nature of the boundary condition enforced at the last scattering surface $\xth$ is dictated by the strength of the MCP interaction with Earth's gravitational or electromagnetic fields. Since the velocity of very light thermalized MCPs is much greater than the terrestrial escape velocity, we impose an \emph{absorbing} boundary condition at $\xth$ for such masses if $\x$ does not couple to Earth's electric field (once such a light thermalized MCP approaches terrestrial depths less than $\xth$ it free streams and escapes the Earth). On the other hand, much heavier MCPs or those that couple strongly to Earth's electric field remain gravitationally or electrically bound, corresponding to imposing a \emph{reflecting} boundary condition at $\xth$ (once reaching depths shallower than $\xth$, an outgoing MCP that is, e.g., gravitationally bound will free stream along a trajectory that causes it to later reenter Earth's environment). For the sake of pedagogy, for each of these regimes we provide two alternative derivations that 1) either explicitly ignore time variations by assuming that steady-state has been reached or 2) involve solving for the time-dependence of the density explicitly. As we will see below, the first of these two will allow us to generalize the solution to realistic terrestrial environments in which the density of nuclear scatterers varies as a function of position by many orders of magnitude. For the reader uninterested in the technical details of these derivations, most of this section may be skipped; the main results are given in the boxed Eqs.~\ref{eq:nxGreen4}, \ref{eq:nxGreen4b}, and \ref{eq:nxsemiref}. 

\addtocontents{toc}{\protect\setcounter{tocdepth}{1}}
\subsection{Steady-State Formalism}
\addtocontents{toc}{\protect\setcounter{tocdepth}{2}}
\label{sec:steadystate}

Here, we provide a semi-analytic treatment for evaluating the terrestrial traffic jam density, in the case that both the diffusion coefficient $D_\x$ and drift velocity $V_g$ vary with depth $x$. However, to simplify the analysis, we take their ratio $D_\x / V_g = T_\oplus / (m_\x \, g)$ to be constant, corresponding to uniform temperature. We comment on the role of temperature gradients in \Sec{tempgrad}, finding that they modify the density near Earth's surface only at the level of $\order{10\%}$. In \Sec{formalism}, we wrote down the fundamental equations governing MCP diffusion  (Eqs.~\ref{eq:continuity} and \ref{eq:convdiff1}), which we rewrite here for convenience,
\be
\label{eq:convdiff1copy}
\partial_t \, n_\x +  \grad \cdot \jv_\x = 0
~~,~~
\grad n_\x - (\Vv_g / D_\x) \, n_\x \simeq -\jv_\x / D_\x 
~.
\ee
In the steady-state limit, the first equation implies that the diffusion current $j_\x$ is constant in space except near regions of discontinuities, where its gradient is ill-defined. Once again approximating the Earth as one-dimensional and taking $V_g / D_\x = \text{constant}$, the second equation of \Eq{convdiff1copy} can be solved with the method of Green's functions,
\be
\label{eq:nxGreen}
n_\x (x) =  \int_x^\infty d x^\p ~ e^{- \frac{V_g \, (x^\p-x)}{D_\x}} ~  \frac{j_\x (x^\p)}{D_\x (x^\p)}
~.
\ee

\np
Thus, evaluating the terrestrial density $n_\x$ requires determining the MCP diffusion current $j_\x$, which is spatially uniform except near regions of discontinuities.  As shown schematically in \Fig{current}, we model the incoming galactic flux of MCPs as flowing unimpeded into Earth, until coming to a sudden halt at $\xvir$ due to repeated scattering in Earth's environment. Therefore, after this point ($x > \xvir$), the MCP flow is governed purely by diffusion, whereas before this point ($x < \xvir$) it is determined both by a combination of diffusion and the inflowing wind of the galactic population. Conservation of total flux then implies that
\be
\label{eq:jmatching}
j_\x (x > \xvir) = j_\x (x < \xvir) + j_\text{vir}
~.
\ee
We then evaluate \Eq{nxGreen} by splitting up the integral corresponding to these two regions. Hence, before the thermalization point $x < \xvir$, the density $n_\x$ of \Eq{nxGreen} is
\be
\label{eq:nxGreen2}
n_\x (x < \xvir) = j_\x (x < \xvir) \int_x^{\xvir} \hspace{-0.2 cm} d x^\p ~   \frac{e^{- \frac{V_g (x^\p-x)}{D_\x}}}{D_\x (x^\p)} \, + \, \Big( j_\x (x < \xvir) + j_\text{vir} \Big) \int_{\xvir}^\infty  \hspace{-0.2 cm} d x^\p  ~ \frac{e^{- \frac{V_g  (x^\p-x)}{D_\x}}}{D_\x (x^\p)}
~,
\ee
where $V_g / D_\x$ is treated as a constant in the argument of the exponential. Similarly, for $x > \xvir$ we have
\be
\label{eq:nxGreen3}
n_\x (x > \xvir) =   \Big( j_\x (x < \xvir) + j_\text{vir} \Big) \, \int_x^\infty d x^\p ~  \frac{e^{- \frac{V_g \, (x^\p-x)}{D_\x}}}{D_\x (x^\p)}
~.
\ee

\np
Note that we need to determine $j_\x (x < \xvir)$ in order to evaluate $n_\x$ in  Eqs.~\ref{eq:nxGreen2} and \ref{eq:nxGreen3}. To proceed, we must therefore enforce the appropriate boundary condition at the last scattering surface $\xth$. For MCPs that do \emph{not} remain gravitationally or electromagnetically bound to the Earth, we model their exiting of the terrestrial environment with an absorbing boundary placed at $\xth$, resulting in $n_\x (\xth) = 0$. From the second equation of \Eq{convdiff1copy}, this boundary condition gives 
\be
\label{eq:jxrelation1}
j_\x (x < \xvir) = - D_\x (\xth) \, \partial_x n_\x (\xth)
~,
\ee
where we used that $j_\x = \text{constant}$ for $x < \xvir$. Using \Eq{nxGreen2} in \Eq{jxrelation1} and solving for $j_\x (x < \xvir)$, we find
\be
\label{eq:j1}
j_\x (x < \xvir)  = - j_\text{vir} ~~ \frac{\int_{\xvir}^\infty d x^\p ~ e^{- (V_g / D_\x) \, (x^\p-\xth)} ~  \frac{1}{D_\x (x^\p)}}{\int_{\xth}^{\infty} d x^\p ~ e^{- (V_g / D_\x) \, (x^\p-\xth)} ~  \frac{1}{D_\x (x^\p)}}
~.
\ee
Having determined $j_\x (x < \xvir)$, Eqs.~\ref{eq:nxGreen2} and \ref{eq:nxGreen3} then determine the unbounded traffic jam density to be
\be
\label{eq:nxGreen4}
\boxed{
\frac{n_\x (x < \xvir)}{\nvir} \Bigg|_\text{unbound} = c_1 \, V_\text{wind} \, \int_x^{\xvir} d x^\p ~  \frac{e^{- \frac{V_g \, (x^\p-x)}{D_\x}}}{D_\x (x^\p)} +  c_2 \, V_\text{wind} \,\int_{\xvir}^\infty d x^\p ~  \frac{e^{- \frac{V_g \, (x^\p-x)}{D_\x}}}{D_\x (x^\p)}
}
\ee
for $x < \xvir$ and
\be
\label{eq:nxGreen4b}
\boxed{
\frac{n_\x (x > \xvir)}{\nvir} \Bigg|_\text{unbound} =  c_2 \, V_\text{wind} \, \int_x^\infty d x^\p ~  \frac{e^{- \frac{V_g  \, (x^\p-x)}{D_\x}}}{D_\x (x^\p)}
}
\ee
for $x > \xvir$. The dimensionless coefficients $c_1$ and $c_2$ are defined as
\be
c_1 =  - \text{Pr}_\text{th}~ \frac{\int_{\xvir}^\infty d x^\p ~ e^{- (V_g / D_\x) \, (x^\p-\xth)} ~  \frac{1}{D_\x (x^\p)}}{\int_{\xth}^{\infty} d x^\p ~ e^{- (V_g / D_\x) \, (x^\p-\xth)} ~  \frac{1}{D_\x (x^\p)}}
~~~,~~~
c_2 = c_1 + \text{Pr}_\text{th}
~,
\ee
where we have included the probability to thermalize, $\text{Pr}_\text{th}$ (\Eq{ProbTherm1}), by hand. 

\np 
We can gain some intuition of the results above when $D_\x$ and $V_g$ are both separately independent of position. Integrating Eq.~\ref{eq:j1}, we see that when diffusion is important, i.e., $m_\x \, g \, (\xvir - \xth) \ll T_\oplus$, $j_\x (x < \xvir) \simeq - j_\text{vir}$, so that the outgoing diffusion flux cancels the incoming galactic flux, as we argued in \Eq{overdensity1}. In the opposite mass limit, there is no flux leaving the Earth, leading to $j_\x(x < \xvir) = 0$. Furthermore, when $D_\x$ and $V_g$ are both independent of position the integrals in Eqs.~\ref{eq:nxGreen4} and \ref{eq:nxGreen4b} can be performed analytically, which yields
\be
\label{eq:absorbanalytic1}
\frac{n_\x (x < \xvir)}{\nvir \, \text{Pr}_\text{th}}  \simeq \frac{V_\text{wind}}{V_g} ~ \left( e^{-\frac{V_g (\xvir - x)}{D_\x}} - e^{- \frac{V_g \Delta x_{_\text{FL}}}{D_\x}}\right) 
\simeq
\begin{cases}
 \frac{V_\text{wind} \, (x - \xth)}{D_\x} & ,~ m_\x \ll \frac{T}{g \, \Delta x_{_\text{FL}}}
 \\
\frac{V_\text{wind}}{V_g} \, e^{-\frac{(\xvir - x) V_g}{D_\x}} & ,~ m_\x \gg \frac{T}{g \, \Delta x_{_\text{FL}}}
\end{cases}
\ee
and
\be
\label{eq:absorbanalytic2}
\frac{n_\x (x>\xvir)}{\nvir \, \text{Pr}_\text{th}} \simeq \frac{V_\text{wind}}{V_g} ~ \left( 1 - e^{- \frac{V_g \, \Delta x_{_\text{FL}}}{D_\x}}\right) 
\simeq
\begin{cases}
 \frac{V_\text{wind} \, \Delta x_{_\text{FL}}}{D_\x} & ,~ m_\x \ll \frac{T}{g \, \Delta x_{_\text{FL}}}
 \\
\frac{V_\text{wind}}{V_g} & ,~ m_\x \gg \frac{T}{g \, \Delta x_{_\text{FL}}}
\end{cases}
~,
\ee
where in the second equality of Eqs.~\ref{eq:absorbanalytic1} and \ref{eq:absorbanalytic2} we have taken the low- or high-mass limit and used the shorthand notation $\Delta x_{_\text{FL}} \equiv \xvir - \xth > 0$. In \Eq{absorbanalytic1}, the first line shows the limit where diffusion dominates over gravitational drift, and the outgoing flux $j_\x (x < \xvir) \simeq -j_\text{vir}$ due to diffusion is supported by a linear density gradient. In the other limit, as shown in the second line, there is no outgoing flux and diffusion is inefficient, such that the particles follow an exponential distribution following the law of atmospheres. Also note that for $x > \xvir$, the first and second lines of \Eq{absorbanalytic2} match the parametric forms given previously in Eqs.~\ref{eq:overdensity1} and \ref{eq:overdensity2}.

\np
Eqs.~\ref{eq:jxrelation1} $-$ \ref{eq:absorbanalytic2} were evaluated for MCPs that are not bound to the Earth. In the other case where these particles do remain gravitationally or electrically bound, we must modify the boundary condition imposed at $\xth$. Instead, for bound particles we enforce a reflecting boundary condition, corresponding to zero flux past the last scattering surface, i.e., $j_\x (\xth) = 0$. Since the diffusion current $j_\x$ is uniform for $x < \xvir$, this reflecting boundary condition along with \Eq{jmatching} implies that
\be
j_\x (x < \xvir) = 0
~~,~~
j_\x (x > \xvir) = j_\text{vir}
~.
\ee
Using this in \Eq{nxGreen2}, the traffic jam overdensity for terrestrially bound particles  is given by
\be
\label{eq:nxsemiref}
\boxed{
\frac{n_\x (x)}{\nvir} \Bigg|_\text{bound} = \text{Pr}_\text{th} \, V_\text{wind} \, \int_{\max{(x, \xvir)}}^\infty  \hspace{-0.2 cm}  d x^\p ~  \frac{e^{- \frac{V_g  (x^\p-x)}{D_\x}}}{D_\x (x^\p)}
}
~.
\ee
Similar to the previous calculation, in the case that $D_\x$  and $V_g$ are independent of position, the integral in \Eq{nxsemiref} can be performed analytically,
\be
\label{eq:reflectanalytic1}
\frac{n_\x (x)}{\nvir \, \text{Pr}_\text{th}} \simeq \frac{V_\text{wind}}{V_g} \times
\begin{cases}
e^{-\frac{(\xvir - x) \, V_g}{D_\x}} &, ~ x<\xvir
\\
1 & ,~ x>\xvir
\end{cases}
~.
\ee
Comparing the first and second lines of \Eq{reflectanalytic1} to the results for an absorbing boundary condition in the second line of Eqs.~\ref{eq:absorbanalytic1} and \ref{eq:absorbanalytic2}, we see that the terrestrial overdensities are independent of the particular boundary condition that is imposed at $\xth$ in the large mass limit. This is as expected, since large masses correspond to enhanced gravitational drift velocities, such that an exponentially small number of particles are able to diffuse against the gravitational field from $\xvir$ to the last scattering surface $\xth$.

\addtocontents{toc}{\protect\setcounter{tocdepth}{1}}
\subsection{Time-Dependent Formalism}
\addtocontents{toc}{\protect\setcounter{tocdepth}{2}}
\label{sec:timdep}

In the previous subsection, we derived semi-analytic expressions for the steady-state traffic jam density by imposing conservation of flux, and found that these expressions reduce to simple analytic ones in the limit that the temperature and density of the terrestrial environment are approximately uniform. In this section, we rederive these analytic results in the time domain, restricting our analysis to the special case $D_\x = \text{constant}$ and  $V_g = \text{constant}$ for simplicity. Aside from serving as a useful cross-check, the derivation below allows us to see the timescale over which the traffic jam density approaches steady-state behavior.  

\np
Let us begin by rederiving the traffic jam density of particles that are not efficiently bound to the Earth. As in the last section, we model the diffusion of such MCPs as a biased one-dimensional random walk starting at position $x = \xvir > \xth > 0$ with an absorber placed at $x = \xth$. To proceed, we note that the diffusion equation of \Eq{convdiff2} also applies to the evolution of the probability per unit length $\mathcal{P}$  of a single diffusing particle~\cite{redner_2001}. For a particle that begins to diffuse through Earth at $\xvir$ at time $t = 0$, the probability per unit length that it is found at a later time $t$ at position $x$ is\footnote{Note that this corrected expression appears in the errata of Ref.~\cite{redner_2001}.}~\cite{redner_2001}
\be
\label{eq:Prob1}
\mathcal{P} (x,t) \, \Big|_\text{unbound} = \frac{1}{\sqrt{4 \pi D_\x \, t}} ~ \bigg( e^{-\frac{(x - \xvir - V_g t)^2}{4 D_\x t}} - e^{- \frac{V_g \, (\xvir - \xth)}{D_\x}} \, e^{-\frac{(x + \xvir - 2 \xth - V_g t)^2}{4 D_\x t}} \bigg)
~.
\ee
In \Eq{Prob1}, the first and second terms are both proportional to Green's functions of the one-dimensional diffusion equation, with the coefficient of the second term fixed to enforce the absorber boundary condition $\mathcal{P} (\xth, t) = 0$. Furthermore, note that for $x > \xth$, this reduces in the $t \to 0$ limit to $\mathcal{P} (x, 0) \simeq  \delta (x - \xvir)$, i.e., the probability density of a particle at the starting point $\xvir$. Let us now generalize \Eq{Prob1} for the diffusion of many MCPs. For a galactic MCP current depositing particles at $\xvir$ at a rate of $\jvir$, the resulting number density at time $t$ is\footnote{To derive \Eq{intProb}, note that a single particle starting a random walk at time $t^\p$ has a number density $\mathcal{P} (x, t-t^\p)/ dA$ at time $t$, where $dA$ is some small transverse area element. The number of such particles starting their random walk within some small initial time interval $d t^\p$ is $\jvir \, dA \, dt^\p$. Hence, integrating over all past times, the total number density of such particles is $n_\x =  \jvir \int_{-\infty}^t dt^\p ~ \mathcal{P} (x, t-t^\p)$. Changing variables $t^\p \to t - t^\p$ and including a factor of $\text{Pr}_\text{th}$ gives the form in \Eq{intProb}.} 
\be
\label{eq:intProb}
n_\x (x,t) = \text{Pr}_\text{th} \, \jvir \, \int_0^t dt^\p ~ \mathcal{P}(x,t^\p)
~.
\ee
Using the explicit form of $\mathcal{P} (x,t)$ in \Eq{Prob1}, we analytically evaluate the above integral in the long-time $t \to \infty$ limit.  From \Eq{Prob1} we see that this limit corresponds to $t \gg  \ell_\text{max} / V_g ~,~ \ell_\text{max}^2 / D_\x$, where $\ell_\text{max} \sim \max (x , \xvir, \xth) - \min (x , \xvir, \xth)$ is the longest characteristic length scale in consideration. Evaluating  \Eq{intProb} at a depth $x$ before ($\xth < x < \xvir$) or after ($x > \xvir > \xth$) the starting point $\xvir$, we find exact agreement with Eqs.~\ref{eq:absorbanalytic1} and \ref{eq:absorbanalytic2}, respectively. 

\np
As before, we instead model the diffusion of MCPs that are efficiently bound to the Earth as a biased one-dimensional random walk starting at position $x = \xvir > \xth > 0$ with a reflector placed at $\xth$. The analysis is nearly identical to the previous example. In this case, the analogue of \Eq{Prob1} for a reflecting boundary condition is~\cite{MOLINI20111841}\footnote{Note that there is a typo in the brackets of Eq.~29 of Ref.~\cite{MOLINI20111841}. This has been corrected in \Eq{Prob1ref}.}
\begin{align}
\label{eq:Prob1ref}
\mathcal{P} (x,t)  \, \Big|_\text{bound} &= \frac{1}{\sqrt{4 \pi D_\x \, t}} ~ \left[ e^{-\frac{(x - \xvir - V_g t)^2}{4 D_\x t}} + e^{- \frac{V_g \, (\xvir - \xth)}{D_\x}} ~ e^{-\frac{(x + \xvir - 2 \xth - V_g t)^2}{4 D_\x t}} \right] 
\nl
&\quad - \frac{V_g}{2 D_\x} \, e^{\frac{V_g \, (x - \xth)}{D_\x}} \left[ 1 - \text{erf} \Big( \frac{x + \xvir - 2 \xth + V_g \, t}{2 \sqrt{D_\x \, t}} \Big) \right]
~.
\end{align}
Integrating over time as in \Eq{intProb}, we find that this results in a terrestrial overdensity identical to that previously derived in \Eq{reflectanalytic1}.

\np
We conclude this section by commenting on the matching condition imposed on $j_\x$ in \Eq{jmatching}. In this section, we independently rederived  Eqs.~\ref{eq:absorbanalytic1}, \ref{eq:absorbanalytic2}, and \ref{eq:reflectanalytic1} without imposing any conditions (other than the reflecting boundary condition) on $j_\x$. Using these results for $n_\x$ in the expression for $j_\x$ in \Eq{convdiff1} and taking $\text{Pr}_\text{th} \simeq 1$, we find
\be
\label{eq:jxmatchderivation}
j_\x^\text{abs} = 
j_\text{vir} \times
\begin{cases}
-e^{-\frac{V_g \, (\xvir - \xth)}{D_\x}}  & , ~ x < \xvir
\\
1 -e^{-\frac{V_g \, (\xvir - \xth)}{D_\x}} & , ~ x > \xvir
\end{cases}
~~,~~
j_\x^\text{ref} = 
j_\text{vir} \times
\begin{cases}
0  & , ~ x < \xvir
\\
1 & , ~ x > \xvir
\end{cases}
~,
\ee
for an absorbing or reflecting boundary condition, respectively. Note that the above expressions agree with and, hence, justify the use of \Eq{jmatching}.

\section{Determining the Overdensities of Cosmological Relic MCPs}
\label{sec:results}

\addtocontents{toc}{\protect\setcounter{tocdepth}{1}}
\subsection{General Results}
\addtocontents{toc}{\protect\setcounter{tocdepth}{2}}
\label{sec:gentrafficjamresults}

The last few sections developed the formalism necessary to estimate the hydrostatic and traffic jam terrestrial overdensities for MCP relics. In this section, we numerically evaluate these overdensities within the larger MCP parameter space, using Eqs.~\ref{eq:nhydrostatic2}, \ref{eq:nxGreen4}, \ref{eq:nxGreen4b}, and \ref{eq:nxsemiref}. Note that in models where MCPs do not couple to Earth's electric field, Eqs.~\ref{eq:nxGreen4} and \ref{eq:nxGreen4b} are valid in the limit that $m_\x \ll 1 \ \GeV$ and \Eq{nxsemiref} strictly applies to $m_\x \gg 1 \ \GeV$, corresponding to solutions of the diffusion equation for absorbing or reflecting boundary conditions, respectively. Hence, our results for the traffic jam density are expected to break down near $m_\x \simeq 1 \ \GeV$. We postpone a more careful treatment to future work since this is only a small subregion of masses that we consider. We also remind the reader that in deriving the semi-analytic expressions of Eqs.~\ref{eq:nxGreen4}, \ref{eq:nxGreen4b}, and \ref{eq:nxsemiref}, we assumed that the ratio $V_g / D_\x \simeq m_\x \, g /  T_\oplus$ is spatially uniform. Since the temperature of the Earth varies within a couple orders of magnitude ranging from the core to the atmosphere, when evaluating the traffic jam overdensity $n_\x (x) / \nvir$ at a depth $x$ we utilize the average quantity $\langle V_g / D_\x \rangle$ which is obtained by averaging $V_g / D_\x$ across the spatial interval spanning $\min(x , \xvir , \xth) - \max(x , \xvir , \xth)$. We investigate the role of temperature gradients later in \Sec{tempgrad}, finding that they have a minor effect near Earth's surface. 

\begin{figure}[t]
\centering
\includegraphics[width=0.6\columnwidth]{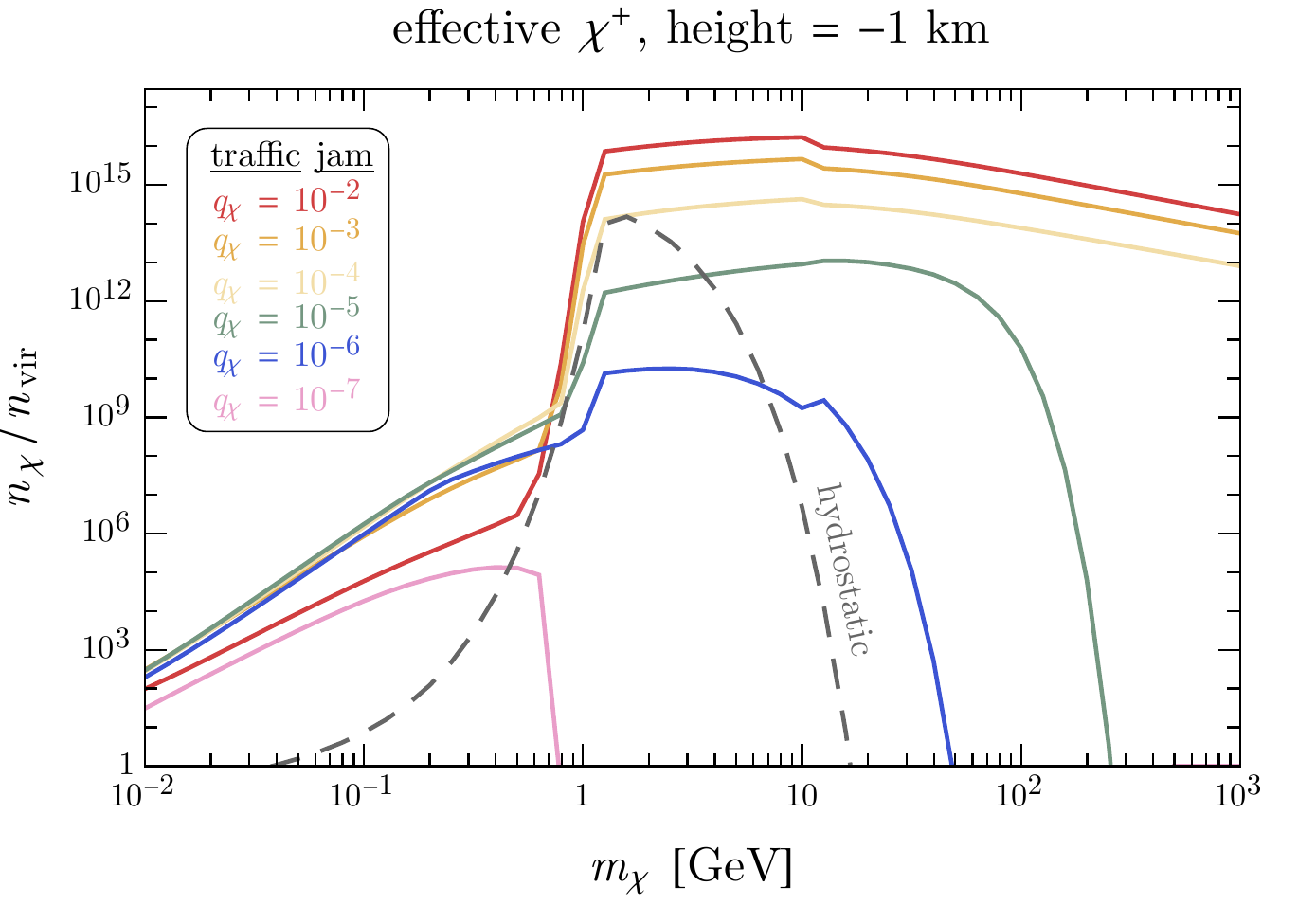}
\caption{The traffic jam (solid colored lines) and bound hydrostatic (dashed gray line) overdensity for positively-charged effective MCP relics, evaluated at a depth of $1 \ \km$, as a function of mass and for various choices of the charge $q_\x$. For the hydrostatic density we assume a temperature at the last scattering surface of $T_\oplus = 300 \ \K$.}
\label{fig:overdensity_relic}
\end{figure}

\np
\Fig{overdensity_relic} shows the separate contributions from the traffic jam and hydrostatic density of effective MCP relics (i.e., in the case that they do not couple efficiently to terrestrial electromagnetic fields), evaluated at a location of $1 \ \km$ underground and as a function of the MCP mass. 
The traffic jam density is shown as solid colored contours for various choices of the MCP coupling. The hydrostatic population is shown in dashed gray, assuming that the terrestrial temperature governing MCP evaporation is $T_\oplus (\xth) = 300 \ \K$ (see the discussion immmediately below \Eq{evap2}) and that the MCP coupling is large enough for these particles to thermalize in the Earth. Since for the largest charges considered additional complications arise from the formation of $\x^- - N^+$ bound states (see discussion below), \Fig{overdensity_relic} shows the density solely for positively-charged MCPs. As discussed in \Sec{static}, we see that the hydrostatic density is strongly peaked near $m_\x \sim 1 \ \GeV$, since much heavier masses are gravitationally attracted to reside closer to the Earth's center while much lighter masses are likely to evaporate over terrestrial timescales. We note that for sufficiently large couplings, the traffic jam density provides the dominant contribution to the terrestrial density at this depth for the full range of masses shown. Evaporation similarly  suppresses the traffic jam density for small masses. For large masses, the density is exponentially suppressed for charges sufficiently small such that MCPs thermalize well below $1 \ \km$ underground. This behavior can be understood by looking at the first line of the simplified expression in \Eq{reflectanalytic1}, which shows that at heights above the point at which MCP thermalizes in the Earth (corresponding to depth $x < \xvir$), the terrestrial density falls off exponentially with the corresponding distance scale $D_\x / V_g \sim T_\oplus / (m_\x \, g)$. Thus, this exponential suppression is increasingly severe for larger masses. 

\np
In \Fig{overdensity_relic}, the dependence of the traffic jam density on the MCP coupling $q_\x$ is reversed for high vs. low masses. For $ m_\x \gg 1 \ \GeV$, larger couplings imply a smaller gravitational drag velocity $V_g$ and thus an enhanced density, as in \Eq{reflectanalytic1}. For $m_\x \ll 1 \ \GeV$, the dependence on $q_\x$ is less trivial. For such masses, we find that the traffic jam density is governed by the diffusion coefficient $D_\x$ evaluated at the position $\xvir$ where a virialized MCP from the galactic halo thermalizes with the Earth. As $q_\x$ is reduced below $10^{-2}$, galactic MCPs tunnel further into the Earth's environment before thermalizing. In this case, $D_\x$ evaluated at $\xvir$ is reduced since the enhanced terrestrial density at $\xvir$ outweighs the smaller MCP coupling, ultimately giving rise to a larger MCP overdensity. This is to be expected, since if $\x$ thermalizes in denser environments, the outgoing diffusion flux is suppressed by the smaller mean free path between collisions, and hence a larger density is required to balance the incoming galactic flux of MCPs. As $q_\x$ is reduced further, the MCP density is eventually suppressed since MCPs thermalize well below a depth of $1 \ \km$. 

\begin{figure}[ht]
\centering
\includegraphics[width=0.49\columnwidth]{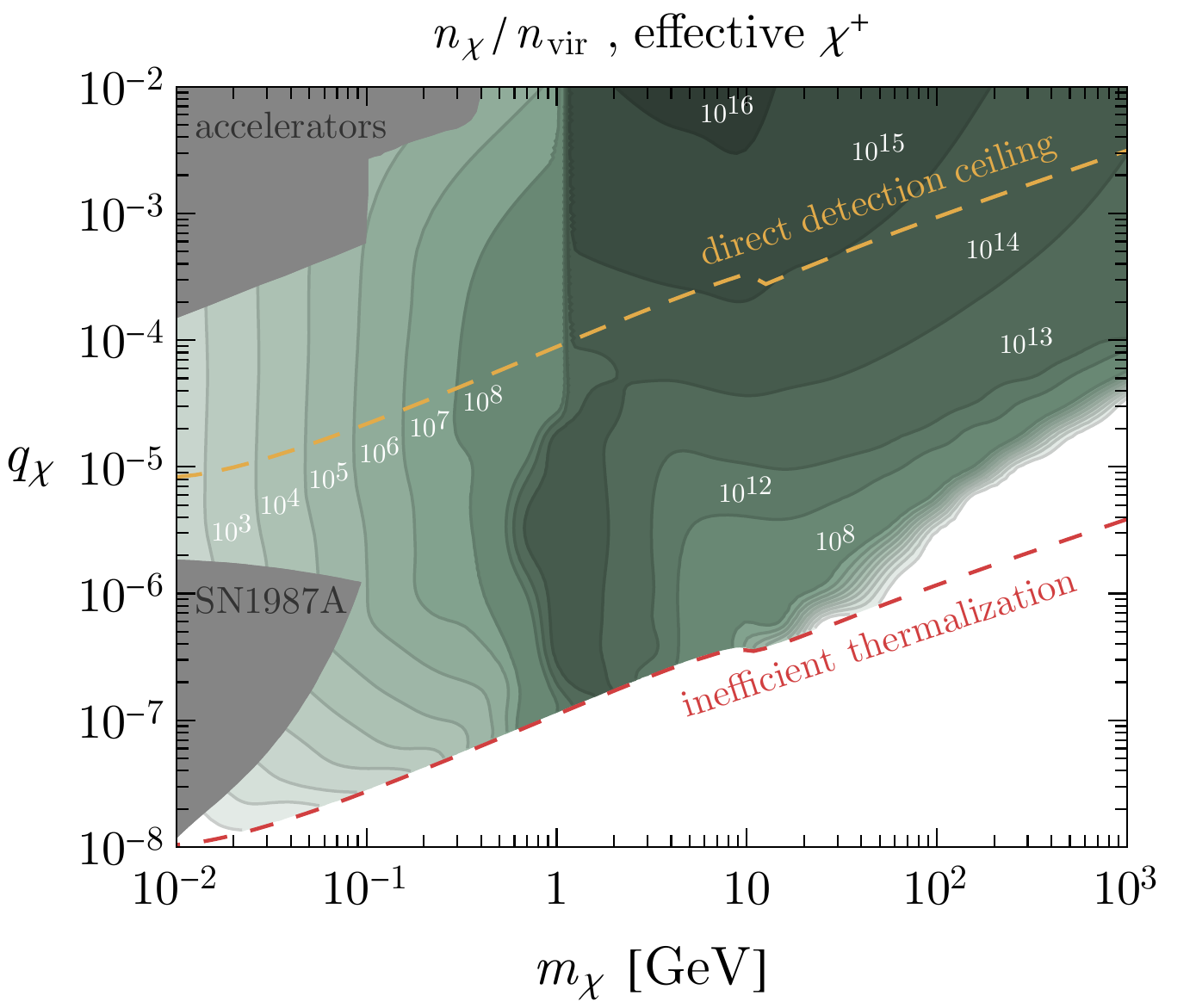}
\includegraphics[width=0.49\columnwidth]{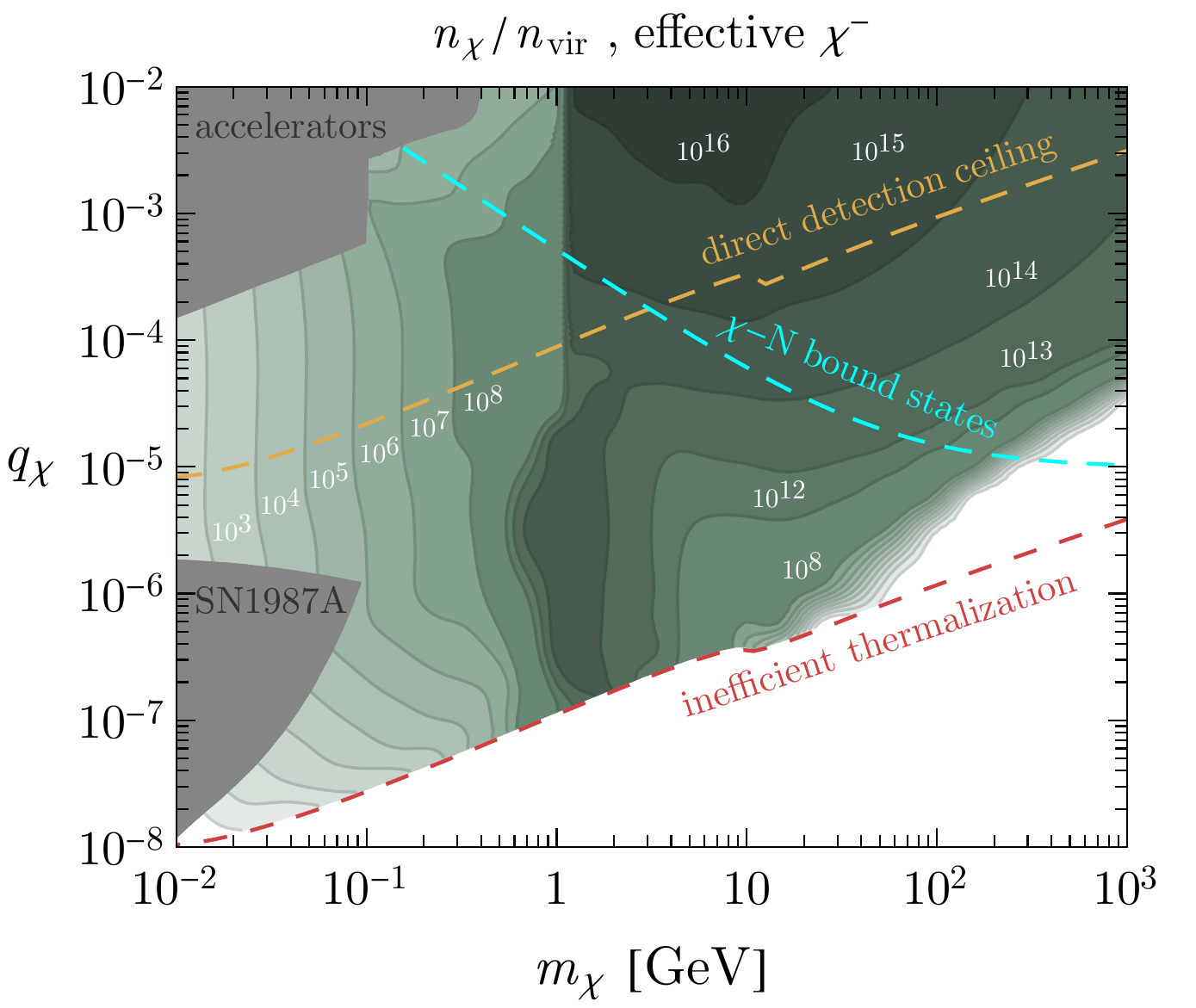}
\par\bigskip
\includegraphics[width=0.49\columnwidth]{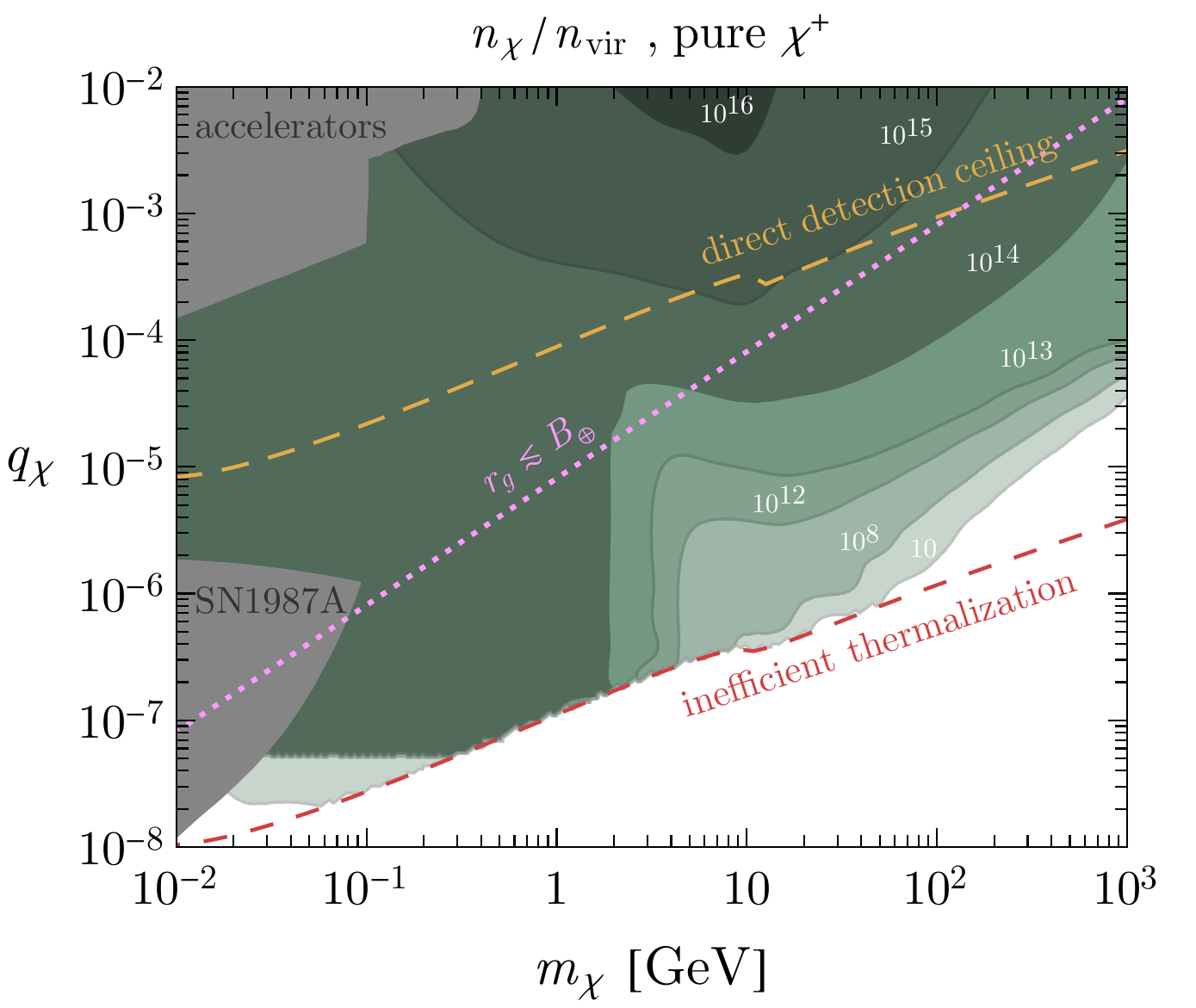}
\caption{The terrestrial overdensity of MCP relics evaluated at a depth of $1 \ \km$ underground, incorporating both the traffic jam and hydrostatic populations, as a function of the MCP mass $m_\x$ and coupling $q_\x$. The shaded green regions correspond to different values of the overdensity $n_\x / \nvir$, as labelled by the white text. In the top-left and top-right panels, these overdensities are evaluated for positively- and negatively-charged effective MCPs, respectively, corresponding to models in which the terrestrial electric field does not bind such particles to the Earth (see \Sec{EMfields}). The bottom panel presents our results for positively-charged pure MCPs where the terrestrial electric field binds these particles to the Earth. In this panel, the incoming flux of MCPs is significantly affected by Earth's magnetic field above the pink dotted line. Also shown in gray are existing constraints on MCPs from accelerator probes and neutrino experiments~\cite{Prinz:1998ua,ArgoNeuT:2019ckq,milliQan:2021lne,Davidson:2000hf,ArguellesDelgado:2021lek} as well as  observations of SN1987A~\cite{Chang:2018rso}. Above the orange and red dashed lines, galactic MCPs rapidly thermalize before traveling a distance of $1 \ \text{m}$ or $R_\oplus / 10$ in Earth's crust, respectively. Thus, the sensitivity of underground direct detection experiments is exponentially suppressed above the orange line, while below the red line, MCPs do not rapidly thermalize in the Earth. For couplings above the dashed cyan line in the top-right panel, negatively-charged MCPs efficiently form bound states with iron nuclei. Thus, in this region of parameter space, the formation of $\x-N$ bound states should drastically modify our estimates. \\ \\ \\ \\}
\label{fig:TrafficJamdensity}
\end{figure}

\np
In \Fig{TrafficJamdensity}, we show the total terrestrial overdensities  as a function of the MCP relic mass and charge, incorporating both the hydrostatic and traffic jam contributions and evaluated $1 \ \km$ underground (the results are nearly unchanged for anywhere near Earth's surface). In the top row, we show the density of positively- (left) and negatively-charged (right) MCPs in models of effective MCPs, in which case the terrestrial electric field inefficiently couples to MCPs (see \Sec{EMfields}). In the bottom panel, we show the density of positively-charged pure MCPs, in which case the terrestrial electric field binds such particles to the Earth. In this case, the region above the pink dotted line denotes couplings for which the MCP gyroradius is $r_g \lesssim R_\oplus$ and hence Earth's magnetic field is expected to alter the incoming flux. Also shown in solid gray are existing constraints~\cite{Prinz:1998ua,ArgoNeuT:2019ckq,milliQan:2021lne,Davidson:2000hf,ArguellesDelgado:2021lek,Chang:2018rso}. We do not show limits that rely on MCPs constituting some fraction of the DM abundance since these are sensitive to the particlar value of $n_\x \propto \nvir$ and are thus irrelevant for MCPs that comprise a very small DM subcomponent.  For instance, underground direct detection experiments are completely insensitive to MCPs that make up much less than $\sim 10^{-8}$ of the galactic DM density since the small flux of unthermalized particles is too small to detect~\cite{Pospelov:2020ktu}. The minimum couplings required for an MCP to thermalize before traveling a distance of $1 \ \text{m}$ or $R_\oplus / 10$ in Earth's crust are shown as the orange or red dashed lines, respectively. Thus, above the orange line the sensitivity of surface-level and underground direct detection experiments is exponentially suppressed, while below this line the sensitivity of conventional experiments depends sensitively on $n_\text{vir}$. Also shown in the top-right panel as a dashed cyan line is the minimum coupling $q_\x \simeq m_e / \muxN$ required for negatively-charged MCPs to efficiently form bound states with iron nuclei~\cite{Pospelov:2020ktu,Berlin:2021zbv}. Hence, above this line the large capture rate of $\x^-$ on terrestrial nuclei should drastically modify our results for the class of models considered here.

\np
For sufficiently large masses and small couplings, galactic MCP relics do not rapidly thermalize in the Earth which exponentially suppresses the magnitude of the terrestrial abundance, corresponding to the region below the dashed red line in \Fig{TrafficJamdensity}. For $q_\x \gtrsim 10^{-4}$ the traffic jam contribution to the density greatly exceeds that of the bound hydrostatic population for all masses. Instead, for $m_\x \simeq 1 \ \GeV$ and $q_\x \lesssim 10^{-4}$ the hydrostatic density is the dominant contribution, consistent with the results of \Fig{overdensity_relic}. We note that there are minor differences in the traffic jam contribution to the density when comparing the top-left and bottom panels, corresponding to positively- and negatively-charged effective MCPs, respectively. In particular, for large couplings negatively-charged MCPs have an enhanced interaction with atomic nuclei, leading to smaller diffusion coefficients and gravitational drag velocities, thus enhancing the local density (see Eqs.~\ref{eq:absorbanalytic2} and \ref{eq:reflectanalytic1}). Our results build upon and correct previous estimates in several important ways: 1) our treatment of the traffic jam population for $m_\x \lesssim 1 \ \GeV$ is qualitatively new and was not considered in previous studies, 2) for $m_\x \gtrsim 1 \ \GeV$ our estimates yield a significantly larger density than previously estimated in, e.g., Ref.~\cite{Pospelov:2020ktu}, due to an improved treatment of MCP-nuclear scattering, and 3) we have shown that Earth's electric field can exponentially increase the local density for sub-GeV pure MCPs.

\addtocontents{toc}{\protect\setcounter{tocdepth}{1}}
\subsection{Effects of Terrestrial Temperature Gradients}
\addtocontents{toc}{\protect\setcounter{tocdepth}{2}}
\label{sec:tempgrad}

In order to simplify our analysis throughout this work we have ignored modifications to the local MCP density arising from terrestrial temperature gradients. Although Ref.~\cite{Leane:2022hkk} recently claimed that such effects are an important contribution to the terrestrial density of sub-GeV relics (specifically for a model of strongly-coupled DM that scatters through heavy mediators), they did not present a quantitative comparison to an analysis that instead assumes a uniform temperature. In this section, we do just that. Our results show that our estimates for the MCP density near Earth's surface are unchanged at the level of $\order{10}\%$. 

\np
As derived in Ref.~\cite{landaukinetics} and discussed in \App{Gamma}, in the low-mass $m_\x \ll m_N$ limit the general form for the MCP diffusion current is
\be
\label{eq:jxlandau}
\jv_\x \simeq \frac{1}{3 n_N} \Big( \frac{m_\x \, \gv}{T_\oplus} - \grad \Big) \Big( n_\x \, \langle v_\text{rel} / \sigma_T \rangle \Big)
~.
\ee
Note that in the case that the thermally-average quantity $\langle v_\text{rel} / \sigma_T \rangle$ does not depend on position (as in the case of uniform temperature), the above expression reduces to \Eq{jx1}. At low masses, we can ignore the contribution from the gravitational field, and \Eq{jxlandau} can be trivially integrated to solve for the MCP density,
\be
\label{eq:nxlandauTraffic0}
n_\x (x) \, \langle v_\text{rel} / \sigma_T \rangle (x) \simeq -3 \int_{\xth}^x dx^\p \, n_N (x^\p) \, j_\x (x)
~,
\ee
where the thermal average of the bracketed quantity is evaluated at a temperature $T_\oplus (x)$ and we have enforced the absorber boundary condition at the last scattering surface, $n_\x (\xth) = 0$, for unbounded MCPs. We can determine the resulting traffic jam density from the matching condition for $j_\x$ in \Eq{jmatching}. In particular, we see that the above expression diverges at large depth, $x \gg \xvir$, unless $j_\x (x > \xvir) = 0$, implying that $j_\x (x < \xvir) = - j_\text{vir}$. Note that this behavior of $j_\x$ agrees with the $V_g \to 0$ limit of the first equation in \Eq{jxmatchderivation}. Using this in \Eq{nxlandauTraffic0} then yields the final form for the traffic jam overdensity in the low-mass limit 
\be
\label{eq:nxlandauTraffic1}
\lim_{m_\x \ll m_N}  \frac{n_\x (x)}{n_\text{vir}} \bigg|_\text{traffic jam} = \frac{3 \, V_\text{wind}}{\langle v_\text{rel} / \sigma_T \rangle (x)} \int_{\xth}^{\min{(x , \xvir)}} \hspace{-1.2cm} dx^\p ~ n_N (x^\p)
~.
\ee

\np
\Eq{jxlandau} can also be used to determine the accumulated hydrostatic density in the presence of temperature gradients for $m_\x \ll m_N$. The procedure is nearly identical to that already shown in \Sec{static}, except that \Eq{jxlandau} is used instead of \Eq{jx1} when enforcing $j_\x = 0$. Proceeding in this manner, we find that \Eq{nhydrostatic2} is modified to
\be
\label{eq:nxlandauStatic1}
\lim_{m_\x \ll m_N} n_\x(r) \bigg|_\text{hydrostatic} = \frac{\frac{1}{3} \, \overline{n}_\text{bound} \, R_\oplus^3 \, e^{- m_\x \, I(r)}}{\int_0^{R_\oplus} dr^\p ~r^{\p \, 2} ~ e^{- m_\x I (r^\p)} \, c_\sigma (r, r^\p)}
~~,~~
c_\sigma (r, r^\p) = \frac{\langle v_\text{rel} / \sigma_T \rangle (r)}{\langle v_\text{rel} / \sigma_T \rangle (r^\p)}
~,
\ee
where $ \overline{n}_\text{bound}$ and $I(r)$ are defined in \Sec{static}. Note that \Eq{nxlandauStatic1} is identical to our previous result in \Eq{nhydrostatic2} aside from the inclusion of $c_\sigma (r,r^\p)$ in the denominator of the expression for $n_\x$. 

\np
It is instructive to compare the results of Eqs.~\ref{eq:nxlandauTraffic1} and \ref{eq:nxlandauStatic1} to our previous analysis that ignored the effect of temperature gradients. For $m_\x \ll 10 \ \GeV$, we find good agreement between the two approaches at the level of $\order{10} \%$ when evaluating the density near Earth's surface. In that sense, ignoring temperature gradients introduces lesser error than, e.g., imperfect modeling of $\text{Pr}_\text{th}$. It is possible that in regions of much larger temperature gradients, such as high up in the ionosphere or in the outskirts of stellar bodies, MCP densities are modified more significantly~\cite{Leane:2022hkk}. We leave a more detailed investigation to future work.

\addtocontents{toc}{\protect\setcounter{tocdepth}{1}}
\subsection{Laboratory-Induced Modifications to the Traffic Jam Density}
\addtocontents{toc}{\protect\setcounter{tocdepth}{2}}
\label{sec:localtrafficjam}

The analysis above calculated the overdensity of MCPs due to the surrounding density of normal matter near Earth's surface. But, what happens to these ambient MCP densities in an underground lab, in which the density of normal matter is much smaller compared to the density of the surrounding rock? Similarly, what is the density of MCPs near precision sensors, whose components are much more dense and cold than the surrounding environment? In this section, we investigate such modifications to the local MCP density arising from variations in the immediate local environment. 

\begin{figure}[t]
\centering
\includegraphics[width=0.6\columnwidth]{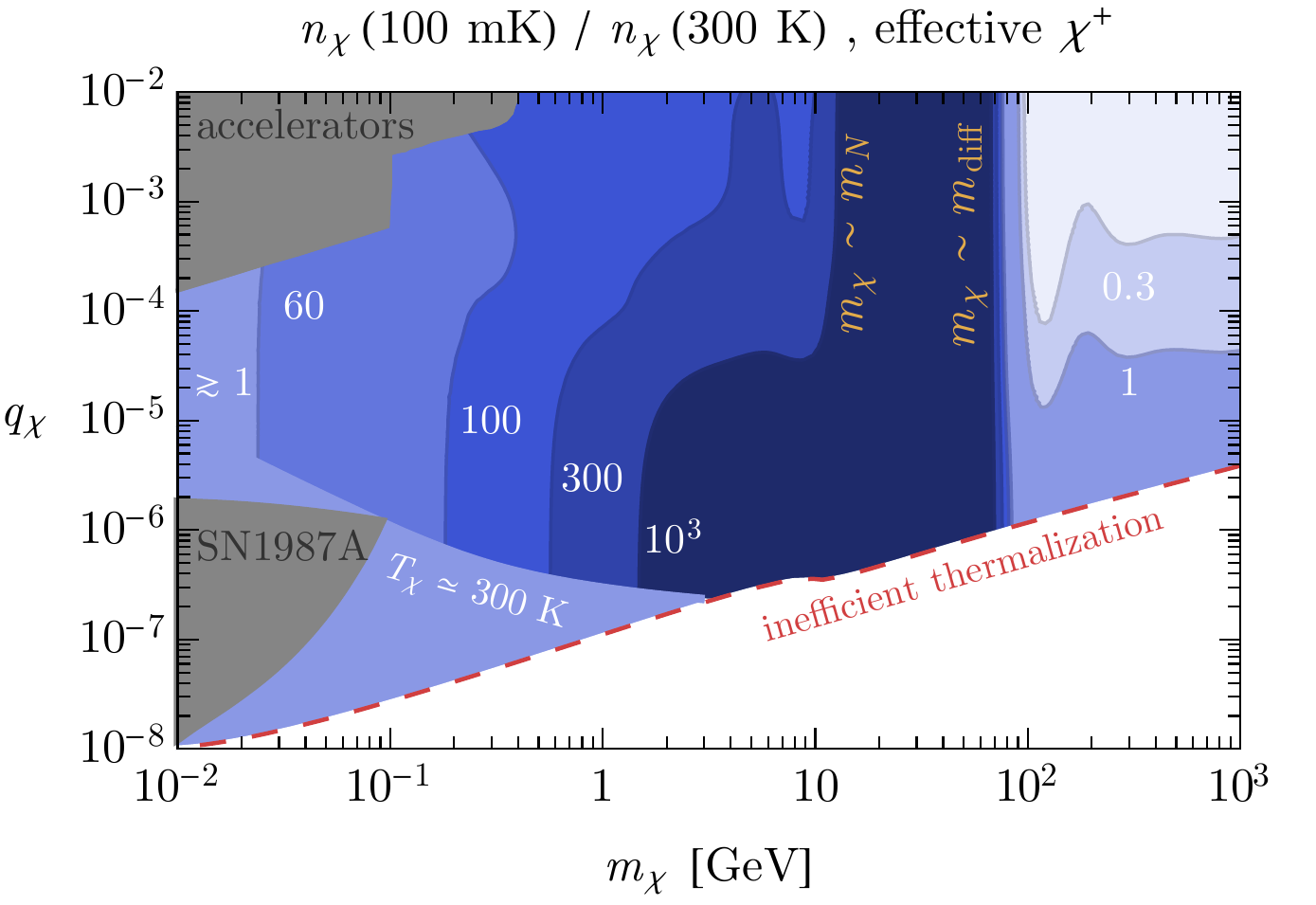}
\caption{In shaded blue with corresponding white labels, enhancement to the local traffic jam density in a $1 \ \text{m}$ sized $100 \ \text{mK}$ cryogenic environment, as compared to room temperature, as a function of the MCP mass $m_\x$ and coupling $q_\x$. These overdensities are evaluated for positively-charged effective MCPs scattering off of oxygen nuclei $N$. Also shown in gray are existing constraints on MCPs from accelerator probes and neutrino experiments~\cite{Prinz:1998ua,ArgoNeuT:2019ckq,milliQan:2021lne,Davidson:2000hf,ArguellesDelgado:2021lek} as well as  observations of SN1987A~\cite{Chang:2018rso}. Above the red dashed line, galactic MCPs rapidly thermalize before traveling a distance of $R_\oplus / 10$ in Earth's crust. In the region labelled $T_\x \simeq 300 \ \K$, room temperature MCPs do not efficiently thermalize with the cryogenic region within a distance of $1 \ \text{m}$. For MCP masses close to $m_N \sim 15 \ \GeV$ or $m_\text{diff} \equiv 100 \ \text{mK} / (g \times 1 \ \text{m}) \sim 80 \ \GeV$, our analytic approximations are not expected to hold. 
}
\label{fig:temp_enhance}
\end{figure}

\np
Let us begin by first investigating the low-mass limit, where the effect of Earth's gravitational field can be ignored. The general result was derived above in \Eq{nxlandauTraffic1}. In particular, assuming that incoming MCPs thermalize at shallower depths than the lab, the position-dependence of the local overdensity scales simply as
\be
\lim_{m_\x \ll m_N}  \frac{n_\x (x > \xvir)}{n_\text{vir}} \propto \frac{1}{\langle v_\text{rel} / \sigma_T \rangle (x)}
~.
\ee
In the left-half of \Fig{temp_enhance}, we plot ratios of this quantity at two different temperatures, comparing it at cryogenic ($T_\x = 100 \ \text{mK}$) and room ($T_\x = 300 \ \K$) temperatures. As discussed in \Sec{scattering}, the cross section $\sigma_T$ is approximately independent of temperature at very low masses. Hence, for the lowest masses shown, the low-temperature enhancement to the MCP density simply scales with temperature as $\sqrt{1/T_\x} \, $. For slightly heavier masses near $m_\x \sim \text{few} \times \GeV$, $\sigma_T$ becomes increasingly large at low temperatures, leading to a greater low-temperature density enhancement. 

\np
Next, we consider the high-mass limit, $m_\x \gtrsim m_N \sim 10 \ \GeV$. As discussed in \App{Gamma}, in this case the general form for the MCP diffusion current is
\be
\label{eq:jxhighmasstempgrad}
\lim_{m_\x \gg m_N} \jv_\x =  \frac{1}{\Gamma_p} \bigg( \gv \, n_\x - \frac{\grad (T_\x n_\x)}{m_\x} \bigg)
~,
\ee
where the temperature $T_\x (x)$ is a function of position. To determine the MCP density, we solve \Eq{jxhighmasstempgrad} in a similar manner as described in \Sec{steadystate}, imposing a reflecting boundary condition at $\xth$. Assuming that incoming MCPs thermalize at shallower depths than the lab, the solution which generalizes \Eq{nxsemiref} is given by
\be
\label{eq:nxRefTempGrad}
\lim_{m_\x \gg m_N} \frac{n_\x (x > \xvir)}{\nvir} =  \text{Pr}_\text{th} \, V_\text{wind} \, \int_x^\infty d x^\p ~ \text{Exp} \bigg[ - \int_x^{x^\p} dx^{\p \p} ~ \frac{g \, m_\x - \partial_{x^{\p \p}} T_\x (x^{\p \p})  }{T_\x (x^{\p \p})} \bigg] ~  \frac{1}{D_\x (x^\p)}
~.
\ee
To evaluate the above expression, we assume that the environment consists of two regions of distinct temperature: outside the lab with temperature $T_\oplus$, and inside the lab with temperature $T_\text{lab} \ll T_\oplus$. Modeling the laboratory as a one-dimensional region centered at the point $x_\text{lab}$ with width $\Delta x_\text{lab}$, we find that the laboratory density of MCPs (compared to the ambient density outside of the lab) reduces to
\be
\label{eq:heavythermvariation}
\lim_{m_\x \gg m_N} \frac{n_\x (x_\text{lab})}{n_\x (\text{outside lab})} 
=  
\begin{cases}
T_\oplus / T_\text{lab} & (m_\x \ll m_\text{diff} \equiv T_\text{lab} / g \, \Delta x_\text{lab})
\\
V_g^{(\oplus)} / V_g^{(\text{lab})} & (m_\x \gg m_\text{diff} \equiv T_\text{lab} / g \, \Delta x_\text{lab})
~,
\end{cases}
\ee
where $V_g^{(\oplus, \text{lab})} = g / \Gamma_p^{(\oplus, \text{lab})}$ is the gravitational drift velocity outside or inside the lab, respectively. In the first and second lines of the expression above, we have taken the limit that the MCP mass is much less or greater than $m_\text{diff} \equiv T_\text{lab} / g \, \Delta x_\text{lab}$, respectively. For $m_\x \gg m_\text{diff}$, the effect of thermal diffusion is negligible compared to the gravitationally induced drift, since in this case the timescale to gravitationally-drift through the laboratory is much shorter than the time spent diffusing the same distance, i.e., $\Delta x_\text{lab} / V_g^{(\text{lab})} \ll \Delta x_\text{lab}^2 / D_\x^{(\text{lab})}$. As a result, the laboratory-induced modification to the MCP density depends solely on the gravitational drift velocity $V_g = g / \Gamma_p$. Instead, for $m_\x \ll m_\text{diff}$, Earth's gravitational field can be ignored such that conservation of MCP pressure $P_\x \simeq T_\x \, n_\x$  (see \Eq{jxhighmasstempgrad}) implies that $n_\x \propto 1 / T_\x$. The modification to the MCP density for a $\Delta x_\text{lab} = 1 \ \text{m}$ sized region at $T_\text{lab} = 100 \ \text{mK}$  is shown for $m_\x \gtrsim m_N$ in the right-half of \Fig{temp_enhance}. In this case, for $m_\x \gtrsim m_N , m_\text{diff}$ the density is reduced at small temperature due to the fact that $\sigma_T$ saturates for very large charges  and the relative velocity is reduced with smaller temperature, which enters through  $\Gamma_p \sim n_N \sigma_T v_\text{rel} \, $. Instead, for $m_N \lesssim m_\x \lesssim m_\text{diff}$, there is a sizable enhancement since the density scales inversely with the temperature. 

\np
In the above example, we considered the effect on the MCP density arising from temperature variations inside a laboratory. However, over what small length-scale does a local variation to the nuclear density become relevant? To investigate this, we assume a uniform temperature environment, such that we can use the formalism previously developed in \Sec{traffic}, and instead we consider the effect of nuclear density variations. Examples of such scenarios include an underground cavern that is equilibrated with the surrounding rock, or a precision sensor that is much denser than the surrounding cryogenic environment. Let us refer to the diffusion coefficient and gravitational drag velocity outside the laboratory as $D_\x^{(\oplus)} , V_g^{(\oplus)} = \const$ and inside the laboratory as $D_\x^{(\text{lab})} , V_g^{(\text{lab})} = \const$, where $D_\x^{(\oplus)} \neq D_\x^{(\text{lab})}$ and $V_g^{(\oplus)} \neq V_g^{(\text{lab})}$ and $D_\x / V_g \equiv D_\x^{(\text{lab})} / V_g^{(\text{lab})} = D_\x^{(\oplus)} / V_g^{(\oplus)}  = T_\x / (m_\x \, g) = \const$, corresponding to a non-uniform nuclear density and uniform temperature between the two regions. We model the laboratory as extending across the finite spatial region $[x_\text{min}^{(\text{lab})} \, , \, x_\text{max}^{(\text{lab})}]$, and, for simplicity, we incorporate the presence of only a single species of nucleus. First, note that for $m_\x \ll m_N$, \Eq{nxlandauTraffic1} implies that for $x > \xvir$, $n_\x$ is independent of the local nuclear density $n_N (x)$. For $m_\x \gg m_N$, we can evaluate the integral in \Eq{nxsemiref} analytically. Inside the lab (i.e., $x_\text{min}^{(\text{lab})} \leq x \leq x_\text{max}^{(\text{lab})}$), we find
\be
\label{eq:labtrafficjam1}
\lim_{m_\x \gg m_N} \frac{n_\x (\text{inside lab})}{\nvir} 
\simeq 
 \text{Pr}_\text{th} \times
\begin{cases}
\frac{\Vwind}{V_g^{(\text{lab})}} & (D_\x / V_g \ll x_\text{max}^{(\text{lab})} - x)
\\
\frac{\Vwind}{V_g^{(\oplus)}} & (D_\x / V_g \gg x_\text{max}^{(\text{lab})} - x)
~.
\end{cases}
\ee
Moreover, we can evaluate the MCP overdensity outside of the laboratory (e.g., $x < x_\text{min}^{(\text{lab})}$), which yields
\be
\label{eq:labtrafficjam2}
\lim_{m_\x \gg m_N}  \frac{n_\x (\text{outside lab})}{\nvir} \simeq 
 \text{Pr}_\text{th} \times
\begin{cases}
\frac{\Vwind}{V_g^{(\text{lab})}} & (x_\text{min}^{(\text{lab})} - x \ll D_\x / V_g \ll x_\text{max}^{(\text{lab})} - x_\text{min}^{(\text{lab})})
\\
\frac{\Vwind}{V_g^{(\oplus)}} & (\text{otherwise})
~.
\end{cases}
\ee
Comparing \Eq{labtrafficjam1} to the second line of \Eq{reflectanalytic1}, we see that for $m_\x \gg m_N$ the MCP overdensity inside the lab tracks the locally expected value\footnote{By the ``locally expected value" we mean that $n_\x (x) / \nvir$ is given by the homogenous result in \Eq{reflectanalytic1} with $D_\x \to D_\x (x)$ and $V_g \to V_g (x)$.} only if the effective length scale $D_\x / V_g = T_\oplus / (m_\x \, g)$ is much smaller than the size of the laboratory. Otherwise, it is partially-determined by the parameters of the external environment. Similarly, from \Eq{labtrafficjam2} we see that the MCP overdensity outside the lab is altered by the presence of the lab if the effective length $D_\x / V_g$ is larger than the distance to the lab in addition to being smaller than the size of the lab itself. Note that this is similar to our finding for temperature variations in \Eq{heavythermvariation}, since $m_\x = m_\text{diff}$ is equivalent to $D_\x^{(\text{lab})} / V_g^{(\text{lab})} = \Delta x_\text{lab}$. In this sense,
\be
\label{eq:DxoverVg}
\frac{D_\x}{V_g} \sim  \order{1} \ \text{m} \times \Big( \frac{T_\oplus}{100 \ \text{mK}} \Big) ~ \Big( \frac{100 \ \GeV}{m_\x} \Big)
\ee
(where we have normalized the temperature to a typical value for cryogenic instruments) is the effective ``resolution length scale" of the traffic jam density over which the MCP population tracks variations in local parameters, as alluded to in the discussion below \Eq{DiffDef}. Hence, in uniform temperature environments, laboratory apparatuses larger than \Eq{DxoverVg} may modify the density of the traffic jam population relative to the external environment. Note that this finding is justified from the point of view of energy conservation; \Eq{DxoverVg} is roughly the distance that an MCP with kinetic energy $T_\oplus$ can climb up the potential energy hill of Earth's gravitational field, i.e., $m_\x \, g \, (D_\x / V_g) \sim T_\oplus$. Thus, thermal-averaging ``washes out" structure on length scales smaller than this value.

\section{Determining the Overdensities of Cosmic-Ray Produced MCPs}
\label{sec:cosmicray}

We now generalize the formalism developed in the previous sections to determine the terrestrial density of MCPs that thermalize in Earth's crust after being produced from the collisions of high-energy cosmic rays in the upper atmosphere. Unlike models of MCP relics, this population of MCPs constitutes an irreducible contribution to the terrestrial density independent of dynamics in the early Universe. Regardless, as we will discuss below, the density of such particles is still subject to the details of the model, since, analogous to the discussion in \Sec{EMfields}, Earth's electromagnetic fields may significantly impact the resulting population after thermalization (before thermalization, the relativistic boost of such particles implies that local electromagnetic fields are of little importance). 

\np
Given the spectrum of such MCPs and their probability to thermalize within the Earth, determining the resulting terrestrial density is a simple generalization of the formalism developed in Secs.~\ref{sec:static} and \ref{sec:traffic}. In particular, we previously approximated the influx of relic MCPs as $\jvir \simeq \nvir \, \Vwind$. For cosmic-ray produced MCPs, we quantify the spectrum as the differential flux $\jcr$ per $\gamma_\x v_\x$, $d \jcr / d (\gamma_\x v_\x)$, where $\gamma_\x$ and $v_\x$ are the boost and velocity of the produced MCP, respectively. To determine the accumulated hydrostatic bound density of MCPs produced from cosmic rays, the accumulation rate of Eqs.~\ref{eq:noplusdot} and \ref{eq:ncap} is modified to
\be
\dot{n}_\text{cap} \simeq \int d (\gamma_\x v_\x) ~  \text{Pr}_\text{th} (\gamma_\x) ~ \frac{d\jcr}{d (\gamma_\x v_\x)}  ~ \frac{4\pi R_\oplus^2}{(4 \pi / 3) \, R_\oplus^3} 
~,
\ee
where $\text{Pr}_\text{th}$ is the probability for an MCP of boost $\gamma_\x$ to stop and fully thermalize within Earth's environment (note that we cannot use \Eq{ProbTherm1} here to determine $\text{Pr}_\text{th}$, since that is only valid for non-relativistic MCPs). The calculation of $\text{Pr}_\text{th}$ and $d \jcr / d (\gamma_\x v_\x)$ will be discussed below. The rest of the analysis to determine the hydrostatic density is identical to that of \Sec{static}.

\np
The results of \Sec{traffic} can also be modified in a straightforward manner in order to determine the traffic jam density $n_\x$ of thermalized MCPs produced from cosmic rays. Indeed, the traffic jam overdensity $(n_\x / \nvir) \big|_\text{relic}$ of relic MCPs from \Sec{traffic} is simply related to the density produced from cosmic rays through 
\be
n_\x = \int d (\gamma_\x v_\x) ~ \frac{d\jcr}{d (\gamma_\x v_\x)} ~ \text{Pr}_\text{th} (\gamma_\x) ~ \frac{(n_\x / \nvir) \big|_\text{relic}}{\Vwind} 
~.
\ee
Note that when integrating over momenta in the integral above, $(n_\x / \nvir) \big|_\text{relic}$ is determined using the same expressions as in \Sec{traffic} (i.e., Eqs.~\ref{eq:nxGreen4}, \ref{eq:nxGreen4b}, and \ref{eq:nxsemiref}), with the thermalization point $\xvir$ calculated as a function of $\gamma_\x v_\x$ (to be determined below), and the last scattering surface $\xth$ determined exactly as in \Sec{traffic}.

\np
The spectrum and stopping power of MCPs produced by cosmic-rays has been calculated previously in Refs.~\cite{Plestid:2020kdm,Harnik:2020ugb,ArguellesDelgado:2021lek}. In this work, we adopt the results of Refs.~\cite{Plestid:2020kdm}, which calculated the secondary production of MCPs from meson decays. For $m_\x \lesssim \order{100} \ \MeV$, pion decays, e.g., $\pi^0 \to \g \,  \x^+ \x^-$, constitute the dominant production mechanism, whereas the decay of vector mesons, e.g., $\rho, \phi, J/\psi \to \x^+ \x^-$, is responsible for the production of heavier MCPs. We conservatively do not include the interstellar medium contribution calculated in Ref.~\cite{Harnik:2020ugb}; in that work the effective range of the Coulombic force was assumed to be galactic and the propagation of the slow MCP flux through the solar wind and Earth's magnetic field requires careful analysis that is beyond the scope of this work. In our analysis, we also use the MCP stopping power as determined in Ref.~\cite{Plestid:2020kdm} to calculate the point $\xvir$ within Earth's crust, demanding that $\x$ thermalizes before traversing a full Earth radius, $R_\oplus$. In particular, we set the thermalization probability to $\text{Pr}_\text{th} (\gamma_\x)  = 0$ when the penetration depth is greater than $R_\oplus$.

\begin{figure}[t]
\centering
\includegraphics[width=0.49\columnwidth]{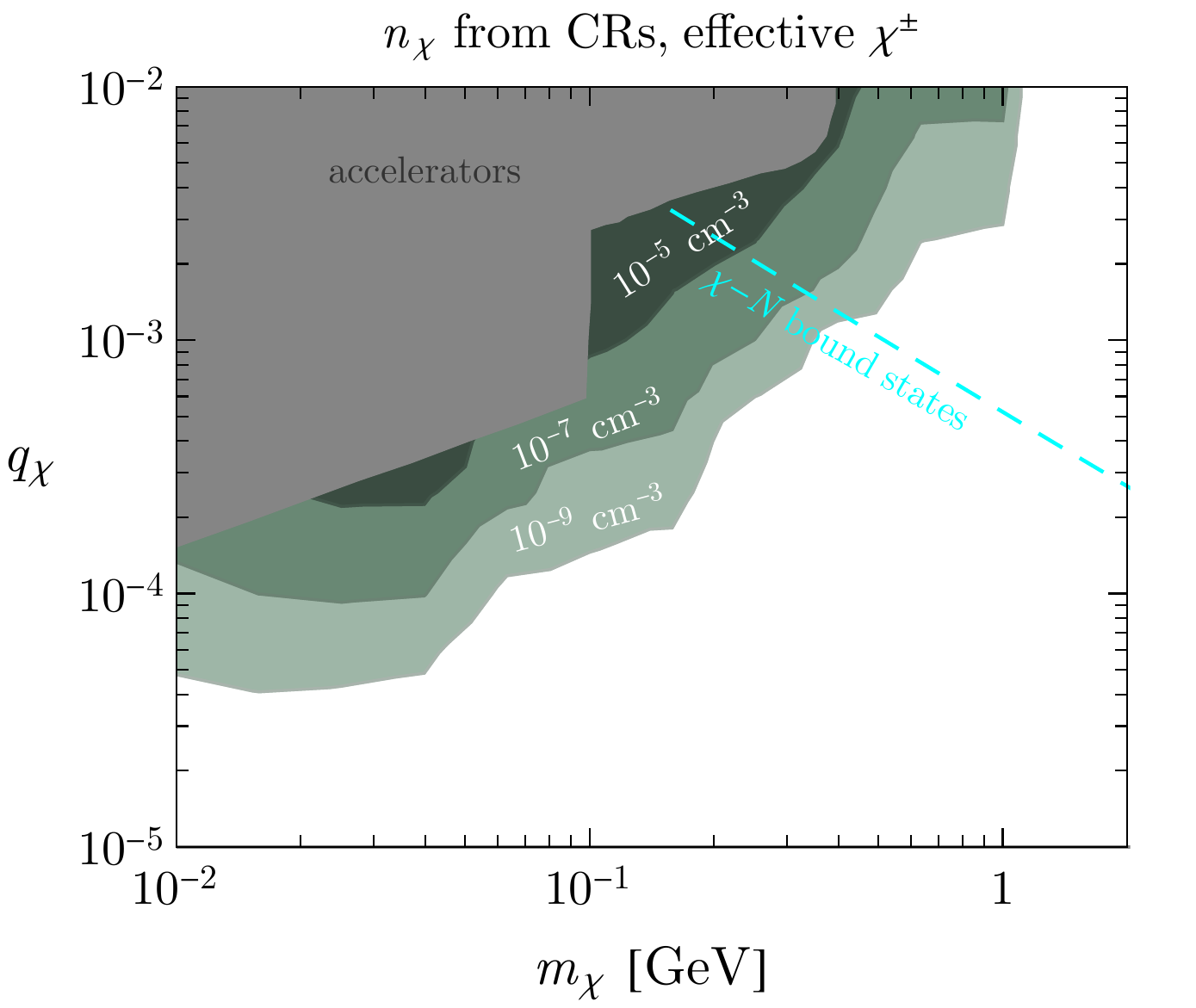}
\includegraphics[width=0.49\columnwidth]{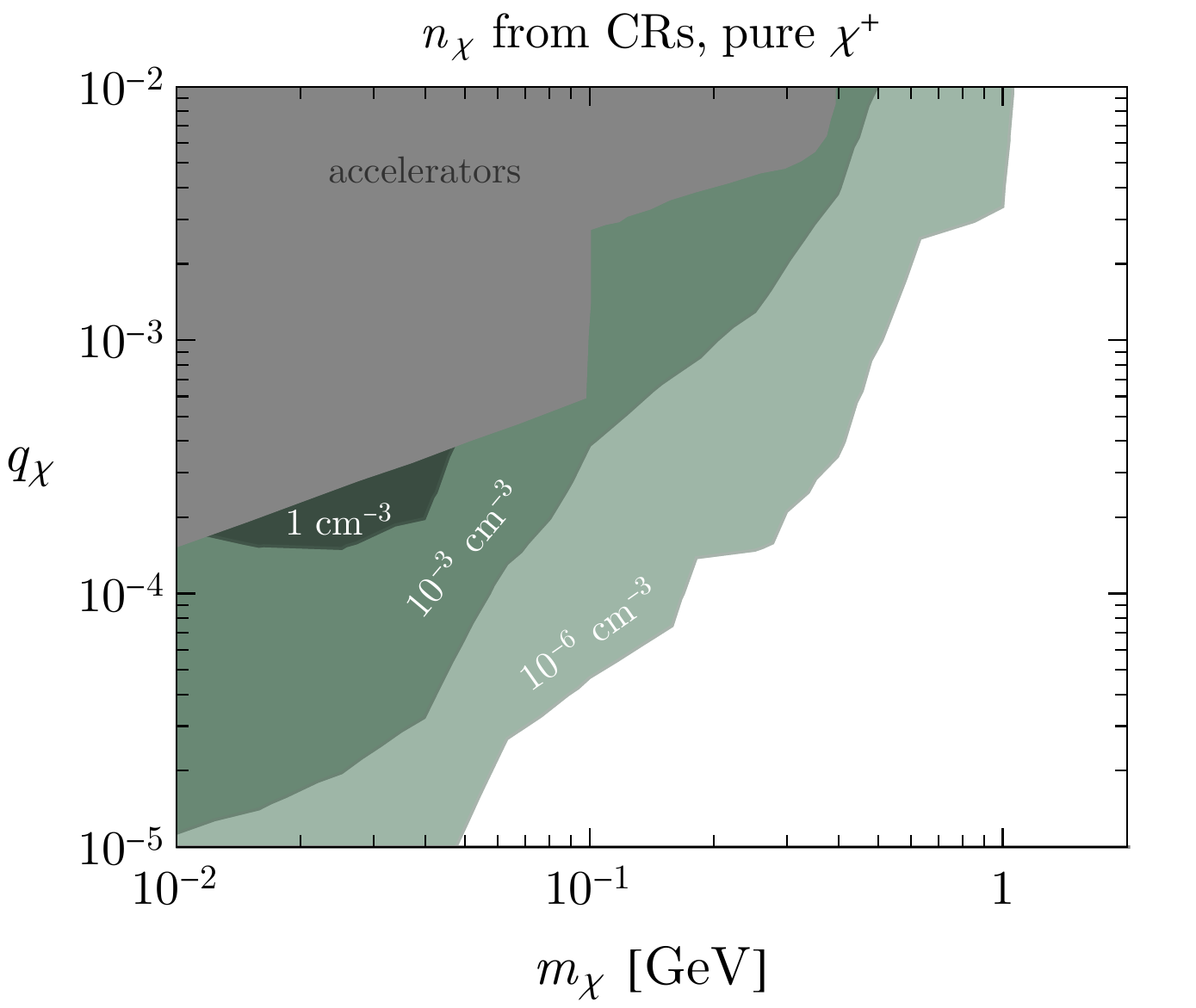}
\caption{Analogous to \Fig{TrafficJamdensity}, the total thermalized density of MCPs evaluated at a depth of $1 \ \km$ underground, but now shown for MCPs produced by the collisions of high-energy cosmic rays. In the left panel, we assume that the terrestrial electric field does not couple to MCPs (as in models where the coupling is generated by a kinetically-mixed light dark photon), resulting in a greatly diminished terrestrial density.   In this case, we average the momentum drag rate $\Gamma_p$ for positively- and negatively-charged MCPs, before evaluating the terrestrial density. In the right panel, we instead assume that MCPs efficiently couple to Earth's electric field, and thus can remain electrically bound to the Earth after thermalizing to terrestrial temperatures. Since in this case negatively-charged MCPs are expelled from the atmosphere, this panel only shows the density of positively-charged MCPs. 
}
\label{fig:CRdensity}
\end{figure}

\np
The number density of thermalized MCPs produced from cosmic rays is shown as a function of the mass and charge in \Fig{CRdensity}, evaluated at a depth of $1 \ \km$ underground (the depth-dependence is not significant for the light masses shown). In the right panel, we assume that positively-charged MCPs are efficiently trapped in Earth's crust by the atmospheric electric field. As discussed in \Sec{EMfields}, this is relevant in models of pure MCPs, where the small millicharge is generated without a kinetically-mixed dark photon. In this case, highly relativistic MCPs produced by cosmic rays are unattenuated by Earth's magnetic field, but positively-charged MCPs remain efficiently trapped by the atmospheric voltage barrier after thermalizing in the crust. As a result, the accumulated bound population dominates over the inflowing traffic jam population for most of the parameter space shown, leading to densities as large as $n_\x \sim 1 \ \cm^{-3}$ for certain masses and couplings. 

\np
In the left panel of \Fig{CRdensity}, we instead assume that $q_\x$ is generated via a dark photon, such as in models of effective MCPs where the effect of Earth's electric field has a negligible effect (see \Sec{EMfields} for additional details). For such models, sub-GeV MCPs are not bound by Earth's electromagnetic or gravitational fields, such that the traffic jam population dominantly contributes to the terrestrial density. However, the evaporation of such particles over short timescales greatly diminishes the thermalized abundance, resulting in densities smaller than $n_\x \sim 10^{-5} \ \cm^{-3}$ throughout the parameter space shown.

\section{Conclusions}
\label{sec:conclusion}

We have considered in detail the accumulation and distribution of exotic millicharged particles (MCPs) in the Earth and its atmosphere. Our main results for the terrestrial density enhancement, as presented in \Fig{TrafficJamdensity}, can be used for further exploration of observational effects associated with MCPs (such as their possible annihilations, scattering in low-threshold dark matter detectors, anomalous heat/charge transport, etc.). In this work, we have presented the best quality to date estimate of the hydrostatic population and the steady flow (traffic jam density) of MCPs. In particular, we have incorporated the intricate dependence of the MCP-atomic scattering cross section on many model and environmental parameters, as well as interactions with terrestrial electromagnetic fields, which significantly complicate the treatment compared to the case of dark matter particles with a constant elastic cross section on atoms. Furthermore, we have developed a semi-analytic formalism necessary for dealing with many aspects of the problem: initial thermalization, gravitationally-biased diffusion, and evaporation of MCPs. This formalism can be applied to specific experimental configurations in order  to account for local variations in temperature and nuclear density and the resulting modifications to the local MCP density.

\np
A potentially important aspect of the treatment, so far ignored in our paper, is the role of MCP self-interactions. For instance, annihilations, whose associated timescale is likely to be rather long, enter as a ``sink" to the MCP density in addition to losses from evaporation. Furthermore, depending on the density, the self-interaction of MCPs may affect the accumulation and distribution in the case that the self-interaction energy significantly contributes to the pressure, such that the MCP population cannot be accurately modeled as an ideal gas. This is most relevant for large densities of effective MCPs, in which case self-interactions are not suppressed by the small degree of $\Ap - \gamma$ kinetic mixing. Both of these effects (annihilation and self-scattering) are, however, higher order in the MCP density, and therefore our results provide a complete treatment in the limit of small densities. 

\np
A detailed understanding of the terrestrial MCP density is an essential tool in order to place meaningful constraints on such models. Beyond this, it would be interesting to apply this understanding to new experimental strategies geared towards trapping or significantly enhancing the MCP density in the lab. For instance, optimized configurations of cryogenic or electromagnetic systems could be used to enhance the sensitivity of existing detection strategies, such as those employing ion traps~\cite{Budker:2021quh}. We leave a larger exploration of such experimental strategies to future work.


\paragraph{Acknowledgements}
The authors would like to acknowledge Rebecca Leane and Juri Smirnov for helpful conversations. This material is based upon work supported by the U.S. Department of Energy, Office of Science, National Quantum Information Science Research Centers, Superconducting Quantum Materials and Systems Center (SQMS) under contract number DE-AC02-07CH11359. Fermilab is operated by the Fermi Research Alliance, LLC under Contract DE-AC02-07CH11359 with the U.S. Department
of Energy. HL is supported by the DOE under Award Number DE-SC0007968, NSF grant PHY1915409, and the Simons Foundation. MP is supported
in part by U.S. Department of Energy (Grant No. DESC0011842). HR is supported in part by NSF Grant
PHY-1720397 and the Gordon and Betty Moore Foundation Grant GBMF7946.

\addtocontents{toc}{\protect\vskip24pt}
\addcontentsline{toc}{section}{Appendix}

\begin{appendices}

\section{Momentum Drag Rate}
\label{app:Gamma}

In this appendix, we discuss the derivation of the scattering rate $\Gamma_p$ provided in, e.g., \Eq{Gamma1}. The starting point is the non-relativistic Boltzmann equation for the scattering process $\x_i N_i \to \x_f N_f$,
\be
\label{eq:Boltz1}
\frac{d f_{\x_i}}{d t} \simeq - \frac{1}{2 m_\x} \, \int \frac{d^3 \pv_{\x_f}}{(2 \pi)^3 \, 2 m_\x} \, \frac{d^3 \pv_{N_i}}{(2 \pi)^3 \, 2 m_N} \, \frac{d^3 \pv_{N_f}}{(2 \pi)^3 \, 2 m_N} ~ (2 \pi)^4 \, \delta^4 (p_{\x_f}^\mu + p_{N_f}^\mu - p_{\x_i}^\mu - p_{N_i}^\mu) ~ \overline{|\mathcal{M}|^2} ~ \mathcal{F}
~,
\ee
where the subscript $i$ or $f$ refers to an initial or final state,  $\overline{|\mathcal{M}|^2}$ is the spin-averaged squared matrix element, and $\mathcal{F} \simeq f_{\x_i} f_{N_i} - f_{\x_f} f_{N_f}$ is the phase space factor. 

\Eq{Boltz1} has been evaluated in the small MCP mass $m_\x \ll m_N$ limit in Sec.~1.11 of Ref.~\cite{landaukinetics}. As discussed there, in the limit that light MCPs scatter rapidly with nuclei, their  phase space can be expanded around a Maxwell-Boltzmann distribution $f_\x^0$, i.e., $f_\x (\pv_\x) \simeq f_\x^0 (p_\x) + \Delta f_\x (\pv_\x)$, where $\Delta f_\x / f_\x^0 \ll 1$. In this case, \Eq{Boltz1} evaluates to~\cite{landaukinetics}
\be
\label{eq:Boltz2}
\lim_{m_\x \ll m_N} \frac{d f_{\x_i}}{d t} \simeq - n_N \, v_\text{rel} \,  \sigma_T \, \Delta f_{\x_i}
~,
\ee
where $n_N$ is the nuclear number density, $v_\text{rel}$ is the initial state relative velocity, and $\sigma_T$ is the transfer cross section. The LHS of \Eq{Boltz2} is then expanded as
\be
\label{eq:chainrule}
\frac{d f_{\x_i}}{d t} = \frac{\partial f_{\x_i}}{\partial t} + \vv_{\x_i} \cdot \frac{\partial f_{\x_i}}{\partial \xv} + \gv \cdot \frac{\partial f_{\x_i}}{\partial \vv_{\x_i}}
~,
\ee
where $\vv_{\x_i}$ is the MCP velocity. In the steady-state limit, the first term in \Eq{chainrule} can be dropped, $\partial f_{\x_i}  / \partial_t \simeq 0$. Using this in \Eq{Boltz2},
\be
\lim_{m_\x \ll m_N} \Delta f_{\x_i} \simeq - \frac{1}{n_N \, \sigma_T \, v_\text{rel}} \Big( \vv_{\x_i} \cdot \frac{\partial f_{\x_i}^0}{\partial \xv} + \gv \cdot \frac{\partial f_{\x_i}^0}{\partial \vv_{\x_i}} \Big) 
~,
\ee
where on the RHS we approximated the MCP phase space as Maxwell-Boltzmann $f_{\x_i}^0$. The MCP number current can then be evaluated as $\jv_\x \simeq \int d^3 \pv_{\x} \, (2 \pi)^{-3} \, \Delta f_{\x} \, \vv_{\x}$, which yields
\be
\label{eq:jxlight}
\lim_{m_\x \ll m_N} \jv_\x = \frac{1}{3 n_N} \, \Big( \frac{m_\x \, \gv}{T_\x} - \grad \Big) ~ \Big( n_\x \, \langle v_\text{rel} / \sigma_T \rangle \Big)
~.
\ee

Instead, in the large mass limit, we can approximate the MCP phase space as being described by a boosted Maxwellian (i.e., consisting solely of a monopole and dipole moment), as in Sec.~1.12 of Ref.~\cite{landaukinetics}. This has also been done in past work studying the effect of DM-nuclear collisions on cosmological observations~\cite{Dvorkin:2013cea,Boddy:2018wzy,Becker:2020hzj,Dvorkin:2020xga}. In particular, using the results provided in the appendix of Ref.~\cite{Dvorkin:2020xga}, the first integrated moment of the RHS of \Eq{Boltz1} is approximately 
\be
\lim_{m_\x \gg m_N} \int \frac{d^3 \pv_{\x_i}}{(2 \pi)^3} ~ \vv_{\x_i} \, \frac{d f_\x}{d t} = - \frac{m_N}{m_\x} \, n_\x \, n_N \int d^3 \vv_\text{rel} ~ f_\text{rel} (\vv_\text{rel} - \Vv_\x) \, v_\text{rel} \, \vv_\text{rel} \, \sigma_T
~,
\ee
where the MCP bulk velocity is $\Vv_\x = j_\x / n_\x$, the relative velocity distribution is defined as $f_\text{rel} (\vv_\text{rel}) = (2 \pi)^{-3/2} \, \bar{v}_0^{-3} ~ e^{- |\vv_\text{rel}|^2 / 2 \sigma_\text{rel}^2}$, and $\bar{v}_0 = \sqrt{T_\x / m_\x + T_N / m_N}\, $. In the limit that the bulk MCP velocity is much smaller than the relative particle velocity $V_\x \ll v_\text{rel}$, $f_\text{rel} (\vv_\text{rel} - \Vv_\x)$ can be expanded around small $V_\x$ in the expression above.\footnote{Although for incoming virialized mCPs $V_\x \sim V_\text{wind}$ is comparable to $\bar{v}_0 \sim V_\text{wind}$, we will use the linearized form of the momentum drag rate, which we expect to introduce only $\order{1}$ errors.} Once again using \Eq{chainrule} to evaluate the LHS, we have
\be
\lim_{m_\x \gg m_N} n_\x \, \partial_t \Vv_\x + n_\x \, (\Vv_\x \cdot \grad) \, \Vv_\x + \frac{\grad P_\x}{m_\x} - n_\x \, \gv = - \frac{m_N}{3 m_\x \bar{v}_0^2} \, n_\x \, n_N \, \langle \sigma_T \,  v_\text{rel}^3 \rangle \, \Vv_\x
~,
\ee
In the tight-coupling limit, the first two terms on the LHS are smaller than the term on the RHS in the expression above, such that the MCP number current is approximated as
\be
\label{eq:jxheavy}
\lim_{m_\x \gg m_N} \jv_\x = \frac{3 m_\x \, \bar{v}_0^2 \, n_\x}{m_N \, n_N \, \langle \sigma_T v_\text{rel}^3 \rangle} \, \Big( \gv - \frac{\grad P_\x}{m_\x \, n_\x} \Big)
~.
\ee
In the limit that the temperature is spatially uniform and the MCP pressure is that of an ideal gas, i.e., $P_\x \simeq T_\x \, n_\x$, Eqs.~\ref{eq:jxlight} and \ref{eq:jxheavy} give Eqs.~$\ref{eq:convdiff1} - \ref{eq:Gamma1}$.

\end{appendices}

\phantomsection

\vskip24pt

\addtocontents{toc}{\protect\vskip24pt}
\addcontentsline{toc}{section}{References}

\makeatletter

\interlinepenalty=10000

{\linespread{1.05}
\bibliographystyle{utphys}
\bibliography{Terrestrial_Abundance}

\providecommand{\href}[2]{#2}\begingroup\raggedright\begin{thebibliography}{10}

\bibitem{Alexander:2016aln}
J.~Alexander {\em et~al.}, ``{Dark Sectors 2016 Workshop: Community Report},''
\newblock 8, 2016.
\newblock \href{http://arxiv.org/abs/1608.08632}{{\ttfamily arXiv:1608.08632
  [hep-ph]}}.

\bibitem{Battaglieri:2017aum}
M.~Battaglieri {\em et~al.}, ``{US Cosmic Visions: New Ideas in Dark Matter
  2017: Community Report},'' in {\em {U.S. Cosmic Visions: New Ideas in Dark
  Matter}}.
\newblock 7, 2017.
\newblock \href{http://arxiv.org/abs/1707.04591}{{\ttfamily arXiv:1707.04591
  [hep-ph]}}.

\bibitem{Dubovsky:2003yn}
S.~L. Dubovsky, D.~S. Gorbunov, and G.~I. Rubtsov, ``{Narrowing the window for
  millicharged particles by CMB anisotropy},''
  \href{http://dx.doi.org/10.1134/1.1675909}{{\em JETP Lett.} {\bfseries 79}
  (2004) 1--5}, \href{http://arxiv.org/abs/hep-ph/0311189}{{\ttfamily
  arXiv:hep-ph/0311189}}.

\bibitem{dePutter:2018xte}
R.~de~Putter, O.~Dor\'e, J.~Gleyzes, D.~Green, and J.~Meyers, ``{Dark Matter
  Interactions, Helium, and the Cosmic Microwave Background},''
  \href{http://dx.doi.org/10.1103/PhysRevLett.122.041301}{{\em Phys. Rev.
  Lett.} {\bfseries 122} no.~4, (2019) 041301},
  \href{http://arxiv.org/abs/1805.11616}{{\ttfamily arXiv:1805.11616
  [astro-ph.CO]}}.

\bibitem{Kovetz:2018zan}
E.~D. Kovetz, V.~Poulin, V.~Gluscevic, K.~K. Boddy, R.~Barkana, and
  M.~Kamionkowski, ``{Tighter limits on dark matter explanations of the
  anomalous EDGES 21 cm signal},''
  \href{http://dx.doi.org/10.1103/PhysRevD.98.103529}{{\em Phys. Rev. D}
  {\bfseries 98} no.~10, (2018) 103529},
  \href{http://arxiv.org/abs/1807.11482}{{\ttfamily arXiv:1807.11482
  [astro-ph.CO]}}.

\bibitem{Buen-Abad:2021mvc}
M.~A. Buen-Abad, R.~Essig, D.~McKeen, and Y.-M. Zhong, ``{Cosmological
  constraints on dark matter interactions with ordinary matter},''
  \href{http://dx.doi.org/10.1016/j.physrep.2022.02.006}{{\em Phys. Rept.}
  {\bfseries 961} (2022) 1--35},
  \href{http://arxiv.org/abs/2107.12377}{{\ttfamily arXiv:2107.12377
  [astro-ph.CO]}}.

\bibitem{Lee:1977ua}
B.~W. Lee and S.~Weinberg, ``{Cosmological Lower Bound on Heavy Neutrino
  Masses},'' \href{http://dx.doi.org/10.1103/PhysRevLett.39.165}{{\em Phys.
  Rev. Lett.} {\bfseries 39} (1977) 165--168}.

\bibitem{Neufeld:2018slx}
D.~A. Neufeld, G.~R. Farrar, and C.~F. McKee, ``{Dark Matter that Interacts
  with Baryons: Density Distribution within the Earth and New Constraints on
  the Interaction Cross-section},''
  \href{http://dx.doi.org/10.3847/1538-4357/aad6a4}{{\em Astrophys. J.}
  {\bfseries 866} no.~2, (2018) 111},
  \href{http://arxiv.org/abs/1805.08794}{{\ttfamily arXiv:1805.08794
  [astro-ph.CO]}}.

\bibitem{Wallemacq:2013hsa}
Q.~Wallemacq, ``{Milli-interacting Dark Matter},''
  \href{http://dx.doi.org/10.1103/PhysRevD.88.063516}{{\em Phys. Rev. D}
  {\bfseries 88} no.~6, (2013) 063516},
  \href{http://arxiv.org/abs/1307.7623}{{\ttfamily arXiv:1307.7623
  [astro-ph.CO]}}.

\bibitem{Wallemacq:2014lba}
Q.~Wallemacq, ``{Milli-interacting dark matter interpretation of the
  direct-search experiments},''
  \href{http://dx.doi.org/10.1155/2014/525208}{{\em Adv. High Energy Phys.}
  {\bfseries 2014} (2014) 525208},
  \href{http://arxiv.org/abs/1401.5243}{{\ttfamily arXiv:1401.5243 [hep-ph]}}.

\bibitem{Wallemacq:2014sta}
Q.~Wallemacq and J.-R. Cudell, ``{Dark antiatoms can explain DAMA},''
  \href{http://dx.doi.org/10.1088/1475-7516/2015/02/011}{{\em JCAP} {\bfseries
  02} (2015) 011}, \href{http://arxiv.org/abs/1411.3178}{{\ttfamily
  arXiv:1411.3178 [hep-ph]}}.

\bibitem{Laletin:2019qca}
M.~Laletin and J.-R. Cudell, ``{Strongly interacting dark matter and the DAMA
  signal},'' \href{http://dx.doi.org/10.1088/1475-7516/2019/07/010}{{\em JCAP}
  {\bfseries 07} (2019) 010}, \href{http://arxiv.org/abs/1903.04637}{{\ttfamily
  arXiv:1903.04637 [hep-ph]}}.

\bibitem{Pospelov:2019vuf}
M.~Pospelov, S.~Rajendran, and H.~Ramani, ``{Metastable Nuclear Isomers as Dark
  Matter Accelerators},''
  \href{http://dx.doi.org/10.1103/PhysRevD.101.055001}{{\em Phys. Rev. D}
  {\bfseries 101} no.~5, (2020) 055001},
  \href{http://arxiv.org/abs/1907.00011}{{\ttfamily arXiv:1907.00011
  [hep-ph]}}.

\bibitem{Pospelov:2020ktu}
M.~Pospelov and H.~Ramani, ``{Earth-bound millicharge relics},''
  \href{http://dx.doi.org/10.1103/PhysRevD.103.115031}{{\em Phys. Rev. D}
  {\bfseries 103} no.~11, (2021) 115031},
  \href{http://arxiv.org/abs/2012.03957}{{\ttfamily arXiv:2012.03957
  [hep-ph]}}.

\bibitem{Leane:2022hkk}
R.~K. Leane and J.~Smirnov, ``{Floating Dark Matter in Celestial Bodies},''
  \href{http://arxiv.org/abs/2209.09834}{{\ttfamily arXiv:2209.09834
  [hep-ph]}}.

\bibitem{Knapen:2017xzo}
S.~Knapen, T.~Lin, and K.~M. Zurek, ``{Light Dark Matter: Models and
  Constraints},'' \href{http://dx.doi.org/10.1103/PhysRevD.96.115021}{{\em
  Phys. Rev. D} {\bfseries 96} no.~11, (2017) 115021},
  \href{http://arxiv.org/abs/1709.07882}{{\ttfamily arXiv:1709.07882
  [hep-ph]}}.

\bibitem{Holdom:1985ag}
B.~Holdom, ``{Two U(1)'s and Epsilon Charge Shifts},''
  \href{http://dx.doi.org/10.1016/0370-2693(86)91377-8}{{\em Phys. Lett. B}
  {\bfseries 166} (1986) 196--198}.

\bibitem{Emken:2019tni}
T.~Emken, R.~Essig, C.~Kouvaris, and M.~Sholapurkar, ``{Direct Detection of
  Strongly Interacting Sub-GeV Dark Matter via Electron Recoils},''
  \href{http://dx.doi.org/10.1088/1475-7516/2019/09/070}{{\em JCAP} {\bfseries
  09} (2019) 070}, \href{http://arxiv.org/abs/1905.06348}{{\ttfamily
  arXiv:1905.06348 [hep-ph]}}.

\bibitem{Gould:1989hm}
A.~Gould and G.~Raffelt, ``{THERMAL CONDUCTION BY MASSIVE PARTICLES},''
  \href{http://dx.doi.org/10.1086/168568}{{\em Astrophys. J.} {\bfseries 352}
  (1990) 654}.

\bibitem{Dziewonski:1981xy}
A.~M. Dziewonski and D.~L. Anderson, ``{Preliminary reference earth model},''
  \href{http://dx.doi.org/10.1016/0031-9201(81)90046-7}{{\em Phys. Earth
  Planet. Interiors} {\bfseries 25} (1981) 297--356}.

\bibitem{atmosphere}
J.~T. Emmert, D.~P. Drob, J.~M. Picone, D.~E. Siskind, M.~Jones~Jr., M.~G.
  Mlynczak, P.~F. Bernath, X.~Chu, E.~Doornbos, B.~Funke, L.~P. Goncharenko,
  M.~E. Hervig, M.~J. Schwartz, P.~E. Sheese, F.~Vargas, B.~P. Williams, and
  T.~Yuan, ``Nrlmsis 2.0: A whole-atmosphere empirical model of temperature and
  neutral species densities,''
  \href{http://dx.doi.org/https://doi.org/10.1029/2020EA001321}{{\em Earth and
  Space Science} {\bfseries 8} no.~3, (2021) e2020EA001321},
  \href{http://arxiv.org/abs/https://agupubs.onlinelibrary.wiley.com/doi/pdf/10.1029/2020EA001321}{{\ttfamily
  https://agupubs.onlinelibrary.wiley.com/doi/pdf/10.1029/2020EA001321}}.
  \url{https://agupubs.onlinelibrary.wiley.com/doi/abs/10.1029/2020EA001321}.
  e2020EA001321 2020EA001321.

\bibitem{Tulin:2013teo}
S.~Tulin, H.-B. Yu, and K.~M. Zurek, ``{Beyond Collisionless Dark Matter:
  Particle Physics Dynamics for Dark Matter Halo Structure},''
  \href{http://dx.doi.org/10.1103/PhysRevD.87.115007}{{\em Phys. Rev. D}
  {\bfseries 87} no.~11, (2013) 115007},
  \href{http://arxiv.org/abs/1302.3898}{{\ttfamily arXiv:1302.3898 [hep-ph]}}.

\bibitem{Colquhoun:2020adl}
B.~Colquhoun, S.~Heeba, F.~Kahlhoefer, L.~Sagunski, and S.~Tulin,
  ``{Semiclassical regime for dark matter self-interactions},''
  \href{http://dx.doi.org/10.1103/PhysRevD.103.035006}{{\em Phys. Rev. D}
  {\bfseries 103} no.~3, (2021) 035006},
  \href{http://arxiv.org/abs/2011.04679}{{\ttfamily arXiv:2011.04679
  [hep-ph]}}.

\bibitem{Dvorkin:2013cea}
C.~Dvorkin, K.~Blum, and M.~Kamionkowski, ``{Constraining Dark Matter-Baryon
  Scattering with Linear Cosmology},''
  \href{http://dx.doi.org/10.1103/PhysRevD.89.023519}{{\em Phys. Rev. D}
  {\bfseries 89} no.~2, (2014) 023519},
  \href{http://arxiv.org/abs/1311.2937}{{\ttfamily arXiv:1311.2937
  [astro-ph.CO]}}.

\bibitem{Boddy:2018wzy}
K.~K. Boddy, V.~Gluscevic, V.~Poulin, E.~D. Kovetz, M.~Kamionkowski, and
  R.~Barkana, ``{Critical assessment of CMB limits on dark matter-baryon
  scattering: New treatment of the relative bulk velocity},''
  \href{http://dx.doi.org/10.1103/PhysRevD.98.123506}{{\em Phys. Rev. D}
  {\bfseries 98} no.~12, (2018) 123506},
  \href{http://arxiv.org/abs/1808.00001}{{\ttfamily arXiv:1808.00001
  [astro-ph.CO]}}.

\bibitem{Becker:2020hzj}
N.~Becker, D.~C. Hooper, F.~Kahlhoefer, J.~Lesgourgues, and N.~Sch\"oneberg,
  ``{Cosmological constraints on multi-interacting dark matter},''
  \href{http://dx.doi.org/10.1088/1475-7516/2021/02/019}{{\em JCAP} {\bfseries
  02} (2021) 019}, \href{http://arxiv.org/abs/2010.04074}{{\ttfamily
  arXiv:2010.04074 [astro-ph.CO]}}.

\bibitem{Dvorkin:2020xga}
C.~Dvorkin, T.~Lin, and K.~Schutz, ``{Cosmology of Sub-MeV Dark Matter
  Freeze-In},'' \href{http://dx.doi.org/10.1103/PhysRevLett.127.111301}{{\em
  Phys. Rev. Lett.} {\bfseries 127} no.~11, (2021) 111301},
  \href{http://arxiv.org/abs/2011.08186}{{\ttfamily arXiv:2011.08186
  [astro-ph.CO]}}.

\bibitem{Chuzhoy:2008zy}
L.~Chuzhoy and E.~W. Kolb, ``{Reopening the window on charged dark matter},''
  \href{http://dx.doi.org/10.1088/1475-7516/2009/07/014}{{\em JCAP} {\bfseries
  07} (2009) 014}, \href{http://arxiv.org/abs/0809.0436}{{\ttfamily
  arXiv:0809.0436 [astro-ph]}}.

\bibitem{Sanchez-Salcedo:2010gfa}
F.~J. Sanchez-Salcedo, E.~Martinez-Gomez, and J.~Magana, ``{On the fraction of
  dark matter in charged massive particles (CHAMPs)},''
  \href{http://dx.doi.org/10.1088/1475-7516/2010/02/031}{{\em JCAP} {\bfseries
  02} (2010) 031}, \href{http://arxiv.org/abs/1002.3145}{{\ttfamily
  arXiv:1002.3145 [astro-ph.CO]}}.

\bibitem{Dunsky:2018mqs}
D.~Dunsky, L.~J. Hall, and K.~Harigaya, ``{CHAMP Cosmic Rays},''
  \href{http://dx.doi.org/10.1088/1475-7516/2019/07/015}{{\em JCAP} {\bfseries
  07} (2019) 015}, \href{http://arxiv.org/abs/1812.11116}{{\ttfamily
  arXiv:1812.11116 [astro-ph.HE]}}.

\bibitem{Stebbins:2019xjr}
A.~Stebbins and G.~Krnjaic, ``{New Limits on Charged Dark Matter from
  Large-Scale Coherent Magnetic Fields},''
  \href{http://dx.doi.org/10.1088/1475-7516/2019/12/003}{{\em JCAP} {\bfseries
  12} (2019) 003}, \href{http://arxiv.org/abs/1908.05275}{{\ttfamily
  arXiv:1908.05275 [astro-ph.CO]}}.

\bibitem{Li:2020wyl}
J.-T. Li and T.~Lin, ``{Dynamics of millicharged dark matter in supernova
  remnants},'' \href{http://dx.doi.org/10.1103/PhysRevD.101.103034}{{\em Phys.
  Rev. D} {\bfseries 101} no.~10, (2020) 103034},
  \href{http://arxiv.org/abs/2002.04625}{{\ttfamily arXiv:2002.04625
  [astro-ph.CO]}}.

\bibitem{Feynman:1494701}
R.~P. Feynman, R.~B. Leighton, and M.~Sands, {\em {The Feynman lectures on
  physics; New millennium ed.}}
\newblock Basic Books, New York, NY, 2010.
\newblock \url{https://cds.cern.ch/record/1494701}.
\newblock Originally published 1963-1965.

\bibitem{Harrison2004}
R.~G. Harrison, ``The global atmospheric electrical circuit and climate,''
  \href{http://dx.doi.org/10.1007/s10712-004-5439-8}{{\em Surveys in
  Geophysics} {\bfseries 25} no.~5, (2004) 441--484}.
  \url{https://doi.org/10.1007/s10712-004-5439-8}.

\bibitem{SIINGH200791}
D.~Siingh, V.~Gopalakrishnan, R.~Singh, A.~Kamra, S.~Singh, V.~Pant, R.~Singh,
  and A.~Singh, ``The atmospheric global electric circuit: An overview,''
  \href{http://dx.doi.org/https://doi.org/10.1016/j.atmosres.2006.05.005}{{\em
  Atmospheric Research} {\bfseries 84} no.~2, (2007) 91--110}.
  \url{https://www.sciencedirect.com/science/article/pii/S0169809506001797}.

\bibitem{An:2013yfc}
H.~An, M.~Pospelov, and J.~Pradler, ``{New stellar constraints on dark
  photons},'' \href{http://dx.doi.org/10.1016/j.physletb.2013.07.008}{{\em
  Phys. Lett. B} {\bfseries 725} (2013) 190--195},
  \href{http://arxiv.org/abs/1302.3884}{{\ttfamily arXiv:1302.3884 [hep-ph]}}.

\bibitem{Redondo:2013lna}
J.~Redondo and G.~Raffelt, ``{Solar constraints on hidden photons
  re-visited},'' \href{http://dx.doi.org/10.1088/1475-7516/2013/08/034}{{\em
  JCAP} {\bfseries 08} (2013) 034},
  \href{http://arxiv.org/abs/1305.2920}{{\ttfamily arXiv:1305.2920 [hep-ph]}}.

\bibitem{Williams:1971ms}
E.~R. Williams, J.~E. Faller, and H.~A. Hill, ``{New experimental test of
  Coulomb's law: A Laboratory upper limit on the photon rest mass},''
  \href{http://dx.doi.org/10.1103/PhysRevLett.26.721}{{\em Phys. Rev. Lett.}
  {\bfseries 26} (1971) 721--724}.

\bibitem{Mirizzi:2009iz}
A.~Mirizzi, J.~Redondo, and G.~Sigl, ``{Microwave Background Constraints on
  Mixing of Photons with Hidden Photons},''
  \href{http://dx.doi.org/10.1088/1475-7516/2009/03/026}{{\em JCAP} {\bfseries
  03} (2009) 026}, \href{http://arxiv.org/abs/0901.0014}{{\ttfamily
  arXiv:0901.0014 [hep-ph]}}.

\bibitem{Caputo:2020bdy}
A.~Caputo, H.~Liu, S.~Mishra-Sharma, and J.~T. Ruderman, ``{Dark Photon
  Oscillations in Our Inhomogeneous Universe},''
  \href{http://dx.doi.org/10.1103/PhysRevLett.125.221303}{{\em Phys. Rev.
  Lett.} {\bfseries 125} no.~22, (2020) 221303},
  \href{http://arxiv.org/abs/2002.05165}{{\ttfamily arXiv:2002.05165
  [astro-ph.CO]}}.

\bibitem{landaukinetics}
E.~M. Lifshitz and L.~P. Pitaevskii, {\em {Landau and Lifshitz Course of
  Theoretical Physics, Volume 10: Physical Kinetics}}.
\newblock 1981.

\bibitem{Dodelson:2003ft}
S.~Dodelson, {\em {Modern Cosmology}}.
\newblock Academic Press, Amsterdam, 2003.

\bibitem{redner_2001}
S.~Redner, \href{http://dx.doi.org/10.1017/CBO9780511606014}{{\em A Guide to
  First-Passage Processes}}.
\newblock Cambridge University Press, 2001.

\bibitem{Bramante:2022pmn}
J.~Bramante, J.~Kumar, G.~Mohlabeng, N.~Raj, and N.~Song, ``{Light Dark Matter
  Accumulating in Terrestrial Planets: Nuclear Scattering},''
  \href{http://arxiv.org/abs/2210.01812}{{\ttfamily arXiv:2210.01812
  [hep-ph]}}.

\bibitem{Catling:2017}
D.~C. Catling and J.~F. Kasting, {\em {Atmospheric evolution on inhabited and
  lifeless worlds}}.
\newblock 2017.

\bibitem{MOLINI20111841}
A.~Molini, P.~Talkner, G.~Katul, and A.~Porporato, ``First passage time
  statistics of brownian motion with purely time dependent drift and
  diffusion,''
  \href{http://dx.doi.org/https://doi.org/10.1016/j.physa.2011.01.024}{{\em
  Physica A: Statistical Mechanics and its Applications} {\bfseries 390}
  no.~11, (2011) 1841--1852}.
  \url{https://www.sciencedirect.com/science/article/pii/S0378437111000884}.

\bibitem{Prinz:1998ua}
A.~A. Prinz {\em et~al.}, ``{Search for millicharged particles at SLAC},''
  \href{http://dx.doi.org/10.1103/PhysRevLett.81.1175}{{\em Phys. Rev. Lett.}
  {\bfseries 81} (1998) 1175--1178},
  \href{http://arxiv.org/abs/hep-ex/9804008}{{\ttfamily arXiv:hep-ex/9804008}}.

\bibitem{ArgoNeuT:2019ckq}
{\bfseries ArgoNeuT} Collaboration, R.~Acciarri {\em et~al.}, ``{Improved
  Limits on Millicharged Particles Using the ArgoNeuT Experiment at
  Fermilab},'' \href{http://dx.doi.org/10.1103/PhysRevLett.124.131801}{{\em
  Phys. Rev. Lett.} {\bfseries 124} no.~13, (2020) 131801},
  \href{http://arxiv.org/abs/1911.07996}{{\ttfamily arXiv:1911.07996
  [hep-ex]}}.

\bibitem{milliQan:2021lne}
{\bfseries milliQan} Collaboration, A.~Ball {\em et~al.}, ``{Sensitivity to
  millicharged particles in future proton-proton collisions at the LHC with the
  milliQan detector},''
  \href{http://dx.doi.org/10.1103/PhysRevD.104.032002}{{\em Phys. Rev. D}
  {\bfseries 104} no.~3, (2021) 032002},
  \href{http://arxiv.org/abs/2104.07151}{{\ttfamily arXiv:2104.07151
  [hep-ex]}}.

\bibitem{Davidson:2000hf}
S.~Davidson, S.~Hannestad, and G.~Raffelt, ``{Updated bounds on millicharged
  particles},'' \href{http://dx.doi.org/10.1088/1126-6708/2000/05/003}{{\em
  JHEP} {\bfseries 05} (2000) 003},
  \href{http://arxiv.org/abs/hep-ph/0001179}{{\ttfamily arXiv:hep-ph/0001179}}.

\bibitem{ArguellesDelgado:2021lek}
C.~A. Arg\"uelles~Delgado, K.~J. Kelly, and V.~Mu\~noz Albornoz,
  ``{Millicharged particles from the heavens: single- and multiple-scattering
  signatures},'' \href{http://dx.doi.org/10.1007/JHEP11(2021)099}{{\em JHEP}
  {\bfseries 11} (2021) 099}, \href{http://arxiv.org/abs/2104.13924}{{\ttfamily
  arXiv:2104.13924 [hep-ph]}}.

\bibitem{Chang:2018rso}
J.~H. Chang, R.~Essig, and S.~D. McDermott, ``{Supernova 1987A Constraints on
  Sub-GeV Dark Sectors, Millicharged Particles, the QCD Axion, and an
  Axion-like Particle},'' \href{http://dx.doi.org/10.1007/JHEP09(2018)051}{{\em
  JHEP} {\bfseries 09} (2018) 051},
  \href{http://arxiv.org/abs/1803.00993}{{\ttfamily arXiv:1803.00993
  [hep-ph]}}.

\bibitem{Berlin:2021zbv}
A.~Berlin, H.~Liu, M.~Pospelov, and H.~Ramani, ``{Low-energy signals from the
  formation of dark-matter\textendash{}nucleus bound states},''
  \href{http://dx.doi.org/10.1103/PhysRevD.105.095028}{{\em Phys. Rev. D}
  {\bfseries 105} no.~9, (2022) 095028},
  \href{http://arxiv.org/abs/2110.06217}{{\ttfamily arXiv:2110.06217
  [hep-ph]}}.

\bibitem{Plestid:2020kdm}
R.~Plestid, V.~Takhistov, Y.-D. Tsai, T.~Bringmann, A.~Kusenko, and
  M.~Pospelov, ``{New Constraints on Millicharged Particles from Cosmic-ray
  Production},'' \href{http://dx.doi.org/10.1103/PhysRevD.102.115032}{{\em
  Phys. Rev. D} {\bfseries 102} (2020) 115032},
  \href{http://arxiv.org/abs/2002.11732}{{\ttfamily arXiv:2002.11732
  [hep-ph]}}.

\bibitem{Harnik:2020ugb}
R.~Harnik, R.~Plestid, M.~Pospelov, and H.~Ramani, ``{Millicharged cosmic rays
  and low recoil detectors},''
  \href{http://dx.doi.org/10.1103/PhysRevD.103.075029}{{\em Phys. Rev. D}
  {\bfseries 103} no.~7, (2021) 075029},
  \href{http://arxiv.org/abs/2010.11190}{{\ttfamily arXiv:2010.11190
  [hep-ph]}}.

\bibitem{Budker:2021quh}
D.~Budker, P.~W. Graham, H.~Ramani, F.~Schmidt-Kaler, C.~Smorra, and S.~Ulmer,
  ``{Millicharged Dark Matter Detection with Ion Traps},''
  \href{http://dx.doi.org/10.1103/PRXQuantum.3.010330}{{\em PRX Quantum}
  {\bfseries 3} no.~1, (2022) 010330},
  \href{http://arxiv.org/abs/2108.05283}{{\ttfamily arXiv:2108.05283
  [hep-ph]}}.

\end{thebibliography}\endgroup
}
\makeatother

\end{document}